\documentclass[final,5p,times,twocolumn]{elsarticle}

\usepackage{amsmath,amssymb}
\usepackage{mathrsfs}

\usepackage{booktabs}
\usepackage[colorlinks=true]{hyperref}
\usepackage[tight, nice]{units}
\usepackage{multirow}
\usepackage{rotating}

\journal{Nucl. Instrum. Methods Phys. Res. A} %2021-07-14 revised version

\begin{document}
\begin{frontmatter}
  \title{Operation and characterization of a windowless gas jet target in high-intensity electron beams}
  %##########################################################################################################
  \author[KPH]{B.S.~Schlimme\corref{CorrespondingAuthor}} \ead{schlimme@uni-mainz.de}
  \author[KPH]{S.~Aulenbacher\fnref{PhD}}
  \author[WWU]{P.~Brand\fnref{PhD,Master}} 
  \author[KPH]{M.~Littich\fnref{PhD}}
  \author[MIT]{Y.~Wang\fnref{PhD}}
  \author[KPH,PRISMA,HIM]{P.~Achenbach}
  \author[HISKP]{M.~Ball}
  \author[StonyBrook,RBRC]{J.C.~Bernauer}
  \author[KPH]{M.~Biroth}
  \author[WWU]{D.~Bonaventura}
  \author[ZAGREB]{D.~Bosnar}
  \author[KPH]{S.~Caiazza} 
  \author[KPH,HIM]{M.~Christmann} 
  \author[StonyBrook]{E.~Cline} 
  \author[KPH,PRISMA,HIM]{A.~Denig} 
  \author[KPH]{M.O.~Distler}
  \author[KPH,PRISMA]{L.~Doria}
  \author[KPH]{P.~Eckert}
  \author[KPH]{A.~Esser}
  \author[MIT]{I.~Fri\v{s}\v{c}i\'c}
  \author[KPH]{S.~Gagneur\fnref{Bachelor}}
  \author[KPH]{J.~Geimer} 
  \author[WWU]{S.~Grieser} 
  \author[KPH]{P.~G\"ulker}
  \author[KPH]{P.~Herrmann}
  \author[KPH]{M.~Hoek}
  \author[KPH]{S.~Kegel}
  \author[MITREC]{J.~Kelsey}
  \author[KPH]{P.~Klag}
  \author[WWU]{A.~Khoukaz} 
  \author[Hampton]{M.~Kohl} 
  \author[JSI]{T.~Kolar\fnref{TK}}
  \author[KPH]{M.~Lau{\ss}}
  \author[WWU]{L.~Le{\ss}mann}
  \author[KPH]{S.~Lunkenheimer}
  \author[JSI]{J.~Marekovi\v{c}}
  \author[KPH]{D.~Markus} 
  \author[HIM]{M.~Mauch} 
  \author[KPH,PRISMA]{H.~Merkel} 
  \author[KPH,ULjubljana,JSI]{M.~Mihovilovi\v{c}} 
  \author[MIT]{R.G.~Milner}
  \author[KPH]{J.~M\"uller}
  \author[KPH]{U.~M\"uller}
  \author[JSI]{T.~Petrovi\v{c}}
  \author[KPH]{J.~Pochodzalla}
  \author[KPH]{J.~Rausch}
  \author[KPH]{J.~Schlaadt}
  \author[KPH]{H.~Sch\"urg}
  \author[KPH,PRISMA]{C.~Sfienti}
  \author[ULjubljana,JSI]{S.~\v{S}irca}
  \author[KPH]{R.~Spreckels} 
  \author[KPH]{S.~Stengel}
  \author[KPH]{Y.~St\"ottinger} 
  \author[KPH]{C.~Szyszka}
  \author[KPH]{M.~Thiel}
  \author[WWU]{S.~Vestrick}
  \author[MITREC]{C.~Vidal}
  \author{\\for the A1 and MAGIX Collaborations}
  \cortext[CorrespondingAuthor]{Corresponding author.}
  \fntext[PhD]{Part of doctoral thesis.}
  \fntext[Master]{Part of master thesis.}
  \fntext[Bachelor]{Part of bachelor thesis.}
  \fntext[TK]{Present address: School of Physics and Astronomy, Tel Aviv University, Tel Aviv 69978, Israel}
  \address[KPH]{Institut f\"ur Kernphysik, Johannes Gutenberg-Universit\"at, D-55099 Mainz, Germany}
  \address[WWU]{Institut f\"ur Kernphysik, Westf\"alische Wilhelms-Universit\"at, D-48149 M\"unster, Germany}
  \address[PRISMA]{PRISMA$^+$ Cluster of Excellence, Johannes Gutenberg-Universit\"at, D-55099 Mainz, Germany}
  \address[HIM]{Helmholtz Institute Mainz, GSI Helmholtzzentrum für Schwerionenforschung, Darmstadt, Johannes Gutenberg-Universit\"at, D-55099 Mainz, Germany}
  \address[HISKP]{Helmholtz-Institut f\"ur Strahlen- und Kernphysik, Rheinische Friedrich-Wilhelms-Universit\"at, D-53115 Bonn, Germany}  
  \address[ZAGREB]{Department of Physics, University of Zagreb, HR-10002 Zagreb, Croatia}
  \address[JSI]{Jo\v{z}ef Stefan Institute, SI-1000 Ljubljana, Slovenia}
  \address[ULjubljana]{Faculty of Mathematics and Physics, University of Ljubljana, SI-1000 Ljubljana, Slovenia}
  \address[MIT]{Laboratory for Nuclear Science, Massachusetts Institute of Technology, Cambridge, Massachusetts 02139, USA}
  \address[MITREC]{MIT Bates Research and Engineering Center, Middleton, Massachusetts, 01949, USA}
  \address[StonyBrook]{Center for Frontiers in Nuclear Science, Department of Physics and Astronomy, Stony Brook University, New York 11794, USA}
  \address[RBRC]{RIKEN BNL Research Center,  Brookhaven National Laboratory, Upton, NY 11973, USA}
  \address[Hampton]{Department of Physics, Hampton University, Hampton, Virginia 23668, USA}
  \hypersetup{pdfauthor={MAGIX and A1 Collaborations}}
  %
  %##########################################################################################################
  \begin{abstract} \addcontentsline{toc}{section}{Abstract}
  %##########################################################################################################
    A cryogenic supersonic gas jet target was developed for the MAGIX
    experiment at the high-intensity electron accelerator MESA.  It
    will be operated as an internal, windowless target in the
    energy-recovering recirculation arc of the accelerator with
    different target gases, e.g., hydrogen, deuterium, helium, oxygen,
    argon, or xenon.  Detailed studies have been carried out at the
    existing A1 multi-spectrometer facility at the electron
    accelerator MAMI.  This paper focuses on the developed handling
    procedures and diagnostic tools, and on the performance of the gas
    jet target under beam conditions. Considering the special features
    of this type of target, it proves to be well suited for a new
    generation of high-precision electron scattering experiments at
    high-intensity electron accelerators.
  \end{abstract}
  %##########################################################################################################
\end{frontmatter}
%
%\clearpage\tableofcontents
%
%%bss \linenumbers
%
%############################################################################################################
\section{Introduction}
\label{sec:introduction}
%############################################################################################################
Electron scattering is a very powerful tool for studying the structure
of nucleons and atomic nuclei~\cite{Walecka:2001}.  Investigations of
form factors, polarizabilities, and of the excitation spectra of
hadrons are being performed to reach a detailed description of
low-energy phenomena on the basis of quantum chromodynamics.  The use
of electron beams as a probe provides a clean reaction mechanism and
allows for a high precision, which often is limited by technical
factors rather than by the interpretation of the measured observables
in physical terms.  When, for instance, the systematic uncertainty of
a cross section measurement should be reduced below the 1-percent
level, a typical limiting factor is related to the considerable
thickness of the target material resulting in energy-loss straggling
and multiple small-angle scattering. In case of targets with gas
enclosed in cells, irreducible backgrounds can be introduced by the
interaction of the electron beam with surrounding structures such as
cell walls or foils.  Therefore, the use of very thin and windowless
targets is highly desirable (see for instance
\cite{Voitsekhovsky:1979}), but poses the challenge of relatively
small luminosities and count rates for beam currents available at
conventional electron accelerators.  A new generation of
high-intensity electron accelerators will allow to overcome this
shortcoming.

One of these new accelerators will be MESA (Mainz Energy-Recovering
Superconducting Accelerator), an Energy Recovery Linac (ERL),
currently under construction in Mainz,
Germany~\cite{Hug:2016,Hug:2020}.  In the ERL mode, the beam is, after
passing the target, guided back into the superconducting
radio-frequency system with a phase shift of 180$^\circ$, so that the
electrons transfer most of their energy back to the accelerating
structures.  This scheme provides a power-efficient beam acceleration
with a maximum beam energy of $E = \unit[105]{MeV}$ and a maximum beam
current of $I = \unit[1000]{\mu A}$ or higher.

% Dhaun-rule
% ==========
The maximum acceptable multiple scattering contribution for the ERL
mode limits the areal thickness of the target to less than
$\rho_\text{areal}\approx \unit[10^{20}]{atoms/cm^2 \cdot
  \nicefrac{1}{Z^2}}$ with $Z$ being the atomic number of the target
material.  In addition, the destructive heat deposition of high beam
currents makes it unfeasible to have the beam passing through target
cell walls.

\begin{figure}[htbp]
  \centering
  \includegraphics[width=0.8\columnwidth]{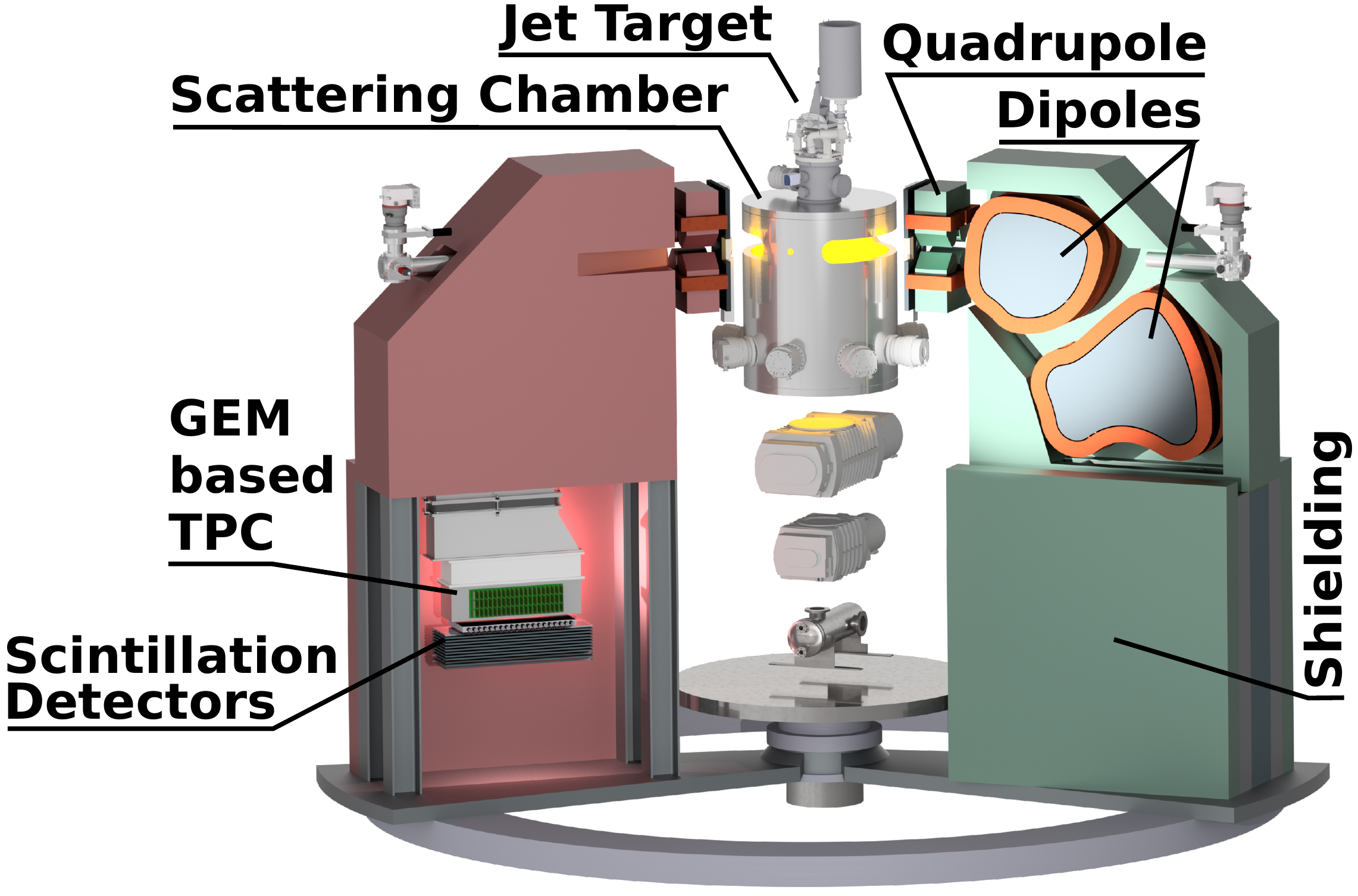}
  \caption{Overview of the planned MAGIX spectrometer setup at the ERL
    arc of the MESA accelerator. The gas jet target will be mounted in
    a scattering chamber with a pumping system connected to it. The
    two magnetic spectrometers in quadrupole-dipole-dipole magnetic
    configuration can rotate around the target. Their focal planes
    will be equipped with TPCs for tracking, and scintillation
    detectors for trigger, timing, and particle identification
    purposes.
    %The radiation shielding reduces the background flux of particles
    %in the detector houses.
    }
  \label{fig:MagixOverview}
\end{figure}

The MAinz Gas Injection target eXperiment MAGIX will employ a
multi-purpose spectrometer system in the ERL arc of MESA and a gas jet
target at its center. The two high-resolution magnetic spectrometers
will be used for the detection of scattered electrons and produced
particles. Their focal planes will be equipped with TPCs (time
projection chambers) with GEM (gas electron multiplier) readout for
tracking~\cite{Caiazza:2020}, and with a dedicated trigger and veto
system consisting of a segmented layer of plastic scintillation
detectors and a flexible number of additional layers and lead
absorbers. Figure~\ref{fig:MagixOverview} shows a schematic of the
planned MAGIX setup.

While the MESA accelerator and the experimental site for the MAGIX
setup are currently under construction, the gas jet target was already
developed and tested at the University of
M\"unster~\cite{Grieser:2018}. In 2017 it could be installed and
successfully operated at the existing A1 multi-spectrometer setup at
the Mainz Microtron (MAMI) electron accelerator in Mainz, Germany.

The optimization of the target system was crucial to realize its full
potential and to open the avenue for innovative precision
experiments. The combination of a low beam energy, a thin and nearly
point-like target and a high beam current is ideally suited to perform
novel precision experiments in nuclear, hadron, and particle physics.
In addition, the target allows for the detection of low-energy recoil
nuclei inside the scattering chamber.

One major motivation is the study of form factors of nucleons and
nuclei in elastic electron scattering at low momentum transfers $Q^2$
reaching down to below $\unit[10^{-4}]{GeV^2}$, allowing, for
instance, a precise determination of the proton charge radius. Key
element for the success of this type of experiments will be the
absence of a target cell and, compared to a liquid target, the very
low target density, minimizing the irreducible background and effects
from energy loss and multiple scattering~\cite{Bernauer:2010,
  Mihovilovic:2016,Mihovilovic:2021}.

With the MAGIX setup, the search for dark photons will be extended to
lower dark photon masses than were probed
before~\cite{Merkel:2011,Merkel:2014}. In this type of experiments a
dark photon $\gamma'$ could radiatively be produced off a nuclear
target $Z$ via the reaction $e^-Z\rightarrow e^-Z\gamma'$, and a
missing mass analysis of coincident $e^+e^-$ pairs from the decay
$\gamma'\rightarrow e^+e^-$ needs to be performed. These searches will
be extended to invisible decay channels, in which a dark photon decays
via $\gamma'\rightarrow\chi\bar\chi$ in a pair of light dark matter
particles, requiring a missing mass analysis of the scattered electron
in coincidence with the recoil nucleus~\cite{Doria:2018, Doria:2019}.

Furhtermore, the study of reactions of astrophysical importance such
as $^{16}$O$(e,e'\alpha)^{12}$C, where the $\alpha$-particle is
detected with a detector arrangement inside the scattering chamber,
will become possible at unprecedented low center-of-mass
energies~\cite{Friscic:2019,Holt:2019}.

This paper is organized as follows: Section~\ref{sec:Setup} provides
an overview of the experimental setup and the gas jet
target. Section~\ref{sec:TargetOperation} describes details of the
handling procedures and operating parameters of the target. The gas
jet is characterized in Sect.~\ref{sec:Characterization}. Its areal
thickness
(Subsect.~\ref{subsec:DensityProfile_HorizontalDisplacement}) and
density (Subsect.~\ref{subsec:TargetDensity}) were deduced from a
comparison to a simulation of elastic scattering events. Clustering of
the gas jet is discussed in Subsect.~\ref{subsec:JetClustering}. A
sophisticated simulation is described in
Subsect.~\ref{subsec:JetSimulation}, and the results are compared to
measured profiles from two different nozzle designs in
Subsect.~\ref{subsec:DifferentNozzles}. Section~\ref{sec:Background}
discusses dedicated background studies for electron scattering
experiments with this target. The stability of the gas jet is
considered in Sect.~\ref{sec:Stability} and the target is compared to
other hydrogen targets at electron accelerators in
Sect.~\ref{sec:OtherTargets}. A summary and an outlook are given in
Sect.~\ref{sec:Summary}.
%
%############################################################################################################
\section{Experimental setup}
\label{sec:Setup}
%############################################################################################################
%------------------------------------------------------------------------------------------------------------
\subsection{MAMI and A1 multi-spectrometer facility}
%------------------------------------------------------------------------------------------------------------
MAMI delivers electrons with energies of up to $E_{\text{beam}} =
\unit[1600]{MeV}$ at beam currents of up to $I_{\text{beam}} =
\unit[100]{\mu A}$~\cite{Herminghaus:1976, Kaiser:2008, Dehn:2011,
  Dehn:2016}.
For experiments at this facility, different types of targets are
available, including solid state targets (single foils and stacks),
liquid hydrogen and deuterium targets, a high pressure helium target,
a polarized $^3$He target, and a waterfall target.

\begin{figure}[htb]
  \centering
  \includegraphics[width=0.6\columnwidth]{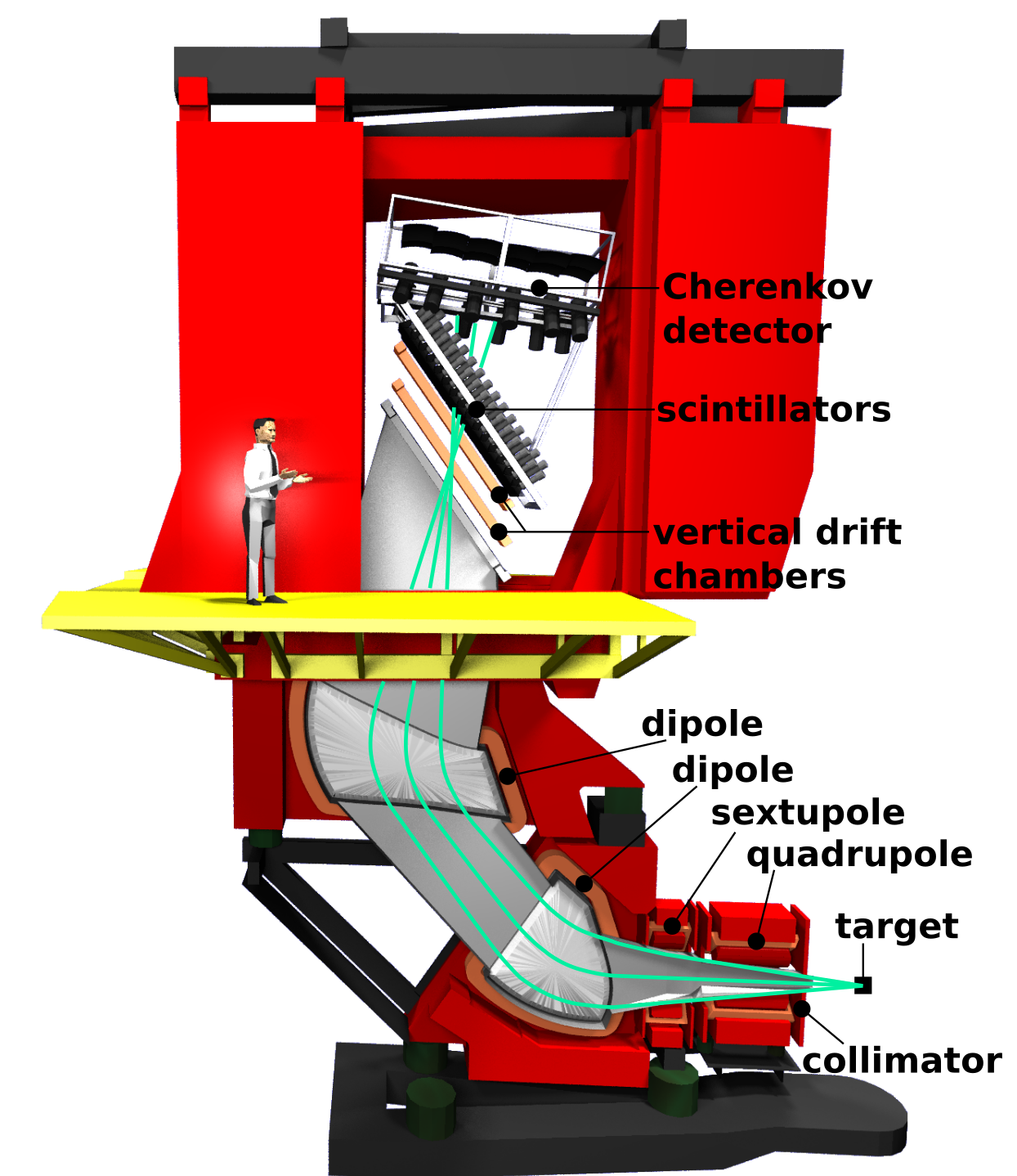}
  \caption{Schematic side view of spectrometer~A at the A1
    multi-spectrometer setup. The magnets are partly cut open to show
    pole pieces and coils. The doors of the shielding house are shown
    in the opened position, exhibiting the detector package. Three
    trajectories for charged particles coming from the target with
    identical momentum are shown. The particles enter the spectrometer
    through a collimator and are guided by the fields of four separate
    magnets to the focal plane. Their momenta and angles at the target
    position are reconstructed from the measured tracks across the
    focal plane.  A Cherenkov detector provides electron
    identification.}
  \label{fig:SpecA}
\end{figure}

At the spectrometer setup of the A1 Collaboration, scattered electrons
from the interaction of the beam electrons with the target can be
detected with the high-resolution magnetic spectrometers A, B, and C,
see Fig.~\ref{fig:SpecA} for a schematic of spectrometer~A.  These
spectrometers have target length acceptances of $\Delta l =
\unit[5]{cm}$, solid angle acceptances of $\Delta \Omega =
\unit[28]{msr}$ (for A and C) and $\Delta \Omega = \unit[5.6]{msr}$
(for B), and momentum acceptances of $\Delta p/p = $ \unit[15]{\%}
(for B), \unit[20]{\%} (for A), and \unit[25]{\%} (for C).  A detailed
description of the spectrometers can be found in
Ref.~\cite{Blomqvist:1998}.  Vertical drift chambers are used for
tracking, scintillation detectors for trigger and timing purposes, and
a threshold gas Cherenkov detector for discrimination between
electrons and pions or muons. The spectrometers have a relative
momentum resolution of $\delta p/p \approx 10^{-4}$ and an angular
resolution at the target of $\delta \theta/\theta \approx
\unit[3]{mrad}$~\cite{Blomqvist:1998}.

\begin{figure}[htbp]
  \centering
  \includegraphics[width=0.8\columnwidth]{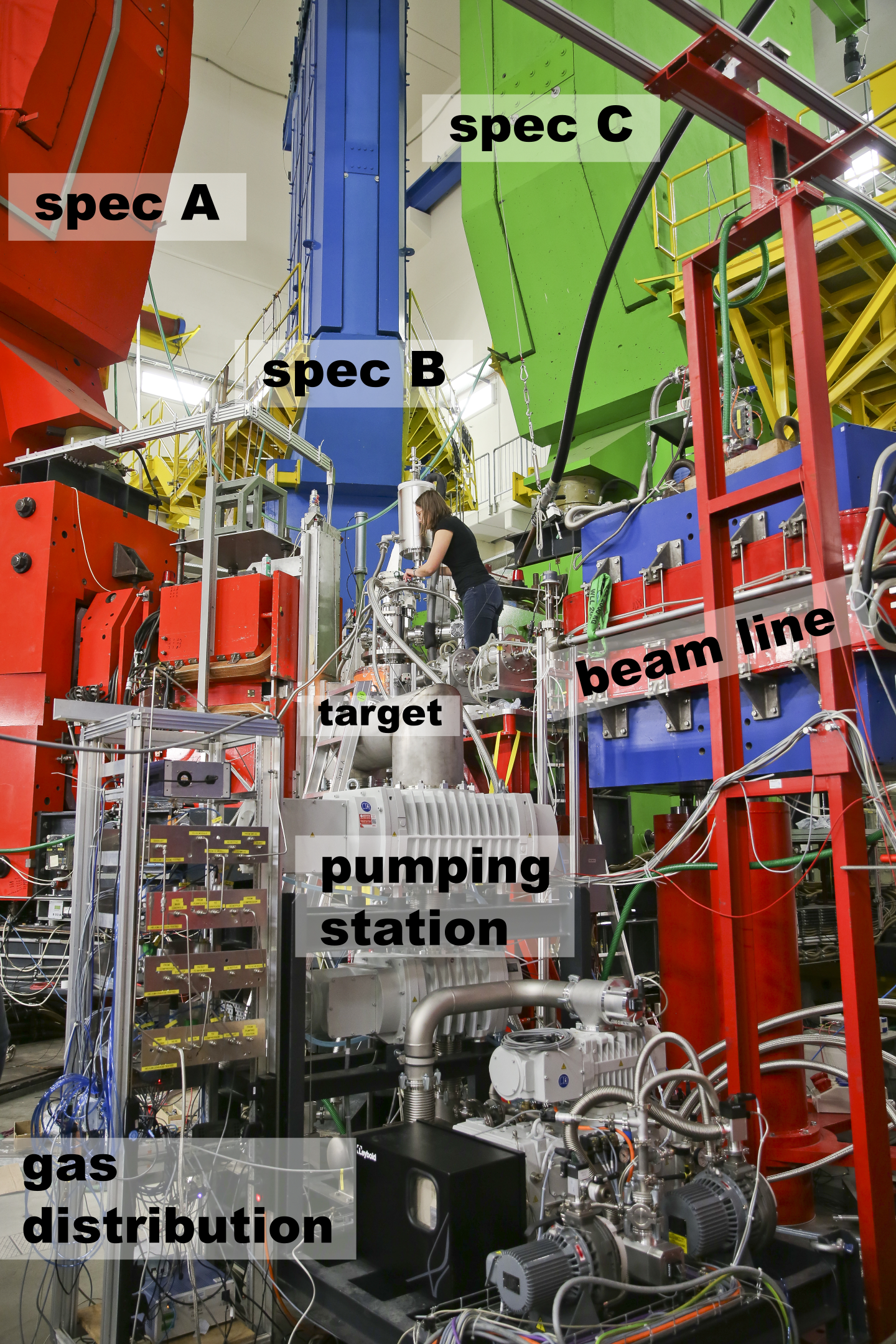}
  \caption{Photograph of the gas jet target mounted at the A1
    multi-spectrometer setup. The electron beam enters the cylindrical
    scattering chamber from the right. Scattered electrons and
    particles emitted in direction of the spectrometers leave the
    chamber through Kapton windows. The three magnetic spectrometers
    can rotate around the target. For the studies presented in this
    paper, spectrometers A (in red) and B (in blue) were
    used. Photograph by T. Zimmermann / JGU.}
  \label{fig:JetAtA1}
\end{figure}

The gas jet target was assembled, tested, and operated at the A1
spectrometer setup with $E_{\text{beam}}$ between $\unit[195]{MeV}$
and $\unit[450]{MeV}$ and at beam currents of up to $I_{\text{beam}} =
\unit[20]{\mu A}$.  A photograph of the experimental setup is shown in
Fig.~\ref{fig:JetAtA1}.
%
%------------------------------------------------------------------------------------------------------------
\subsection{Electron beam size}
\label{sec:ElectronBeamWidth}
%------------------------------------------------------------------------------------------------------------
In order to determine the width of the electron beam on the gas jet
target, a gold-plated tungsten wire with a diameter of $\varnothing =
\unit[100]{\mu m}$ was moved through the beam using the stepper motor
shown in Fig.~\ref{fig:ScatteringChamber}.  Since the turning radius
$r = \unit[106]{mm}$ was relatively large, the rotation of the wire
target resulted in an approximately even, horizontal movement of the
vertically aligned wire for the small displacements of typically
$\Delta x = \unit[20]{\mu m}$.

\begin{figure}[htbp]
  \centering
  \includegraphics[width=\columnwidth]{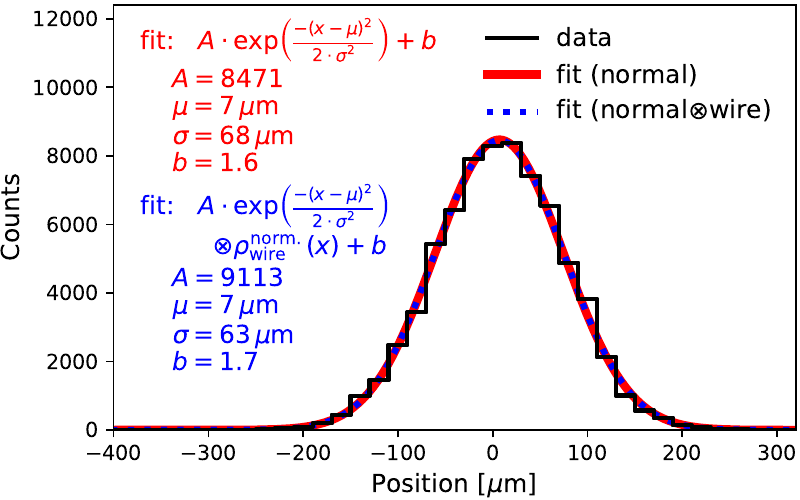}
  %generated: schlimme@a1cool ~/exp/jet2018/MXJet2020Profile $ sh X_wirescan_2018-04-13.sh
  \caption{Scattering events as a function of the wire position for
    data taking periods of $\Delta t = \unit[30]{s}$ at each
    position. A fit with a normal distribution on top of a small
    offset was performed. For a very thin wire, the width parameter
    $\sigma$ can be interpreted as the horizontal electron beam
    width. Including the finite wire size in the fit results in a
    $\unit[10]{\%}$ smaller parameter $\sigma$.}
  \label{fig:WireScanBvsPosition}
\end{figure}

Two spectrometers were set to a kinematic setting corresponding to the
quasi-elastic electron scattering off the heavy nuclei of the wire.
The variation of the count rates with the wire positions was
monitored. From a fit to the distribution of scattering events as a
function of the reconstructed position, the horizontal width of the
electron beam was determined to be of the order of FWHM $\approx
\unit[100]{\mu m}$, see Fig.~\ref{fig:WireScanBvsPosition}. From the
beam spot on the luminescence screen, it was known that the beam was
axially symmetric.
%
%------------------------------------------------------------------------------------------------------------
\subsection{Fast beam wobbler}
\label{sec:WobblerCalibration}
\begin{figure}[htbp]
  \centering
  \includegraphics[width=\columnwidth]{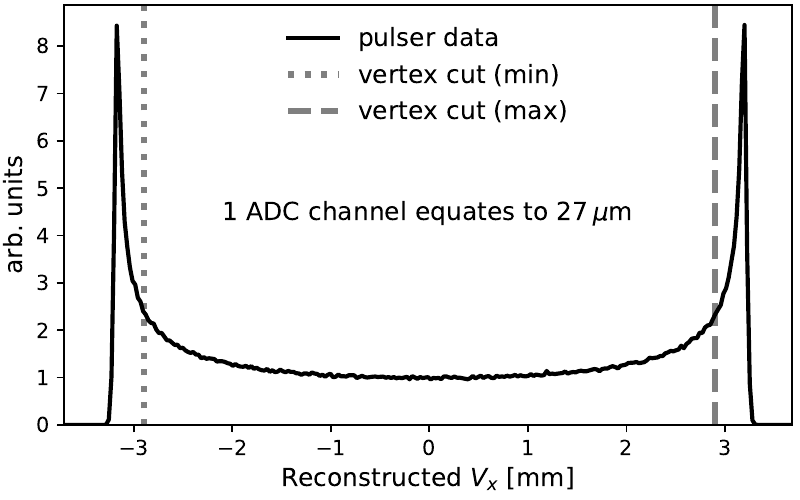}
  %generated: schlimme@a1cool ~/exp/jet2018/MXJet2020Profile $ sh X_wirewobblerscan_2018-04-16.sh
  \caption{Horizontal displacement distribution of the electron beam
    at the target location. The wobbler current was recorded by an
    ADC. The $x$-axis has been calibrated in $\unit[]{mm}$ units.}
  \label{fig:WireWobblerScan_Wobbler_B}
\end{figure}

%------------------------------------------------------------------------------------------------------------
A fast beam wobbler, using air-core coils, is installed $\unit[12]{m}$
upstream of the target~\cite{Wilhelm:1993} to raster the electron beam
horizontally, vertically, or both at the same time. The coils are
energized by alternating sine wave currents with frequencies of $f =
\unit[2.918]{kHz}$ and $f = \unit[2.000]{kHz}$, respectively. The
displacement of the beam at the target location is approximately
proportional to the coil current, which is monitored by means of a
bipolar 8-bit analog digital converter (ADC).  The displacement
distribution, as recorded by the ADC, is shown in
Fig.~\ref{fig:WireWobblerScan_Wobbler_B} for horizontal wobbling.
%expected: ~ 1/sqrt(1-x^2)
The figure also shows the cuts that were applied to the vertex
position in the following analyses to reduce the bias from the
non-uniform distribution.

\begin{figure}[htbp]
  \centering
  \includegraphics[width=\columnwidth]{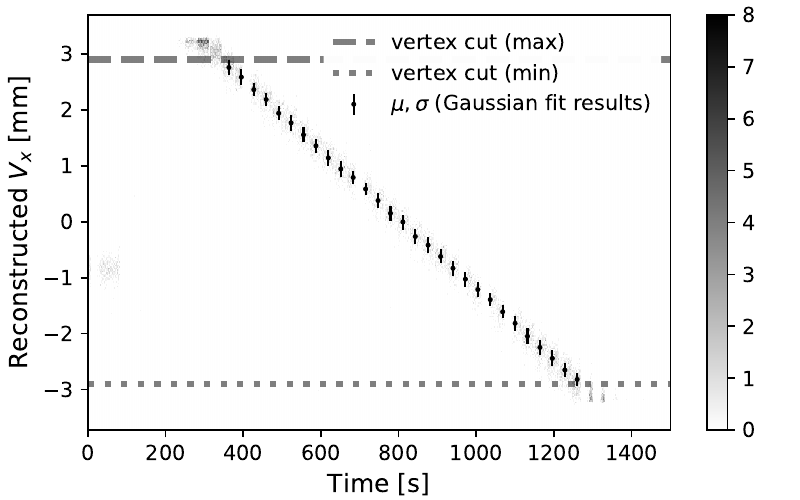}
  %generated: schlimme@a1cool ~/exp/jet2018/MXJet2020Profile $ sh X_wirewobblerscan_2018-04-16.sh
  \caption{Reconstructed horizontal vertex position as a function of
    time for a rastered beam during a stepwise movement of the wire
    target. For each data taking period, the wire position was known,
    and a fit with a normal distribution was applied to the histogram
    of the reconstructed vertex position. These data are used to
    calibrate the wobbler readout conversion from ADC channels to mm,
    and to estimate the precision of the reconstructed beam
    displacement.}
  \label{fig:WireWobblerScan_VVsTime_B}
\end{figure}

To calibrate the wobbler readout in terms of displacement at the
target position, data were taken for different wire target positions,
see Fig.~\ref{fig:WireWobblerScan_VVsTime_B}.  When the displaced beam
covered the wire position, the spectrometers registered quasi-elastic
scattering events.  From the known wire position for each setting, the
acquired wobbler ADC values were related to the displacement.

\begin{figure}[htbp]
  \centering
  \includegraphics[width=\columnwidth]{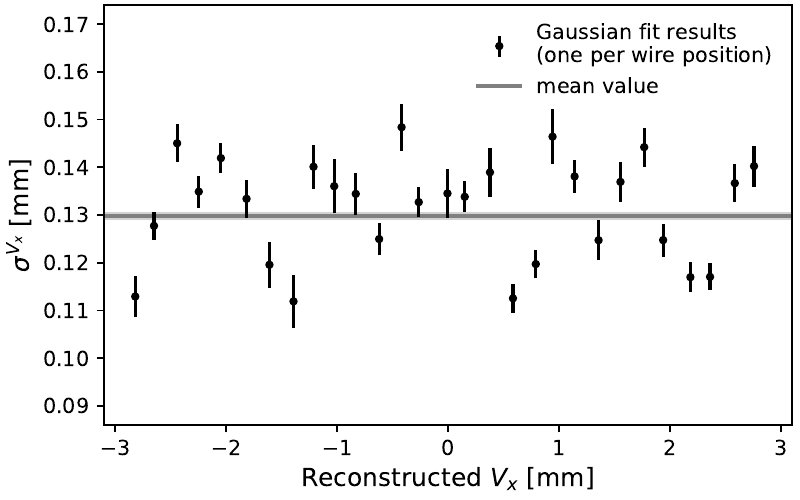}
  %generated: schlimme@a1cool ~/exp/jet2018/MXJet2020Profile $ sh X_wirewobblerscan_2018-04-16.sh
  \caption{Precision of the position reconstruction as a function of
    the horizontal displacement. The average precision for this beam
    energy of $E = \unit[450]{MeV}$ was $\sigma^{V_x} = \unit[130]{\mu
      m}$.}
  \label{fig:WireWobblerScan_VResolution_B}
\end{figure}

The achievable precision of the position reconstruction for a single
scattering event is limited by the finite electron beam width, by the
resolution of the wobbler ADC, and by specific instrumental
restrictions which are detailed in Ref.~\cite{Wilhelm:1993}. The coil
current in the wobbler magnets for a given beam displacement is
proportional to the beam energy. Thus, the range of amplitudes within
the dynamic range of the ADC changes with beam energy.  The typical
precision during the electron scattering studies at a beam energy of
$E = \unit[315]{MeV}$ was $\sigma^{V_x}\approx \unit[150]{\mu m}$, and
slightly better at $E = \unit[450]{MeV}$,
cf.\ Fig.~\ref{fig:WireWobblerScan_VResolution_B}.

%
%------------------------------------------------------------------------------------------------------------
\subsection{Gas jet target}
\label{subsec:GasJetTarget}
\begin{figure}[htbp]
  \centering
  \includegraphics[width=0.6\columnwidth]{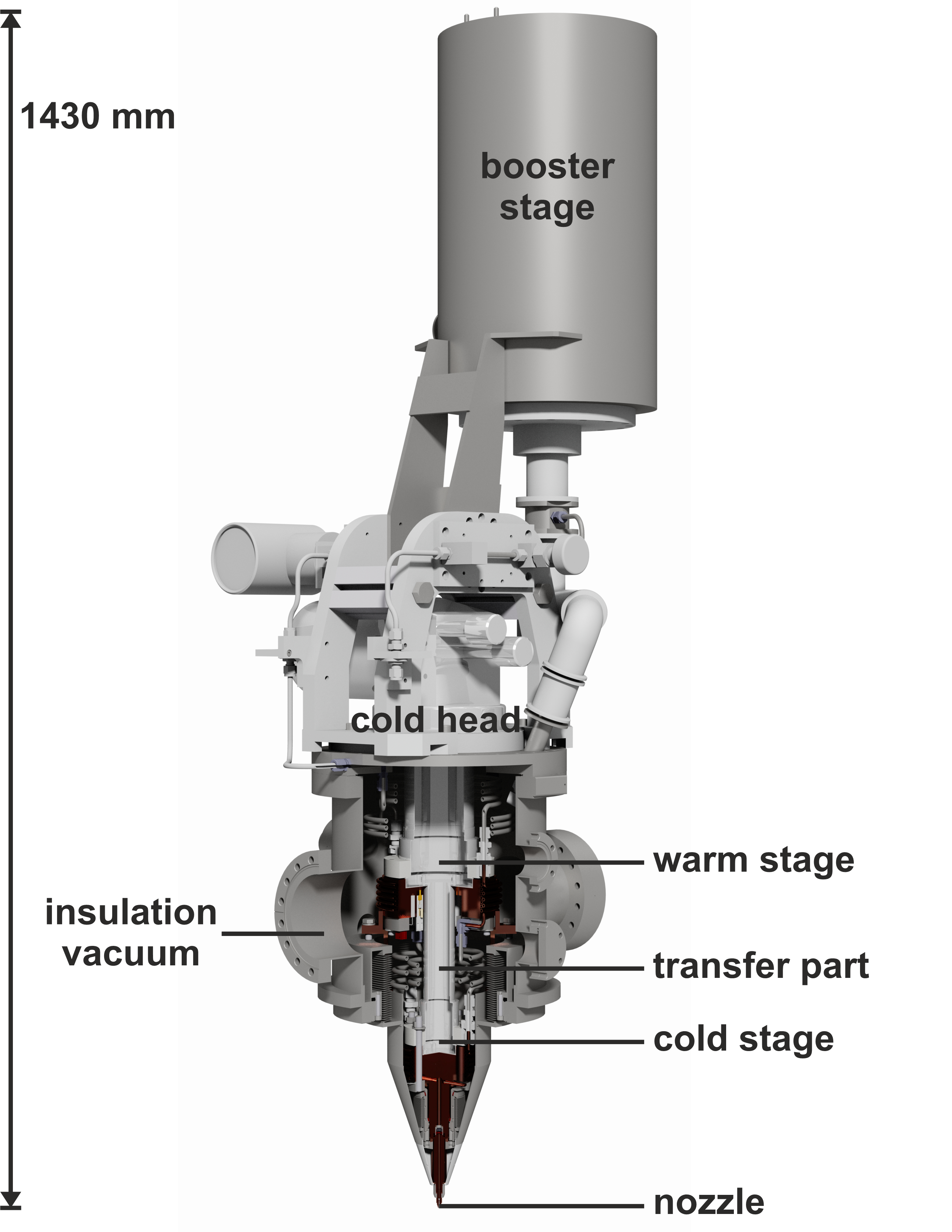}
  \caption{Drawing of the gas jet target for the MAGIX setup. The
    booster stage can be filled with liquid nitrogen to precool the
    gas, which is then guided to a two-stage cold head where it
    reaches a temperature of $T_0 = \unit[40]{K}$ when operated with
    hydrogen. The convergent--divergent nozzle is mounted on top of an
    extension which is surrounded by a conical tip to ensure a high
    angular acceptance for the spectrometers.}
  \label{fig:target}
\end{figure}

%------------------------------------------------------------------------------------------------------------
The target for the MAGIX setup as shown in Fig.~\ref{fig:target} is a
cryogenic supersonic jet target that can be operated with most
gases~\cite{Grieser:2018}.  It was designed to achieve areal
thicknesses of more than $\rho_\text{areal}\approx
\unit[10^{18}]{atoms/cm^2}$ when using hydrogen. This requires a gas
flow rate of up to $q_V = \unit[2400]{l_n/h}$ while the hydrogen gas
has to be cooled down to $T_0 = \unit[40]{K}$. Two systems provide the
necessary cooling power. The target gas first enters a booster stage
that can be filled with liquid nitrogen to precool the gas. Based on a
sophisticated level meter, the nitrogen refill is automated.  It is
then guided through copper windings that surround the two stages of a
cryogenic cold head. To allow a precise temperature regulation, both
stages are equipped with heaters and temperature sensors, where the
measured temperature at the second stage corresponds to the gas
temperature directly in front of the nozzle. Furthermore, the booster
stage and the cold head are surrounded by two separate evacuated
insulation volumes to minimize the heat conduction with the
environment and to minimize the nitrogen consumption within the
booster. The cryogenic gas is then pressed through a
convergent--divergent nozzle where it is accelerated to supersonic
velocities and adiabatically cooled down during the expansion. This
results in a supersonic jet that is ejected vertically from the nozzle
and is caught several \unit{mm} below by a funnel-shaped structure,
the catcher.  The electron beam of the accelerator traverses the gas
jet horizontally inside the scattering chamber, resulting in a very
compact interaction zone. The target including its technical details
and operational parameters is described in Ref.~\cite{Grieser:2018} in
more detail.

Assuming a perfect gas, the velocity of the gas leaving the nozzle can
be calculated by
\begin{equation}\label{eq:gasVelocity}
    v = \sqrt{2c_pT_0} = \sqrt{2\frac{\kappa}{\kappa-1}\frac{RT_0}{M}}
\end{equation}
with the specific heat capacity $c_p$, the heat capacity ratio
$\kappa$ (with $\kappa=5/3$ for cryogenic hydrogen at \unit[40]{K}),
the universal gas constant $R$, the gas temperature $T_0$ at the
nozzle inlet and the molar mass $M$ of the gas molecules
\cite{Taeschner:2013}. Thus, the use of gases at cryogenic
temperatures has two advantages.  A lower gas temperature results in a
lower velocity and therefore in a larger areal
thickness. Additionally, working points close to the phase transition
open the possibility for the gas to cross the vapor pressure curve
during the expansion in the nozzle outlet.  Then, the vanishing
relative velocities at low temperatures in combination with the high
probability of three-body collisions within the nozzle lead to an
accelerated formation of clusters consisting of thousands of molecules
\cite{Pauly:2000}.  These clusters are much heavier than the
surrounding gas, so that they do not change their direction when
scattering on single gas molecules. This leads to well-shaped cluster
beams with a very small and constant angular divergence.

Cluster beams are routinely used in cluster-jet targets where the
interaction point is located up to several meters behind the
nozzle~\cite{Brauksiepe:1996, Dombrowski:1997, Taschner:2011,
  PANDA-Target-TDR:2012, Grieser:2019}, whereas in the MAGIX setup,
this point is directly behind the nozzle. When operating this target
at cryogenic temperatures, cluster production reduces the divergence
of the beam, resulting in an increase in areal thickness for the same
gas flow rate and a reduction of the required catcher diameter for the
same catcher efficiency. Moreover, larger catcher--nozzle distances
become possible and consequences of this advantage are described in
Sect.~\ref{sec:Background}.

For the operation of a cryogenic jet target with a high gas flow rate,
a crucial component is the convergent--divergent nozzle and especially
its divergent outlet shape. The latter defines the expansion of the
target gas and has a large influence on the beam shape and temperature
after expansion and thus also on the probability of cluster
production. Therefore, different nozzles have been tested and
optimized using numerical simulations as described in
Subsects.~\ref{subsec:JetSimulation} and
\ref{subsec:DifferentNozzles}.
%
%------------------------------------------------------------------------------------------------------------
\subsection{Gas flow system}
\begin{figure}[htbp]
  \centering
  \includegraphics[width=0.6\columnwidth]{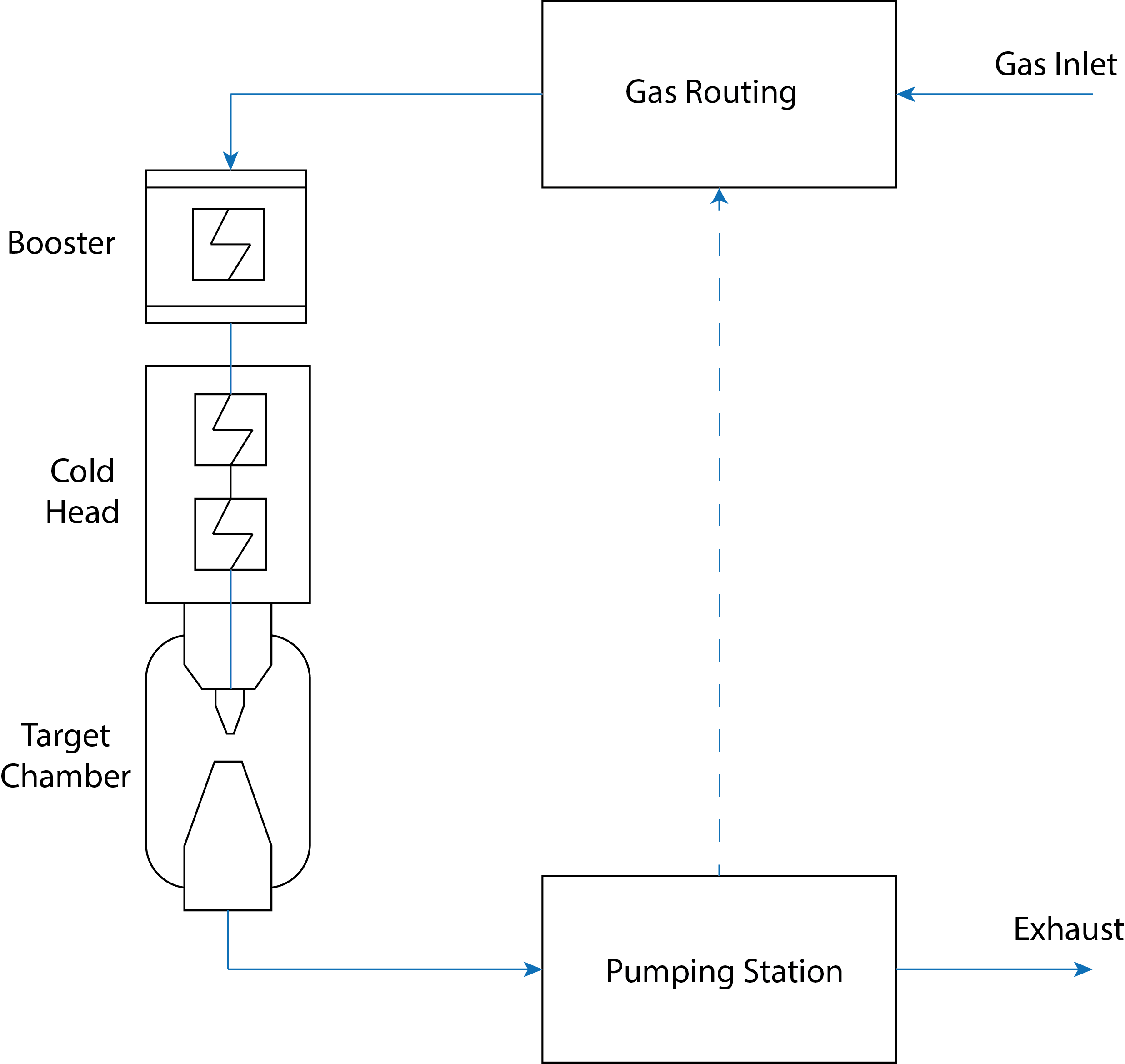}
  \caption{Simplified gas flow diagram of the jet target.}
  \label{fig:gasSystem}
\end{figure}

%------------------------------------------------------------------------------------------------------------
A simplified sketch of the gas flow is shown in
Fig.~\ref{fig:gasSystem}. A bundle of twelve hydrogen bottles
($\unit[50]{l} \times \unit[200]{bar}$ each, \unit[99.999]{\%} purity)
was used at the inlet, which lasted for about two days at the maximum
gas flow rate.

The gas routing system includes mass flow controllers, pressure
sensors, temperature sensors, magnetic as well as sampling valves, a
hardware state monitor, and a manual interlock button.  The entire
system is remotely controlled and the measurement of the gas pressure
at key positions ensures a high level of control. Sensors which are
not delivering a digital signal are included in the system by applying
analog to digital conversion. The flow rate of the gas is set in the
range between $q_V = \unit[50]{l_n/h}$ and $\unit[5000]{l_n/h}$ by a
Brooks Instrument SLA5851 controller with an accuracy of $\delta
q_V/q_V =$ \unit[$\pm$ 0.9]{\%} for a set point between \unit[20]{\%}
and \unit[100]{\%} of the full scale and an accuracy of $\delta
q_V/q_V =$ \unit[$\pm$ 0.18]{\%} of the full scale below
\unit[20]{\%}, corresponding to $\delta q_V = \unit[\pm 9]{l_n/h}$
uncertainty for low flow rates.

To enable the use of rare and expensive target gases, the setup will
be extended with a recirculation and purification system.
%
%------------------------------------------------------------------------------------------------------------
\subsection{Scattering chamber}
\label{sec:VacuumChamber}
\begin{figure}[htbp]
  \centering
  \includegraphics[width=0.6\columnwidth]{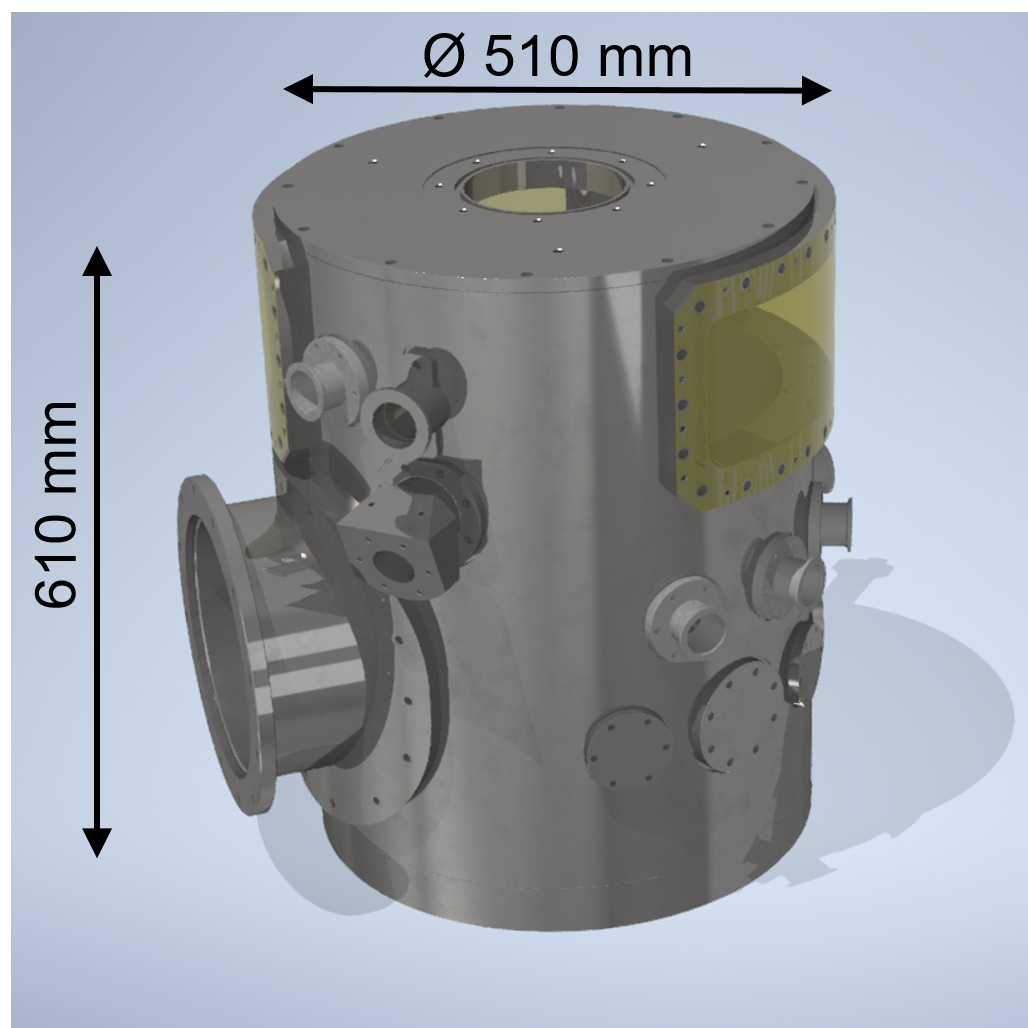}
  \caption{Drawing of the scattering chamber. Kapton windows,
    several feedthroughs, and the flange for the turbomolecular pump
    at the left side are visible. To minimize moisture leakage through
    the windows, an additional layer of aramid foil is placed on top
    of the Kapton.}
  \label{fig:ScatteringChamber}
\end{figure}

\begin{figure}[htbp]
  \centering
  \includegraphics[width=\columnwidth]{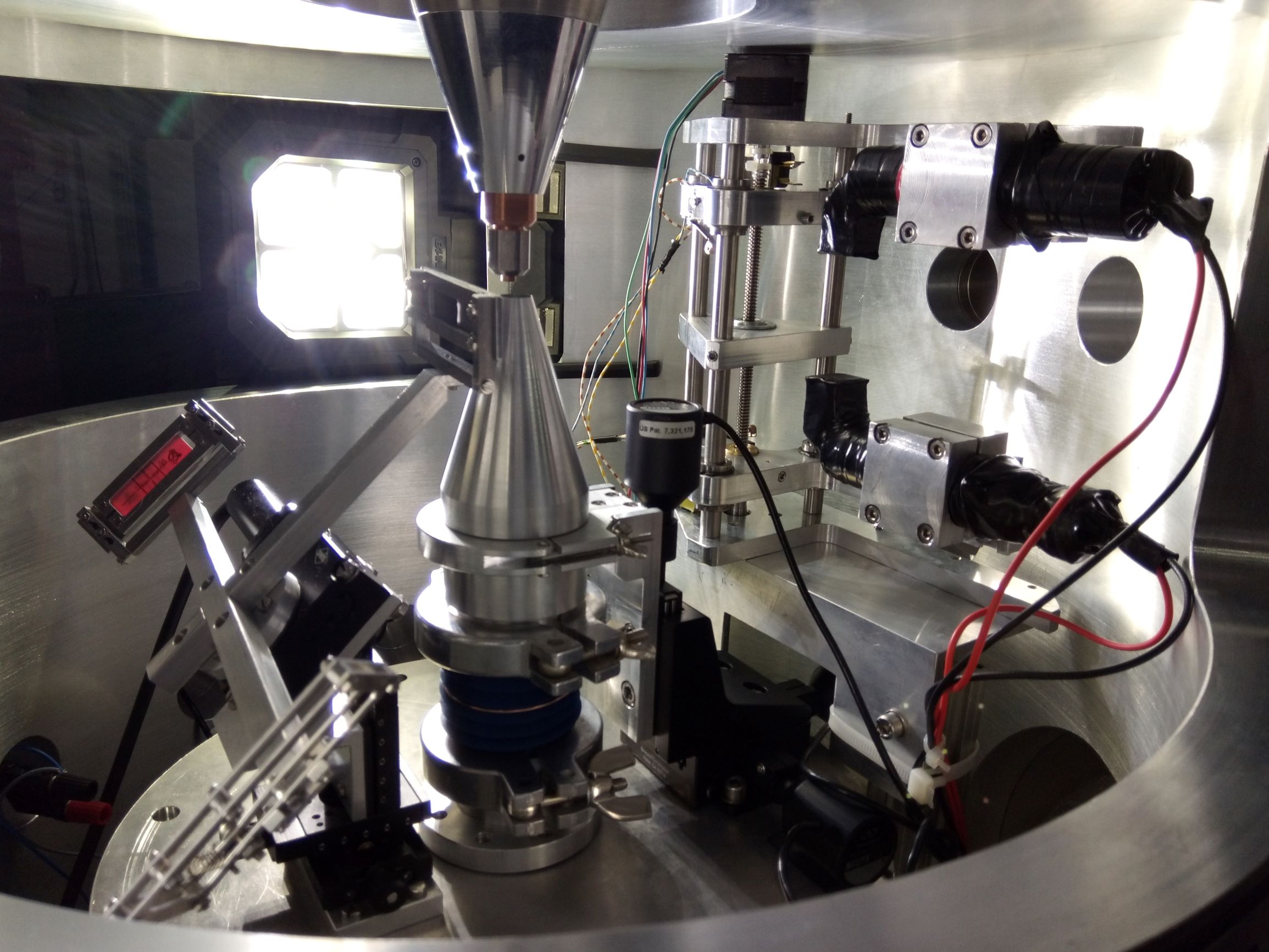}
  \caption{Photograph of the inside of the scattering
    chamber. Target nozzle (top) and catcher (bottom) are located in
    the center. The beam enters from the right. The catcher can be
    aligned in all directions, it is attached to the vacuum system by
    a flexible silicon bellow. Solid state targets are mounted
    downstream on a holder which can be rotated by a stepper
    motor. Veto scintillation detectors are placed upstream and can be
    aligned vertically.}
  \label{fig:InsideScatteringChamber}
\end{figure}

%------------------------------------------------------------------------------------------------------------
The aluminum scattering chamber for the gas jet target is depicted in
Fig.~\ref{fig:ScatteringChamber}. In order to align the chamber with
the beam pipe, a frame underneath allows adjusting the height as well
as the tilt.  Kapton windows with a height of $\unit[120]{mm}$ allow
the detection of particles with scattering angles between $\theta =
21.5^\circ$ and $135^\circ$ on the left side of the exit beamline, and
between $\theta = 11^\circ$ and $122.5^\circ$ on the right side. Ten
universal feedthroughs of different sizes for sensor cables are placed
around the chamber. In addition, a glass view port to the pivot point
allows the installation of a camera. A DN 260 flange on the side of
the chamber is reserved for a turbomolecular pump. A plate on top of
the chamber is holding the gas jet target via three fine-threaded bars
for a precise adjustment. The tip of the target, confined by a
membrane bellow, is lowered into the chamber, see
Fig.~\ref{fig:InsideScatteringChamber}. The catcher can be
independently and remotely aligned in all directions, driven by three
motorized micrometer screws.

Additional solid state targets are mounted on a rotatable holder which
can be moved into the electron beam by a stepper motor: a tungsten
wire to determine the width of the electron beam, an aluminum oxide
screen to monitor the position of the beam, and a carbon foil for
calibration measurements.

%
%------------------------------------------------------------------------------------------------------------
\subsection{Vacuum system}
\label{sec:VacuumSystem}
\begin{figure}[htbp]
  \centering
  \includegraphics[width=0.7\columnwidth]{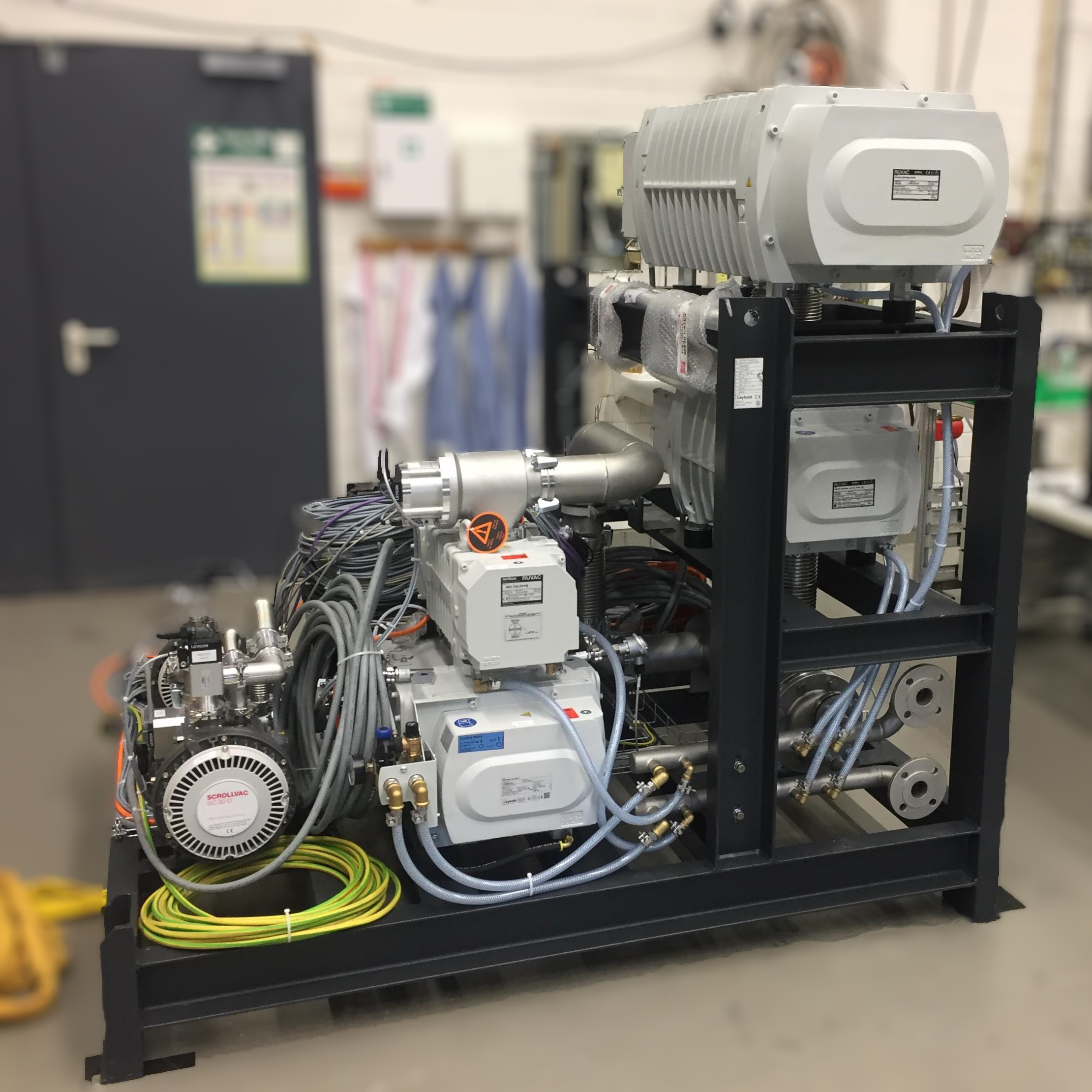}
  \caption{Photograph of the pumping station after its delivery. It
    consists of three roots pumps, a screw pump, two scroll pumps, a
    turbomolecular pump (not shown) and a gas cooler. The pumps as
    well as the valves are controlled remotely. Temperature and
    pressure sensors provide continuous readout of relevant
    parameters. Most of the pumps are powered using external frequency
    converters (not shown). Cooling is mainly performed using a common
    water cooling circuit.}
  \label{fig:pumps}
\end{figure}

%------------------------------------------------------------------------------------------------------------
To remove the target gas efficiently from the scattering chamber and
to keep the residual pressure inside the chamber low, an assembly of
pumps is used, see Fig.~\ref{fig:pumps}. A set of roots pumps is
connected to the catcher: one Leybold {\small RUVAC WH7000FU}, one
Leybold {\small RUVAC WH2500FU}, and one Leybold {\small RUVAC
  WH700FU}.  Between the second and the third pump, a gas cooler is
installed.  The rated pumping speeds, specified for nitrogen, are $S =
\unit[7000]{m^3/h}$, $\unit[2500]{m^3/h}$, and $\unit[710]{m^3/h}$,
respectively.  As roughing pump, a Leybold {\small DRYVAC DV650C}
screw pump with a rated pumping speed of $S = \unit[650]{m^3/h}$, is
connected. Purge gas can be injected into this pump to allow for a
safe operation with oxygen as a target gas. When using hydrogen or
helium, no purge gas is used, and a Leybold {\small SCROLLVAC SC30D}
scroll pump with a rated pumping speed of $S = \unit[30]{m^3/h}$ is
added, because the screw pump is not efficient for light gases.  In
addition, a Leybold {\small TURBOVAC MAG W2800C} turbomolecular pump
is directly mounted to the scattering chamber, with a Leybold {\small
  SCROLLVAC SC15D} scroll pump as roughing pump.

Without gas ballast, a pressure of $p\approx \unit[10^{-6}]{mbar}$ is
typically reached inside the scattering chamber.  Since a significant
amount of gas enters the scattering chamber during operation, a larger
gas pressure builds up inside. Amongst other things, this results in a
reduction of the pumping speed for hydrogen of the turbomolecular pump from
$S\approx \unit[2100]{l/s}$ to, for example, $S\approx\unit[180]{l/s}$ at
$p=\unit[0.2]{mbar}$.  Consequences for the applicable gas flow rate
are discussed in Subsect.~\ref{subsec:ApplicableGasFlow}.

%
%------------------------------------------------------------------------------------------------------------
\subsection{Beam halo suppression system}
\label{subsec:HaloBlocker}
%------------------------------------------------------------------------------------------------------------
During beam tests with the gas jet target, scattering reactions of
electrons from the beam halo with nozzle and catcher were identified.
To suppress this kind of background, a collimator and a veto detector
were implemented.
%
% - - - - - - - - - - - - - - - - - - - - - - - - - - - - - - - - - - - - - - - - - - - - - - - - - - - - - -
\subsubsection{Collimator}
\begin{figure}[htbp]
  \centering
  \includegraphics[width=0.21\textwidth]{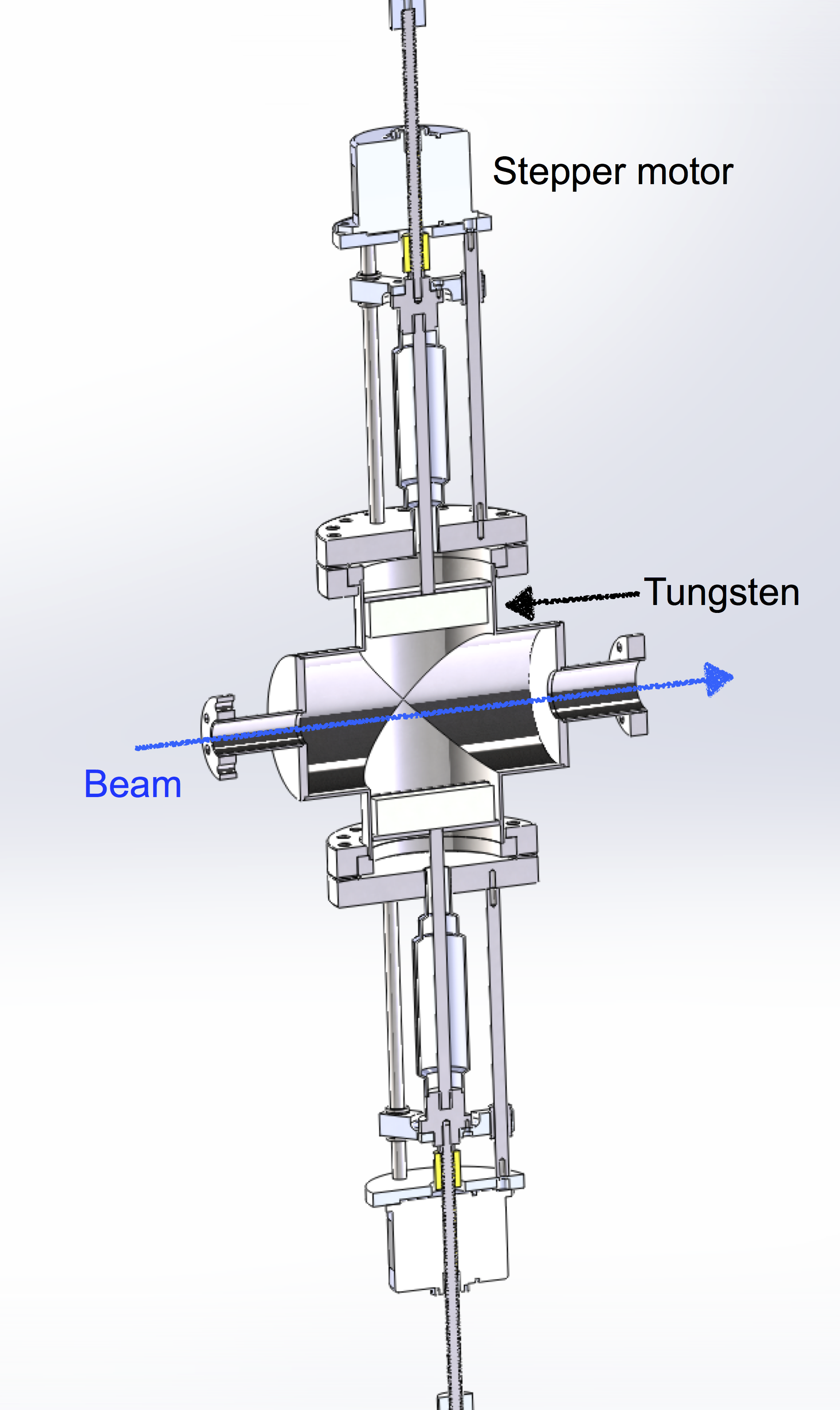}
  \caption{Section view of the collimator assembly. The movable
    tungsten absorbers are located above and below the beam pipe.}
  \label{fig:collimator}
\end{figure}

% - - - - - - - - - - - - - - - - - - - - - - - - - - - - - - - - - - - - - - - - - - - - - - - - - - - - - -
The upstream collimator consists of two vertically movable absorbers
made of tungsten above and below the beam pipe with a thickness of $d
= \unit[13]{cm}$ along the beam direction. The collimator can block
electromagnetic showers of up to $E = \unit[1.5]{GeV}$ primary
energy. The absorbers are each connected with a shaft to a stepper
motor that controls the vertical position, as shown in
Fig.~\ref{fig:collimator}.

%
% - - - - - - - - - - - - - - - - - - - - - - - - - - - - - - - - - - - - - - - - - - - - - - - - - - - - - -
\subsubsection{Veto detector}
% - - - - - - - - - - - - - - - - - - - - - - - - - - - - - - - - - - - - - - - - - - - - - - - - - - - - - -
The veto detector is positioned upstream of the gas jet target inside
the scattering chamber, as can be seen in
Fig.~\ref{fig:ScatteringChamber}. It is designed to reject scattering
reactions from residual beam halo electrons hitting the catcher or the
nozzle. During the experiment, the positions of the collimator and the
veto detectors were adjusted to cover the front of the nozzle and the
catcher as much as possible without introducing additional background.

\begin{figure}[htbp]
  \centering
  \includegraphics[height=0.155\textheight]{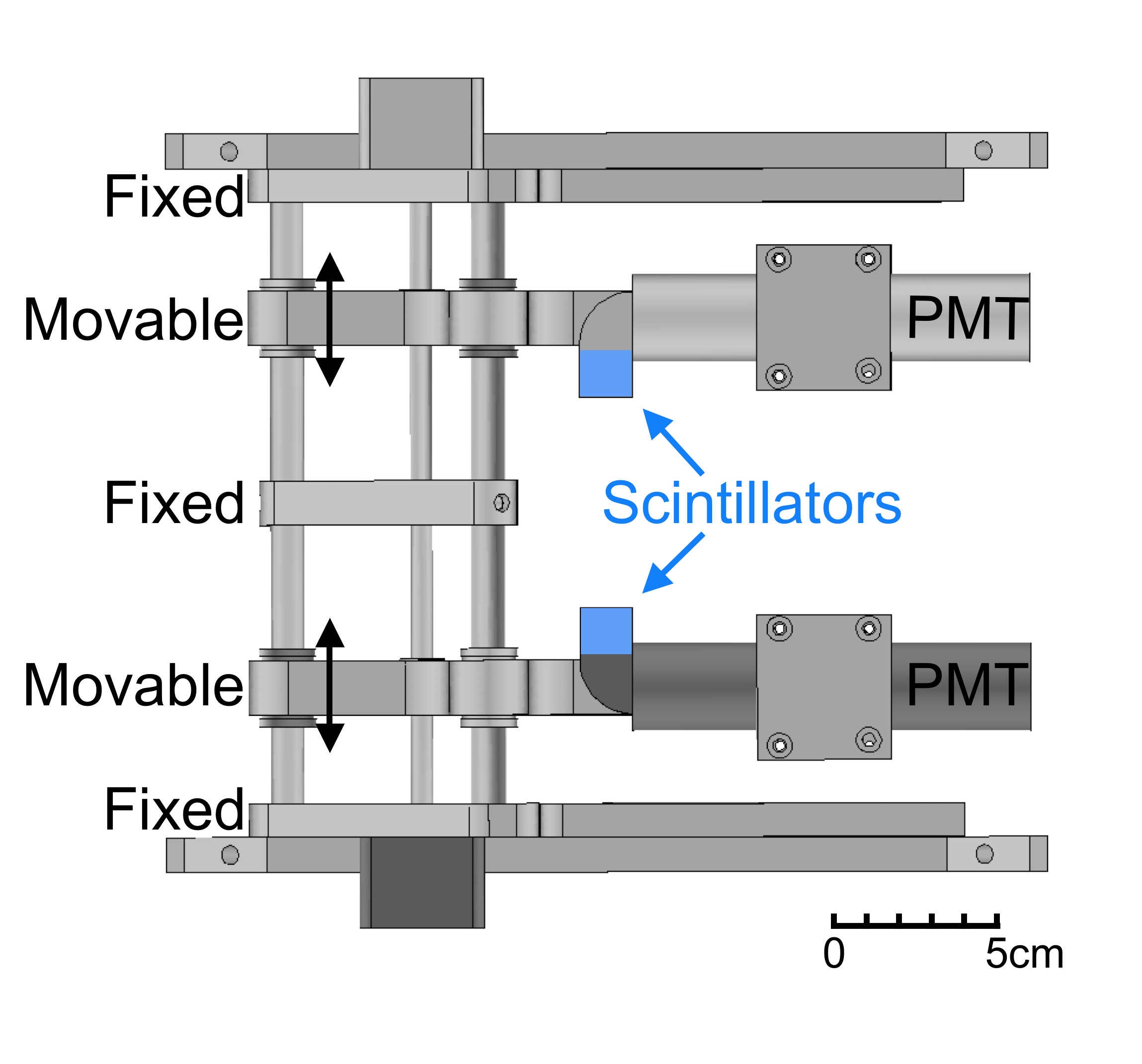}    
  \includegraphics[height=0.13\textheight]{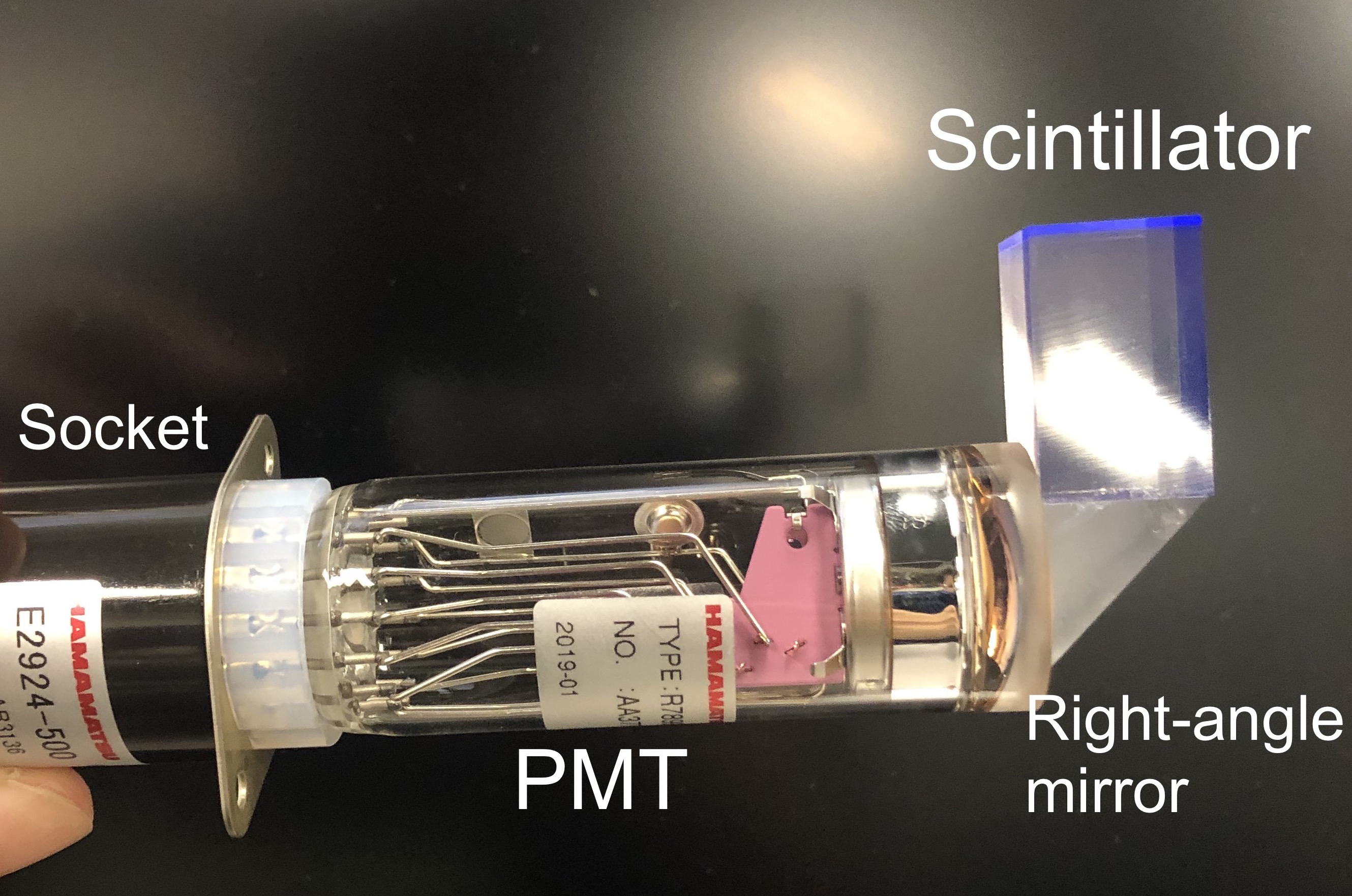}
  \caption{Left: Schematic of the veto system. The beam pipe is
    located in the center. Right: Photograph of one detector arm with
    its components.}
  \label{fig:veto}
\end{figure}

The detector consists of a frame, a moving mechanism, and the active
parts, as shown in Fig~\ref{fig:veto}.  The aluminum frame consists of
three fixed plates and two moving platforms, one each in a symmetric
upper and lower section. The plates are connected by three metal
rods. The platforms are guided by three linear shafts along the
vertical direction and are moved with stepper motors from Lin
Engineering 3518 series. These have 200 steps per revolution,
translating to $\unit[10]{\mu m}$ per step. One detector arm is
mounted on each platform, consisting of a scintillator, a right-angle
prism, and a photomultiplier tube (PMT), as shown in
Fig.~\ref{fig:veto}. The scintillator is of type EJ-212 from Eljen
Technology with dimensions of $\unit[15]{mm} \times
\unit[15]{mm}\times \unit[20]{mm}$. Its light emission peaks at a
wavelength $\lambda\approx$ \unit[430]{nm}. This light is guided by
the prism to the PMT with an efficiency of \unit[99]{\%}. The PMTs are
of type R7899 from Hamamatsu with an effective circular area of
\unit[22]{mm} diameter. The PMTs have ten dynode stages and provide a
gain of $2.0\times 10^6$. Typical rise times are \unit[1.6]{ns} and
pulse lengths are \unit[17]{ns}.
%
%------------------------------------------------------------------------------------------------------------
\subsection{Target slow control}
\label{subsec:SlowControl}
%------------------------------------------------------------------------------------------------------------
The gas jet target setup has 547 controllable parameters in total,
requiring a full-scale control system. For this purpose, EPICS
(Experimental Physics and Industrial Control System)~\cite{EPICS} is
used. This is an open source, flexible and scalable system which can
be supplemented by modules supplied by the users. For visualization
and as graphical user interface, Control System Studio (CSS) is
used~\cite{CSS}. CSS is a tool for monitoring and operating
large-scale control systems, that is based on the Eclipse source code
development system.

Most devices are controlled by separate front-end computers
(input-output controllers), currently seven Raspberry Pi with a Debian
based Linux system. The communication is mostly achieved by the
standard Asynchronous Driver Support (asynDriver4-31) and StreamDevice
(V2-6) modules of EPICS~\cite{StreamDevice,ASYN}. For the controller
of the pumping station an additional StreamDevice BUS API is needed to
control Profibus devices.
%
%------------------------------------------------------------------------------------------------------------
\subsection{Analog sensor readout}
\label{sec:readoutADC}
%------------------------------------------------------------------------------------------------------------
The analog sensor readout system enables digitizing the output
voltages and currents of the analog sensors used in the system. The
readout consists of an analog sensor interface, an analog-to-digital
converter (ADC), and a Raspberry Pi. The hardware drivers controlling
the readout are embedded in EPICS.

The analog sensor interface is a custom-built circuit board that
provides a voltage of up to \unit[24]{V} and enables a channel-wise
input configuration by jumpers. The input options are single-ended,
differential voltage measurement, or on-board high-precision
current-voltage conversion. The output voltage is routed to the ADC
board.

The ADC board was custom-built for MESA and MAGIX to share a common
hardware platform for slow control and data acquisition.  This board
provides a relay switch, six digital LVTTL lines, and eight analog
differential inputs accepting true negative signals. The analog part
is based on two 24-bit delta-sigma analog-to-digital converters of
type ADS1248~\cite{TI-ADS1248}. The native full-scale input range of
$V_{\text{pp}} = \pm$ \unit[2.048]{V} is expanded using a differential
voltage divider to \unit[$\pm$ 10]{V}. The resistors share a common
housing and are temperature matched. The input sensitivity can be
increased by a programmable-gain amplifier in powers of two within the
range between $\times 1$ and $\times 128$. The input channels are
sampled during an acquisition time of \unit[50]{ms} that enables an
attenuation of the power grid frequency of \unit[($50 \pm 1$)]{Hz} by
\unit[66]{dB} using finite impulse response filters. External
parameters, such as supply voltages and the board temperature, are
acquired with the maximum sampling frequency of \unit[2]{kHz}. All
values are smoothed with an adjustable moving average filter. The
board enables the integration of analog sensors into the EPICS slow
control system. With the board's capability of digitizing voltage as
well as current outputs it can be connected to a wide range of
different sensors.
%
%############################################################################################################
\section{Operation of the gas jet target}
\label{sec:TargetOperation}
%############################################################################################################
%------------------------------------------------------------------------------------------------------------
\subsection{Alignment of nozzle and catcher}
%------------------------------------------------------------------------------------------------------------
\begin{figure}[htbp]
  \centering
  \includegraphics[angle=270, width=\columnwidth]{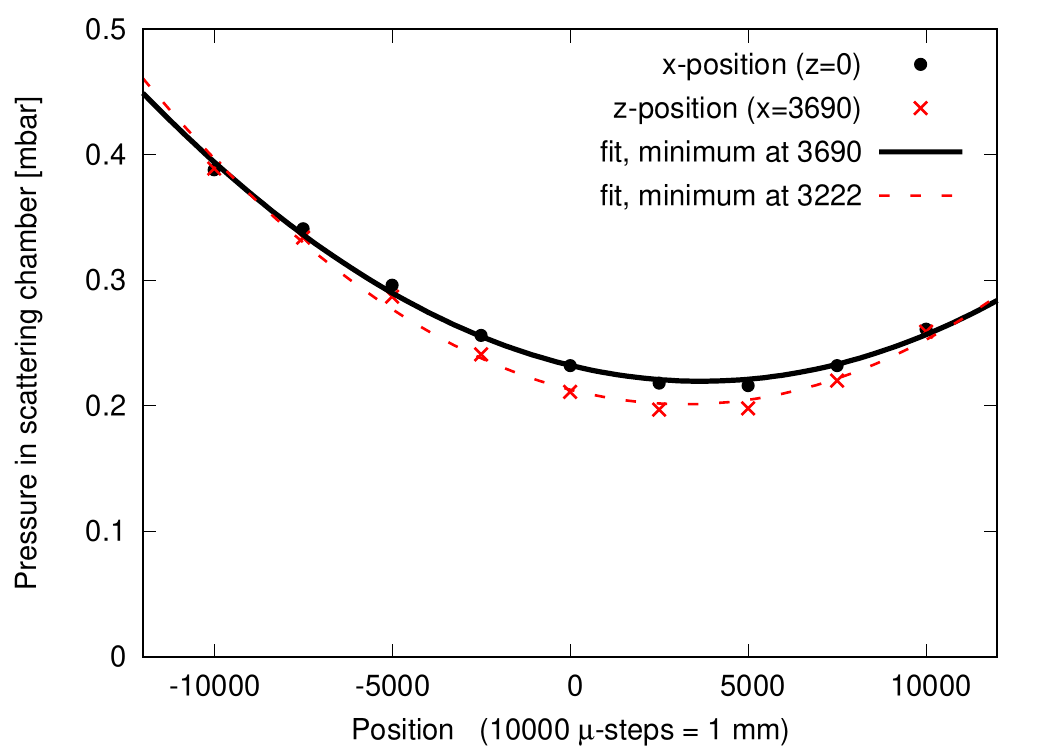}
  %generated: schlimme@freestyle:~/git/MXJetTargetAtA1/work$ gnuplot play_plotcatcherlog_bss.gnuplot;
  %epstopdf play_catcherlog.eps ResidualPressureVsPosition.pdf
  \caption{Residual pressure inside the scattering chamber as a
    function of the horizontal catcher position for a gas flow rate of
    $\unit[1200]{l_n/h}$. The optimum position relative to the target
    nozzle is at the pressure minimum. The $z$ coordinate denotes the
    direction along the beamline, the $x$ coordinate is perpendicular
    to it, both have arbitrary offsets.}
  \label{fig:PressureVsPosition}
\end{figure}
First, the target nozzle and the catcher were aligned optically at
room temperature in a position at the center of the spectrometers
using theodolites. It was later observed, that the nozzle tip moved
upwards by $\Delta y \approx \unit[2]{mm}$ and horizontally by $\Delta
x, \Delta z\approx \unit[\nicefrac{1}{4}]{mm}$ due to thermal
contraction during the cooling of the target.  Therefore, the warm
nozzle had to be aligned in such a way that it moved to the
approximate center position during cooling. The remaining vertical
offset from the center position could be determined by using the
target camera, which simultaneously delivered a centered view of the
screen with its scale and the target nozzle.  The horizontal offset
could be determined using the electron beam during data taking, see
Subsect.~\ref{subsec:DensityProfile_HorizontalDisplacement}.  To
account for a remaining misalignment of the nozzle to the catcher, the
catcher position was adjusted along both horizontal directions with
the stepper motors.  For its optimum position relative to the target
nozzle, a minimum pressure inside the scattering chamber was reached,
see Fig.~\ref{fig:PressureVsPosition}.  In order to better align the
cold target in the future, metal windows with view ports will be
attached to the scattering chamber.
%
%------------------------------------------------------------------------------------------------------------
\subsection{Thermodynamic conditions}
\label{subsec:thermodynamics}
%------------------------------------------------------------------------------------------------------------
During the operation of the gas jet target with hydrogen, the
temperature $T_0$ is typically chosen to be \unit[40]{K} and the gas
flow rate $q_V$ is varied depending on the required areal thickness.
The gas pressure $p_0$ at the nozzle inlet depends on $T_0$ and $q_V$
as well as the cross-sectional area $A^*$ at the narrowest part of
the nozzle. These variables follow the relation~\cite{Taeschner:2013}
\begin{equation}
\label{eq:gasFlow}
    q_V = A^* \frac{p_0}{\sqrt{M T_0}} \frac{T_N}{p_N} \left( \frac{2}{\kappa+1} \right)^{\frac{\kappa+1}{2(\kappa-1)}} \sqrt{\kappa R}\,,
\end{equation}
with the normal temperature $T_N = \unit[273.15]{K}$ and pressure $p_N
= \unit[1.01325]{bar}$. For a specific combination of temperature and
gas flow rate, the nozzle has to be designed such that on the one hand
a sufficient pressure is achieved to form a supersonic jet, and on the
other hand the maximum allowable operating pressure of the system is
not exceeded. Typical pressures are in the range of $p_0 =$
\unit[4]{bar} to \unit[15]{bar}.  When operating the gas jet target,
the variables $p_0$, $T_0$, and $q_V$ are measured and their mutual
relation can be checked. Differences occur, e.g., when impurities lead
to nozzle choking.

When the areal thickness at the nozzle exit is assumed to be
homogeneous, it can be estimated by
\begin{equation}
\label{eq:thickness_simple}
     \rho_{\text{areal}} = 4N \frac{q_V}{\pi d v} \frac{p_N N_A}{T_N R}
\end{equation}
from the nozzle outlet diameter $d$, the number of atoms per molecule
$N$, the Avogadro constant $N_A$, the gas flow rate $q_V$ and the gas
velocity $v$ as given in Eq.~\ref{eq:gasVelocity}. For hydrogen ($N =
2$), an areal thickness of $\rho_{\text{areal}} = \unit[5 \times
  10^{18}]{atoms/cm^2}$ is reached, when using $d = \unit[0.1]{cm}$
and $q_V = \unit[2400]{l_n/h}$ at $T_0 = \unit[40]{K}$. Due to the
finite divergence of the gas jet, this value has to be interpreted as
an upper limit.
%
%------------------------------------------------------------------------------------------------------------
\subsection{Gas flow rate during electron beam studies}
\label{subsec:ApplicableGasFlow}
\begin{figure}[htbp]
  \centering
  \includegraphics[width=\columnwidth]{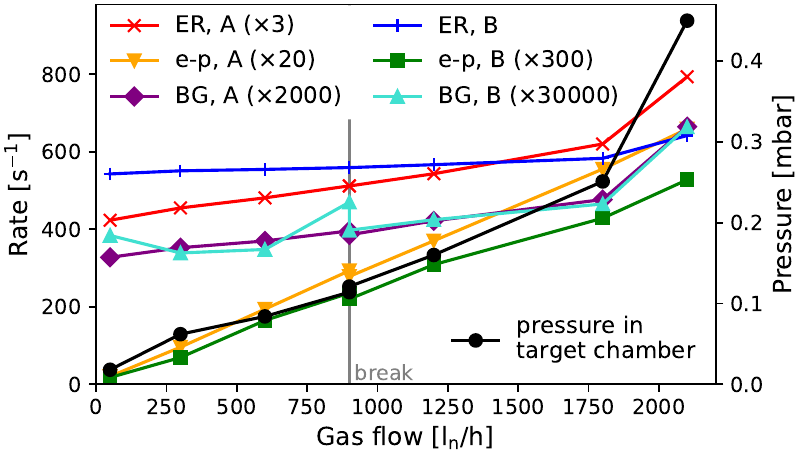}
  %cola@a1analyse ~/bss $ python RateVsFlow_new.py
  %cola@a1analyse ~/bss $ python3 RateVsFlow_supernew.py
  \caption{Flow rate-dependent observables during the electron beam
    study with the hydrogen jet target: Processed event rate (ER) of
    spectrometers A and B, rate of elastic e-p scattering events as
    selected by a cut on the missing mass peak region, and rate of
    background events (BG) from a missing mass sideband
    analysis. Individual scaling factors have been applied. The
    pressure inside the target chamber is shown as well. The data
    points were linearly connected to guide the eye.}
  \label{fig:RateVsFlow}
\end{figure}

%------------------------------------------------------------------------------------------------------------
Figure~\ref{fig:RateVsFlow} shows the measured scattering rates for a
gas flow rate between $q_V = \unit[50]{l_n/h}$ and
$\unit[2200]{l_n/h}$ during the electron beam studies.  While the rate
of the reaction of interest, elastic \mbox{e-p} scattering, is
proportional to the gas flow rate, the rate of background related
events is almost constant up to $q_V \approx \unit[1800]{l_n/h}$. The
nature of the background is studied in detail in
Sect.~\ref{sec:Background}.  Its major contributions are scattering
events of electrons off the nozzle and the catcher. The handling of
the data requires a background subtraction, and the related systematic
uncertainty scales with the fraction of background to signal
events. This makes a high gas flow rate highly desirable. However,
with increasing flow rate, the gas pressure increases as seen in the
figure.
The applicable flow rate is limited by the residual pressure inside
the scattering chamber. One limiting value is $p \approx
\unit[0.5]{mbar}$, the maximum for the attached turbomolecular pump,
another is $\unit[1]{mbar}$, the maximum for the backflow of target
gas into the beamline towards the accelerator. For most of the studies
with the electron beam, the gas jet target was operated at $q_V =
\unit[1200]{l_n/h}$ without the booster stage and with a residual
pressure inside the scattering chamber of $p\approx \unit[0.2]{mbar}$.
Optimization of the pumping power as well as the scattering chamber,
nozzle and catcher geometries should make it possible to operate the
gas jet target at $q_V = \unit[2400]{l_n/h}$ with a lower residual
pressure in the future.  With the experience gained at the A1
spectrometer setup, the conditions for the MAGIX setup could be
extrapolated and a vacuum system with eight smaller pumps was
designed, which features a better characteristic at high pressure and
will - according to calculations based on the pumping speed curves
provided by the manufacturers - allow to improve the residual pressure
inside the scattering chamber by one order of magnitude.
%
%############################################################################################################
\section{Characterization of the gas jet}
\label{sec:Characterization}
%############################################################################################################
The hydrogen gas jet was studied with a gas flow rate of up to $q_V =
\unit[1500]{l_n/h}$. The electron beam was rastered on the target, and
the number of elastic scattering events in the spectrometers were
measured as a function of the beam displacement. The results are of
particular interest for the alignment of the electron beam with the
gas jet and for the optimization of the position and the dimension of
the catcher.  Furthermore, the profile measurements allow to benchmark
simulations of the jet properties, and to test target nozzle designs.
%
%------------------------------------------------------------------------------------------------------------
\subsection{Density profile and horizontal displacement}
\label{subsec:DensityProfile_HorizontalDisplacement}
\begin{figure}[htbp]
  \centering
  \includegraphics[width=0.76\columnwidth]{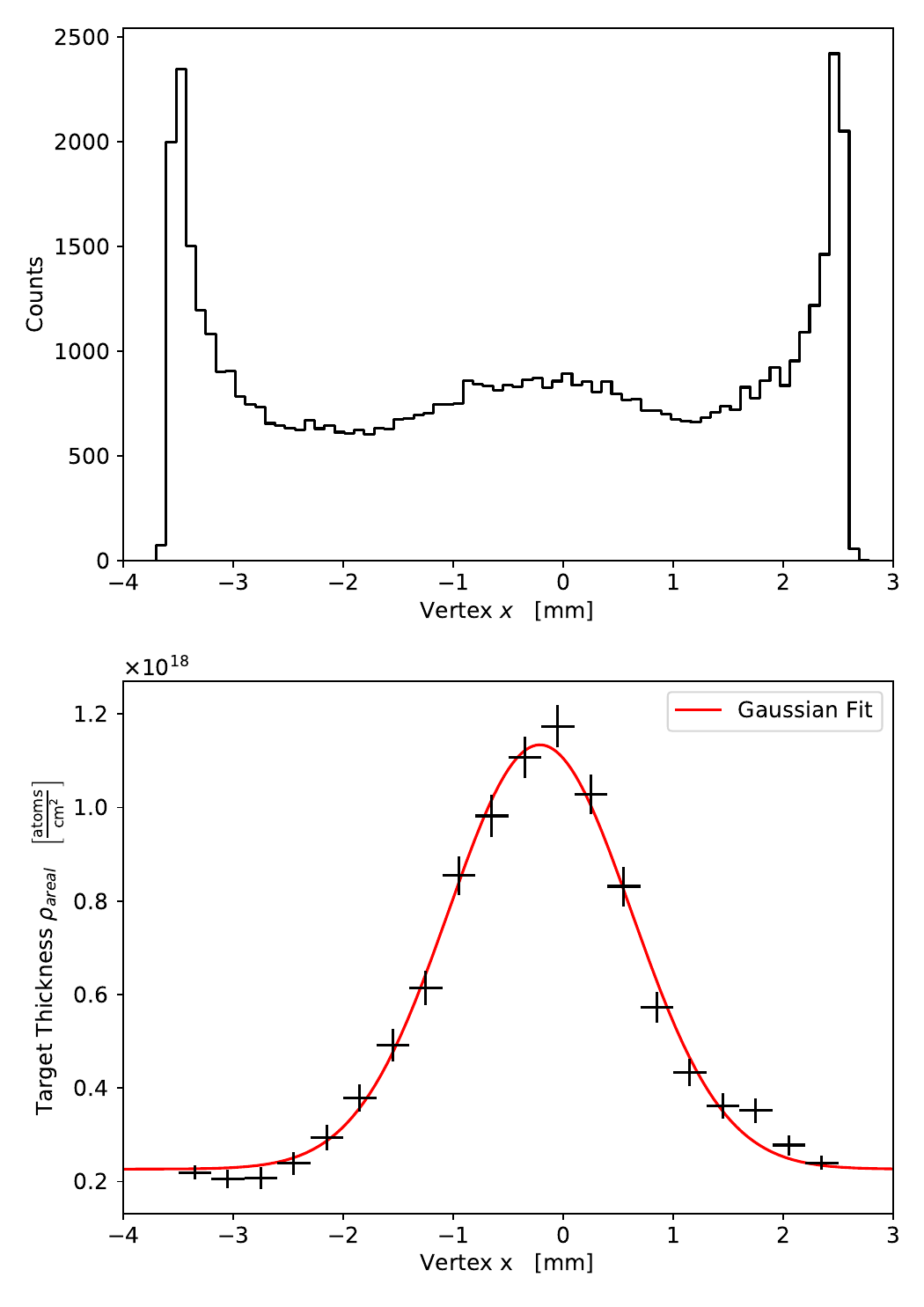}
  \caption{Top: Measured vertex distribution of scattering events. The
    spectrum is dominated by the wobbler displacement of the electron
    beam. The small bump in the center is caused by reactions with the
    gas jet. Bottom: Calculated areal thickness profile showing a
    clear peak with a Gaussian shape. From a fit to the data,
    parameters such as peak position, maximum, and width were
    extracted~\cite{Brand:2019}.}
  \label{fig:Profile2018}
\end{figure}

%------------------------------------------------------------------------------------------------------------
The measured vertex distribution of scattering events as presented in
the upper panel of Fig.~\ref{fig:Profile2018} is dominated by the
wobbler displacement of the beam and exhibits only a small bump in the
center. For the data analysis, the vertex displacement, the scattering
angle and the missing mass were reconstructed.  The missing mass
\begin{equation}
    m_{\mathrm{miss}} = \sqrt{(k_e -k_e' + p_p)^2}
\end{equation}
with the 4-momenta $k_e$ of the incoming electron (given by the beam),
$k_e'$ of the scattered electron (reconstructed by the spectrometer
data), and $p_p$ of the initial proton (assumed to be at rest), can be
reconstructed event-wise for selecting elastic scattering events. For
elastic e-p scattering, it peaks at the proton mass $m_p$, with
radiative effects resulting in a radiative tail at larger values.
Scattering events originating at the nozzle exit and the catcher
entrance were removed by identifying the elastic e-p scattering events
in the missing mass spectrum.
For each vertex position, the number of elastic e-p scattering events
was determined as a function of the scattering angle and corrected for
the spectrometer acceptances from simulations. The luminosity for the
measured integrated beam current at each vertex position was then
extracted by scaling the count rate to the known cross
section~\cite{Bernauer:2010}. After correcting for the wobbler
displacement function, the areal thickness profile was calculated, see
the lower panel of Fig.~\ref{fig:Profile2018}. It exhibits a Gaussian
shape, from which a constant background contribution ($\unit[(0.23 \pm
  0.01)\cdot 10^{18}]{atoms/cm^2}$), which corresponds to the
scattering on residual gas inside the scattering chamber, the peak
position, the peak maximum ($\unit[(0.91 \pm 0.03)\cdot
  10^{18}]{atoms/cm^2}$ above background) as well as the peak width
($\sigma=\unit[(0.833 \pm 0.024)]{mm}$) were extracted.  With a target
length acceptance along the beamline of approximately $\unit[5]{cm} /
\sin(\theta_{\text B}) = \unit[19]{cm}$ for the spectrometer angle
$\theta_{\text B}=15^\circ$ and assuming an axially symmetric gas jet
beam, a volume density of $\unit[(4.3 \pm 0.2)\cdot
  10^{18}]{atoms/cm^3}$ at the jet center with a background gas
density on a $\unit[1\cdot 10^{16}]{atoms/cm^3}$ level is estimated.

%
%------------------------------------------------------------------------------------------------------------
\subsection{Gas jet density}
\label{subsec:TargetDensity}
%------------------------------------------------------------------------------------------------------------
Assuming that the gas jet has a two-dimensional Gaussian shape
perpendicular to the flow direction, the maximum volume density in the
beam center can be expressed by
\begin{equation}
    \rho = N \frac{q_V}{2\pi\sigma^2 v} \frac{p_N N_A}{T_N R}
\end{equation}
with $N=2$ for hydrogen, the jet width $\sigma$, and the gas velocity
$v$ from Eq.~\ref{eq:gasVelocity}. Integrating along the direction of
the electron beam leads to the maximum areal thickness of
\begin{equation}
\label{eq:thickness_gauss}
    \rho_{\text{areal}} = N \frac{q_V}{\sqrt{2\pi\sigma^2} v} \frac{p_N N_A}{T_N R}\
\end{equation}
which provides a better approximation than
Eq.~\ref{eq:thickness_simple}, but requires the knowledge of the jet
width~\cite{Brand:2019}. Assuming a hydrogen flow rate of $q_V =
\unit[2400]{l_n/h}$ at a temperature of $T_0 = \unit[40]{K}$ and a
typical jet width of $\sigma = \unit[1]{mm}$, this results in a
maximum volume density of approximately $\rho = \unit[6.3 \times
  10^{18}]{atoms/cm^3}$ corresponding to a maximum areal thickness of
$\rho_{\text{areal}} = \unit[1.6 \times 10^{18}]{atoms/cm^2}$.
%
%------------------------------------------------------------------------------------------------------------
\subsection{Gas jet clustering}
\label{subsec:JetClustering}
\begin{figure}[htbp]
  \centering
  \includegraphics[angle=0, width=\columnwidth]{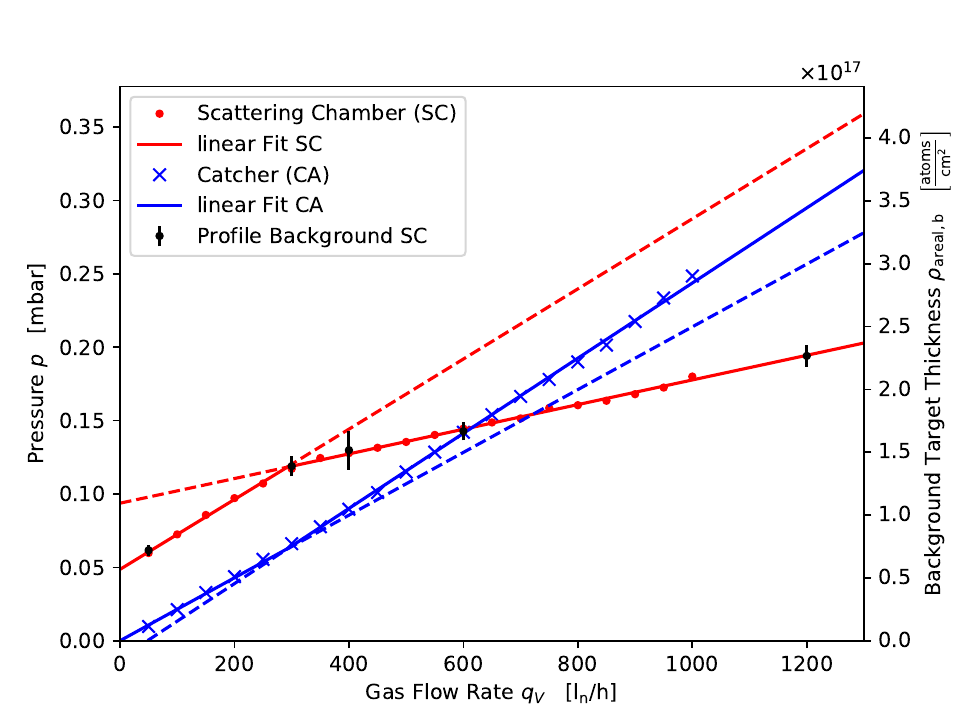}
  \caption{Pressure inside the scattering chamber and the catcher as a
    function of the hydrogen gas flow rate.  The measured background
    target thickness, which corresponds to e-p scattering not on the
    gas jet directly but on residual gas inside the scattering
    chamber, scales with the pressure inside the chamber
    accordingly. The catcher position was at its optimum,
    cf.\ Fig.~\ref{fig:PressureVsPosition}. At a flow rate $q_V >
    \unit[300]{l_n/h}$ the smaller slope of the pressure inside the
    chamber and the larger slope inside the catcher indicate a less
    divergent gas jet.}
  \label{fig:PressureVsFlowRate}
\end{figure}

%------------------------------------------------------------------------------------------------------------
The divergence of the gas jet, which is also influenced by the onset
of clustering formation, has observable effects on the gas pressure as
a function of the flow rate.  In Fig.~\ref{fig:PressureVsFlowRate},
this is shown for the residual pressure inside the scattering chamber
and the pressure inside the catcher. The change of both of the slopes
at $q_V\approx \unit[300]{l_n/h}$ indicates a more focused gas jet for
higher flow rates, so that a larger fraction of it is collected by the
catcher. This can be interpreted as a change of state of the gas jet.

%
%------------------------------------------------------------------------------------------------------------
\subsection{Simulation of the gas jet}
\label{subsec:JetSimulation}
%------------------------------------------------------------------------------------------------------------
Numerical simulations were performed to model the expansion of the gas
into the scattering chamber~\cite{Brand:2019}. These simulations also
allow to optimize the shape of the nozzle outlet which has a large
impact on the jet divergence.

The results have been obtained with the solver \emph{rhoCentralFoam}
in OpenFOAM~\cite{Greenshields:2009}. This solver is able to simulate
the compressible, supersonic fluids in a convergent--divergent jet
target nozzle, where the speed of sound is reached at the end of the
convergent part. Due to conditions close to the phase transition from
gas to liquid, the gas properties are modeled with the Peng-Robinson
equation-of-state~\cite{Peng:1976}:
\begin{equation}
    p = \frac{R_s T}{v-b} - \frac{ a \, \alpha(T)}{v^2 + 2vb - b^2}
\end{equation}
with the specific gas constant $R_s=R/M$, the specific volume
$v=1/\rho$, the constants
\begin{equation}
    a = \frac{0.45724 \, R_s^2 T_c^2}{p_c} \quad \text{and} \quad b = \frac{0.0778 \, R_s T_c}{p_c}
\end{equation}
and the function
\begin{equation}
    \alpha(T) = \left( 1+\lambda \left( 1-\sqrt{T/T_c} \right) \right)^2
\end{equation}
with
\begin{equation}
    \lambda = 0.37464 + 1.54226\,\omega - 0.26992\,\omega^2\,.
\end{equation}
The values of the state variables depend on the temperature $T_c$ and
pressure $p_c$ at the critical point, and the acentric factor
$\omega$, that describes the molecular shape.

\begin{figure}[htbp]
  \centering
  \includegraphics[width=\columnwidth]{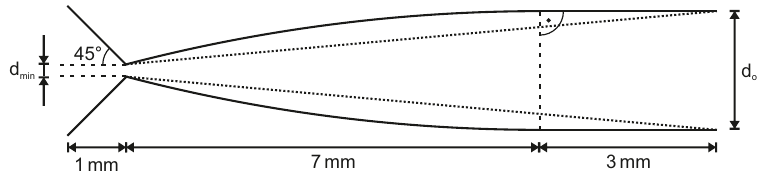}
  \caption{Two different design types of the nozzle outlet considered
    in the simulations. The linear nozzle outlet, that constantly
    diverges over the complete outlet length of \unit[10]{mm}, is
    drawn with the dotted line. The cup-shaped nozzle is drawn with
    solid lines. Its outlet forms an arc on the first \unit[7]{mm} and
    a constant diameter for the last \unit[3]{mm}.}
  \label{fig:Nozzle}
\end{figure}

The simulations have been performed for six different nozzles
geometries with two different designs for the nozzle
outlet~\cite{Brand:2019}. All simulated nozzles start with a linearly
converging inlet with a half-opening angle of $45^\circ$ and a length
of \unit[1]{mm}. The inlet ends with the narrowest inner diameter,
which was $d_{\text{min}} = \unit[0.5]{mm}$ for the original nozzle
used during the electron beam studies in 2018. It is then followed by
the outlet with a length of \unit[10]{mm}, which diverges linearly
towards the outlet diameter of $d_\mathrm{o} = \unit[1]{mm}$. This
nozzle type was also simulated with a smaller narrowest inner diameter
of $d_{\text{min}} = \unit[0.2]{mm}$ and a larger outlet diameter of
$d_\mathrm{o} = \unit[2]{mm}$. In addition to that, a cup-shaped
outlet was simulated where the outlet diameter diverges on the first
\unit[7]{mm} and then stays constant for the last
\unit[3]{mm}. Figure~\ref{fig:Nozzle} shows the two different design
types that have been simulated with different diameters.

\begin{figure}[htbp]
  \centering
  \includegraphics[width=\columnwidth]{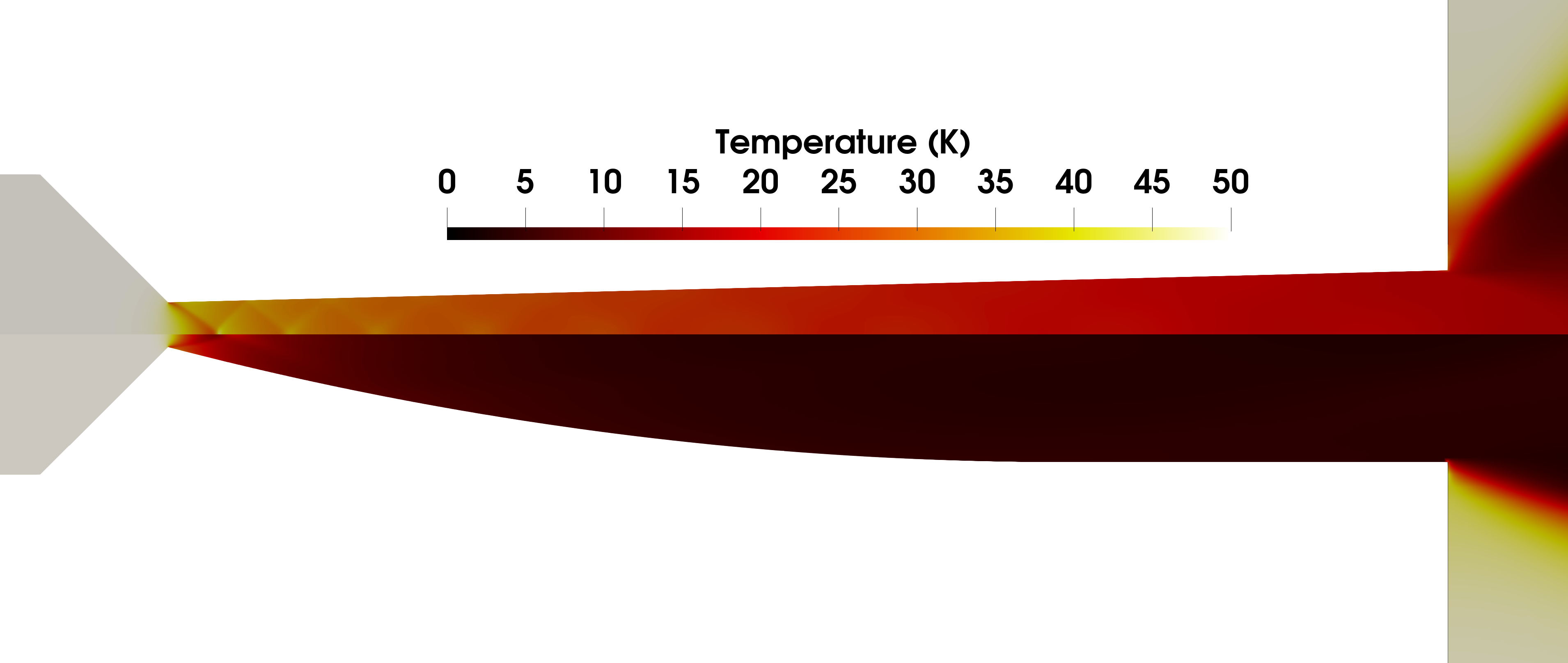}
  \caption{Comparison of the simulated temperature distribution within
    the linear nozzle with a narrowest inner diameter of
    $d_{\text{min}} = \unit[0.5]{mm}$ and an outlet diameter of
    $d_\mathrm{o} = \unit[1]{mm}$ (top) and a cup-shaped nozzle that
    diverges from $d_{\text{min}} = \unit[0.2]{mm}$ to $d_\mathrm{o} =
    \unit[2]{mm}$ (bottom). The oscillations at the narrowest inner
    diameter are typical shock waves for supersonic fluids. On the
    right a small part of the scattering chamber volume is shown. Both
    simulations have been performed for a gas flow rate of $q_V =
    \unit[1500]{l_n/h}$ and an inlet temperature of $T_0 =
    \unit[50]{K}$~\cite{Brand:2019}.}
  \label{fig:NozzleSimTemp}
\end{figure}

These simulations show that two effects have a major influence on the
target performance. The first one is the dependence of the expansion
within the nozzle on the divergence as can be seen in
Fig.~\ref{fig:NozzleSimTemp}. A greater divergence of the nozzle
causes a faster expansion and faster temperature decrease. Secondly,
the exact outlet shape is crucial, as the expansion within the
cup-shaped nozzles progresses more rapidly as compared to the linear
shape. A cup-shaped nozzle with a small $d_{\text{min}}$ and large
$d_\mathrm{o}$ leads to lower temperatures at the nozzle exit. A small
$d_{\text{min}}$ has the additional advantage of a higher pressure
that is needed for the same volume flow rate, so that the conditions
are closer to the vapor pressure curve. In combination with the lower
temperature, this increases the probability for cluster production and
thus for a more directed jet.
\begin{figure}[htbp]
  \centering
  \includegraphics[width=\columnwidth]{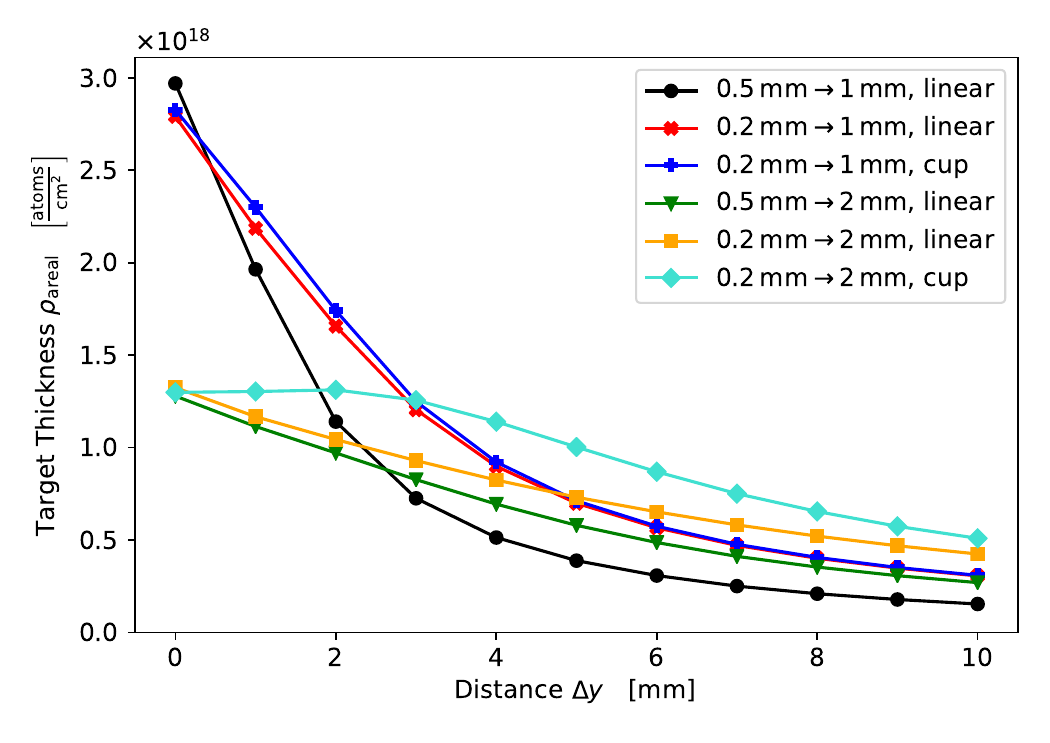}
  \caption{Comparison of the simulated areal thickness for six
    different nozzle designs as a function of the distance to the
    nozzle outlet for a gas flow rate of $q_V = \unit[1500]{l_n/h}$
    and a temperature of $T_0 = \unit[50]{K}$. Nozzles with a smaller
    outlet diameter have a larger thickness near the outlet, but
    exhibit a steeper decrease with distance~\cite{Brand:2019}. The
    points were interpolated linearly to catch the
    eye~\cite{Brand:2019}.}
  \label{fig:NozzleSimDensity}
\end{figure}
The simulated areal thickness as a function of the distance to the
nozzle outlet for different nozzle designs is summarized in
Fig.~\ref{fig:NozzleSimDensity}. In agreement with
Eq.~\ref{eq:thickness_simple}, the thickness for nozzles with smaller
$d_\mathrm{o}$ is larger directly behind the nozzle, but already at a
distance of $\Delta y = \unit[3]{mm}$ the cup-shaped nozzle with a
larger $d_\mathrm{o}$ leads to the largest thickness. This is caused
by a smaller divergence due to the faster expansion within the nozzle,
so that a compromise in the design has to be found. Since the vertex
position is typically in a distance of $\Delta y \approx \unit[4]{mm}$
behind the nozzle outlet, a cup-shaped nozzle was produced and tested
with the electron beam.
%
%------------------------------------------------------------------------------------------------------------
\subsection{Comparison of simulations with experiment}
\label{subsec:DifferentNozzles}
\begin{figure}[htbp]
  \centering
  \includegraphics[width=\columnwidth]{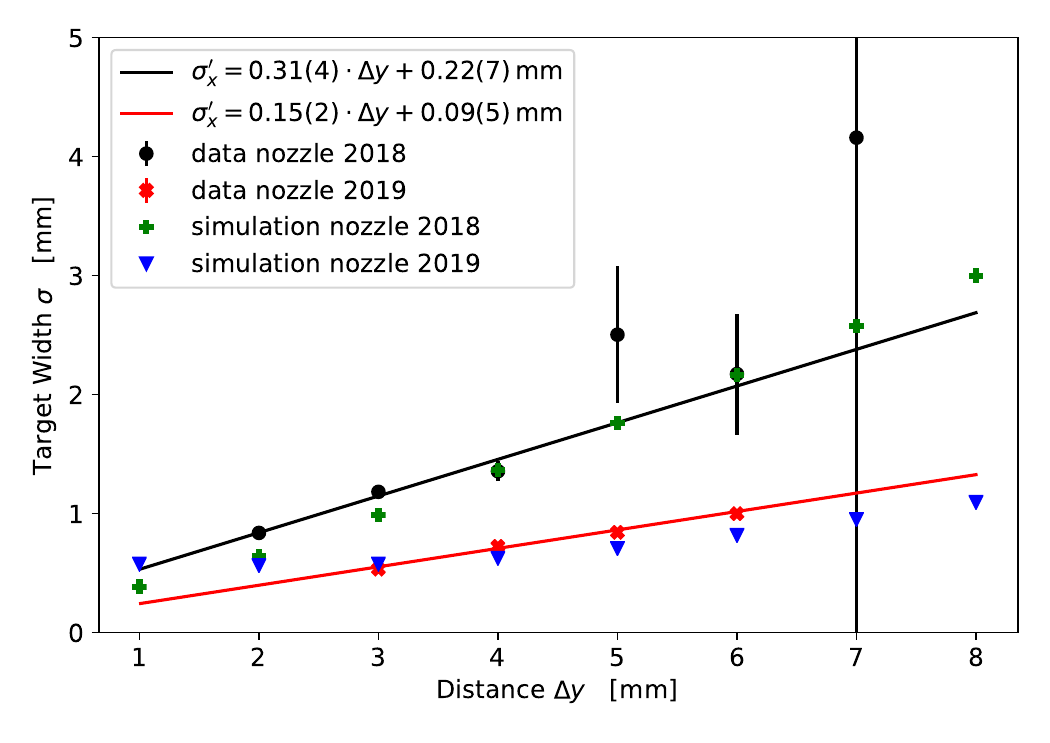}
  \caption{Measured target widths and simulations. The nozzle used in
    2018 had a linear shape with $d_\text{min} = \unit[0.5]{mm}$ and
    $d_\mathrm{o} = \unit[1]{mm}$, while the nozzle used in 2019 had a
    cup shape with $d_\text{min} = \unit[0.2]{mm}$ and $d_\mathrm{o} =
    \unit[2]{mm}$. The experimental conditions were slightly
    different, in 2018 the target was operated with $q_V =
    \unit[1500]{l_n/h}$ at $T_0 = \unit[50]{K}$ while in 2019 it was
    operated with $q_V = \unit[1200]{l_n/h}$ at $T_0 =
    \unit[40]{K}$. These conditions were considered in the
    simulations.}
  \label{fig:NozzleDataVsSim}
\end{figure}

%------------------------------------------------------------------------------------------------------------
In Fig.~\ref{fig:NozzleDataVsSim} the experimentally determined target
widths from electron beam studies in 2018 and 2019 are compared to the
simulations~\cite{Brand:2019}. In 2018, a linear nozzle with
$d_{\text{min}} = \unit[0.5]{mm}$ and $d_\mathrm{o} = \unit[1]{mm}$
was used, in 2019 a cup-shaped nozzle with $d_\text{min} =
\unit[0.2]{mm}$ and $d_\mathrm{o} = \unit[2]{mm}$. The data are in
agreement with the simulations. The cup-shaped nozzle reduces the jet
divergence by a factor of two, which leads to twice the areal
thickness at the same gas flow rate. With a fit of a linear function
to the data, the half-opening angle of the gas jet was extracted. It
reduces from $17.2^\circ$ to $8.5^\circ$ when using the cup-shaped
nozzle, although the outlet diameter of this nozzle is larger.

%
%############################################################################################################
\section{Background induced by electron beam halo}
\label{sec:Background}
%############################################################################################################
The areal thickness of the gas jet is much lower than the one of the
nozzle and the catcher. In addition, the atomic number is much higher
for the nozzle (mainly copper, $Z = 29$) and the catcher (aluminum, $Z
= 13$) than for hydrogen as target gas ($Z = 1$). These elements are
located a few \unit{mm} above and below the electron beam,
respectively. As a consequence, the presence of an electron beam halo
can lead to significant background in scattering experiments, even if
the flux of electrons in the halo is many orders of magnitude smaller
than in the main core of the beam. Such a halo originates in part from
the interaction of the beam electrons with residual gas inside the
accelerator, so that it cannot be suppressed completely.

As a countermeasure, a beam halo suppression system was developed, as
presented in Subsect.~\ref{subsec:HaloBlocker}. The idea behind the
multi-step approach was to use the collimator, positioned several
meters in front of the target, to absorb a large fraction of beam halo
electrons. Since some of them may leak through the collimator or
rescatter off the jaws in forward direction, the active shielding at a
few \unit{cm} in front of the target should cover the nozzle and the
catcher.  In this study, the beam halo veto detector was not used. The
focus was the alignment of the collimator system with respect to the
target position, which is mandatory in order to achieve minimal,
stable, and reproducible background conditions.
%
%------------------------------------------------------------------------------------------------------------
\subsection{Background introduced by the collimator setup}
%------------------------------------------------------------------------------------------------------------
As the gas jet, when leaving the nozzle, is divergent, the electron
beam should be positioned as close as possible to the nozzle to
maximize the areal thickness. The positioning of the catcher close to
the nozzle is desirable in order to improve the target gas
removal. However, if nozzle and catcher are too close to each other,
small variations of the beam position would lead to a significant
change of the background situation.  Shielding the target accurately
by means of the collimator becomes more delicate for small relative
distances.
%Rescattering of halo electrons occurs at the jaws of the collimator.
In conclusion, there is an optimum collimator position for a given
distance between the catcher and the nozzle.

\begin{figure}[htbp]
  \centering
  \includegraphics[width=\columnwidth]{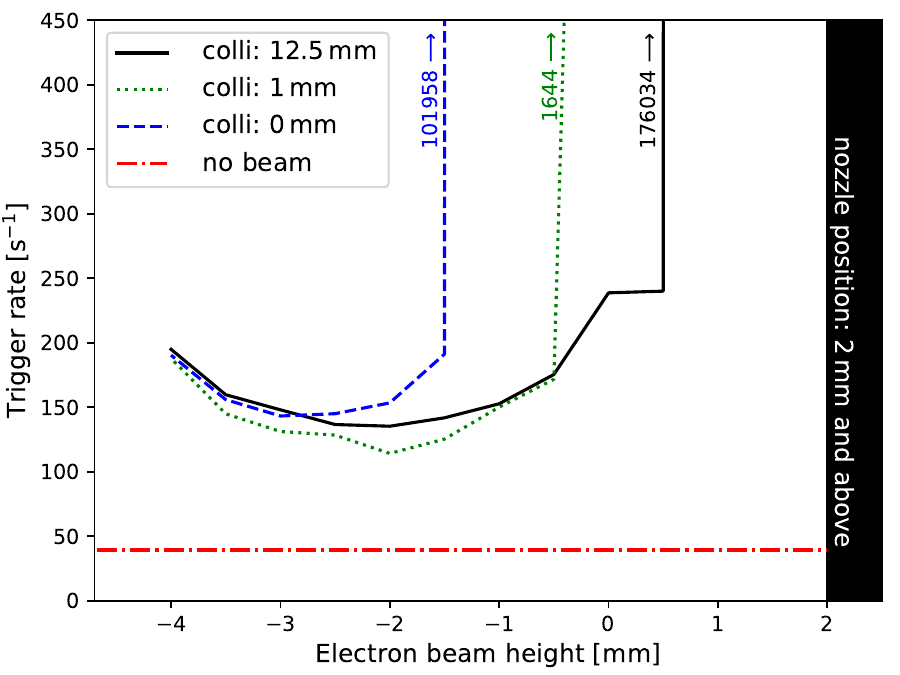}
  %generated: cola@a1analyse ~/jet2019/ana $ python 2020-01-23_HaloReductionStudy_Reduced.py
  \caption{Background event rates of spectrometer B for
    different positions of the electron beam and the upper collimator,
    which shadows the target nozzle.  When the electron beam position
    was as close as $\Delta y \approx \unit[1-1.5]{mm}$ to the nozzle
    or the collimator, the rates increased distinctively.}
  \label{fig:HaloReductionStudy}
\end{figure}

\begin{figure}[htbp]
  \centering
  \includegraphics[width=\columnwidth]{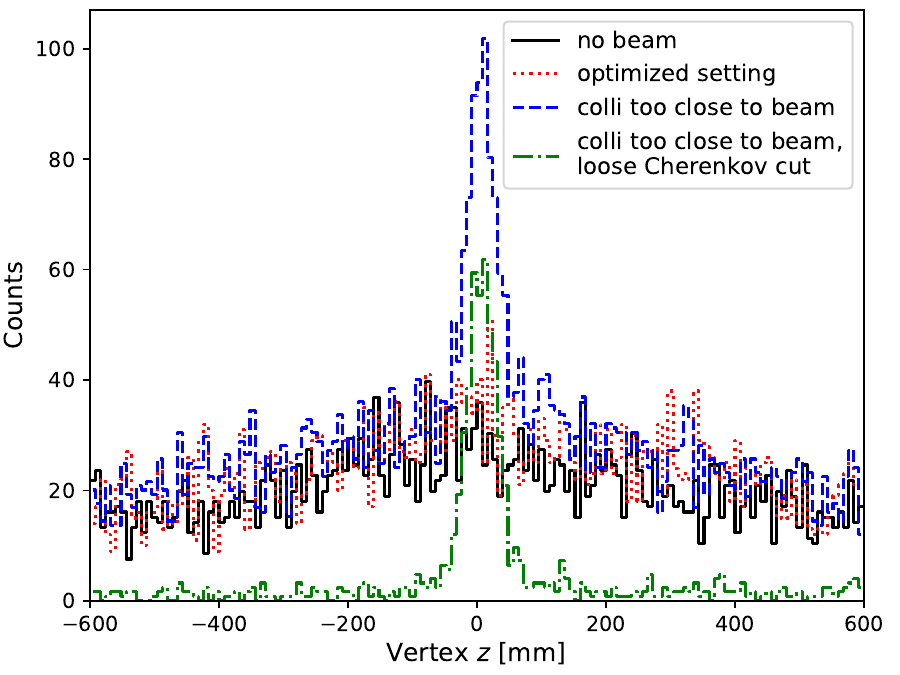}
  %cola@a1cool ~/jet2019/ana $ python 2020-01-23_HaloReductionStudy_Specs.py
  \caption{Reconstructed vertex position for selected data, corrected
    for different run lengths and data acquisition dead times.  When
    the beam position is too close to the collimator, a peak at $z=0$
    appears. A signal in the Cherenkov detector identifies these
    particles as electrons, indicating rescattering of beam halo
    electrons.}
  \label{fig:HaloReductionStudy_Specs_Vz}
\end{figure}

In an alignment study, scattering rates were measured using the
spectrometers without the gas jet for different beam and collimator
positions. A minimum distance between the electron beam from the
collimator, the nozzle, and catcher of $\unit[1.5]{mm}$ was
determined, at which no significant background was being produced by
the beam hitting these elements, see
Figs.~\ref{fig:HaloReductionStudy} and
\ref{fig:HaloReductionStudy_Specs_Vz}.
%
%------------------------------------------------------------------------------------------------------------
\subsection{Background reduction by the collimator setup}
\begin{figure}[bhtp]
  \centering
  \includegraphics[width=\columnwidth]{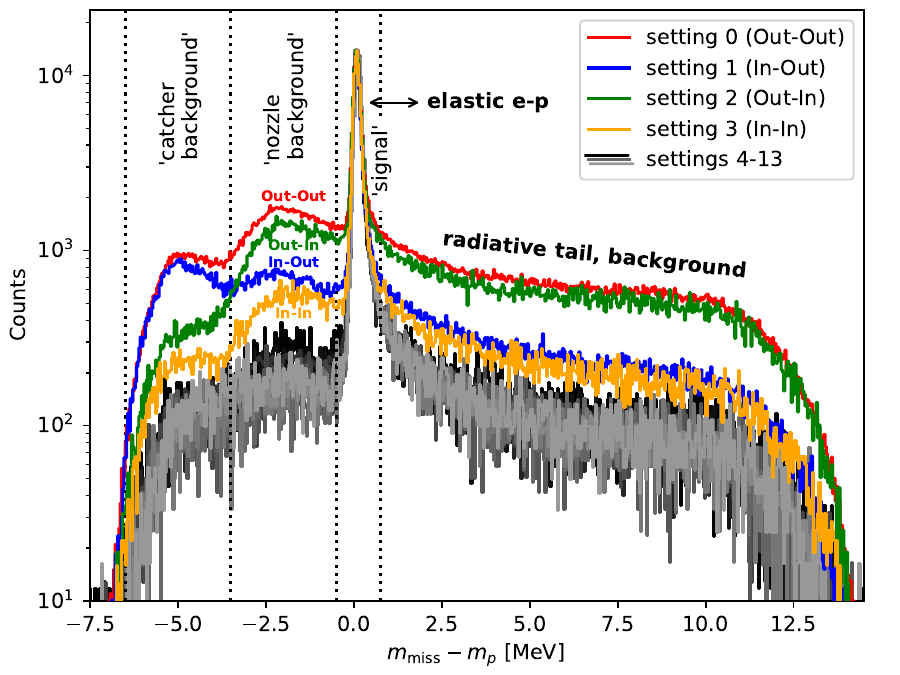}
  %generated: cola@a1analyse ~/jet2019/ana $ python 2020-02-01_DiffContribs.py
  \caption{ Missing mass spectrum measured with spectrometer B at
    $\theta_{\text B} = 15^\circ$ and a beam energy of \unit[315]{MeV}
    for different collimator settings. The target was operated with
    $\unit[1200]{l_n/h}$. The spectra were scaled to the same height
    of the elastic e-p scattering peak at $\unit[0]{MeV}$. The labels
    for settings 0 to 3 indicate the top and bottom collimator
    positions, the additional settings correspond to a fine-tuning of
    the collimator positions which is illustrated in
    Fig.~\ref{fig:Illustration}. The missing mass regions of strong
    backgrounds originating from the catcher and the nozzle are
    indicated.}
  \label{fig:DiffContribs_B_mmX}
\end{figure}

%------------------------------------------------------------------------------------------------------------
Figure~\ref{fig:DiffContribs_B_mmX} shows a missing mass spectrum
measured with spectrometer B for different collimator settings. The
elastic \mbox{e-p} peak is clearly visible at $\unit[0]{MeV}$ and at
higher missing mass values its radiative tail is present. Satellite
peaks of significant height are located at negative missing mass
values, when none of the collimator jaws is used to block the electron
beam halo (Out-Out).  Moving the lower collimator in (Out-In) shadows
the catcher (thin aluminum) from the beam halo and the number of
events in the first peak is significantly reduced. Moving the upper
collimator in (In-Out) shadows the target nozzle (thick copper) from
the beam halo and the second peak is reduced.  Moving in both
collimators (In-In) reduces the background in both regions.

\begin{figure}[htbp]
  \centering
  \includegraphics[width=\columnwidth]{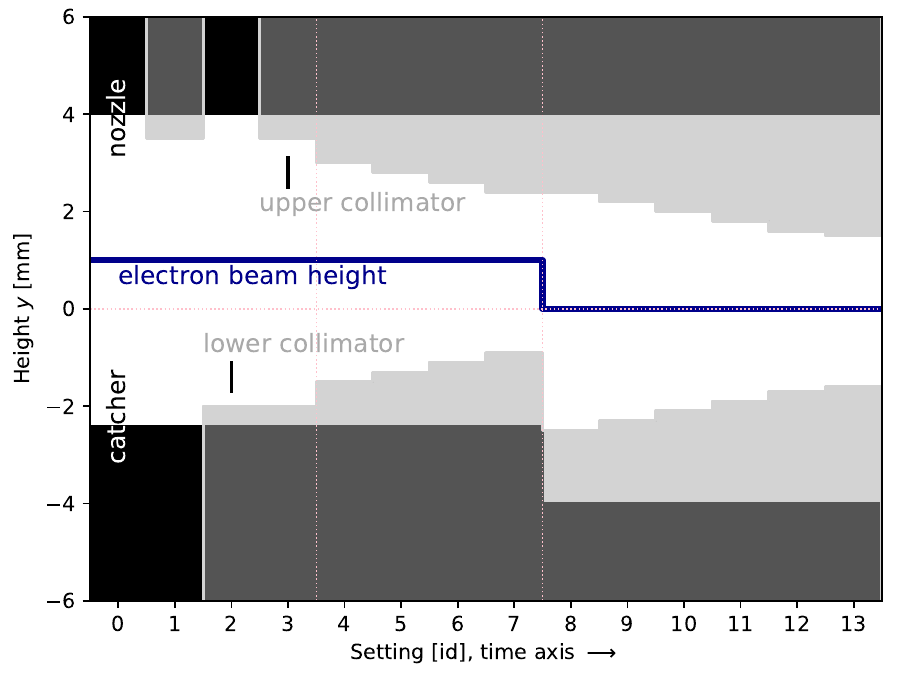}\\
  %generated: cola@a1analyse ~/jet2019/ana $ python 2020-02-0102_Illustration.py
  \includegraphics[width=\columnwidth]{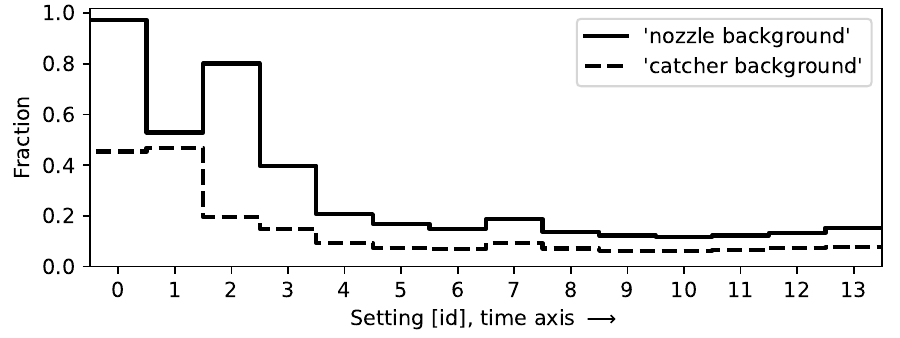}
  %generated: cola@a1analyse ~/jet2019/ana $ python 2020-02-01_DiffContribs.py
  \caption{Top: Illustration of different settings during the study of
    electron beam halo induced background. The height of the nozzle
    tip relative to the target coordinate system was fixed at
    \unit[4]{mm}. The vertical positions of the catcher and the
    electron beam were changed with setting 8. The positions of the
    upper and lower collimators were changed in small steps. Bottom:
    Number of events in the two background regions as defined in
    Fig.~\ref{fig:DiffContribs_B_mmX} relative to the number of events
    in the signal region. Setting 13 was used for the subsequent beam
    studies.}
  \label{fig:Illustration}
\end{figure}

Satellite peaks could be cut from the spectrum in the analysis of e-p
scattering. However, their radiative tails and their wide energy loss
distributions also contribute to the events in the e-p peak
region. Such a background cannot be removed event-by-event, but its
distribution needs to be subtracted from the spectrum.  As a measure
for the blocking effectivity, the number of events in the missing mass
regions associated to catcher and nozzle backgrounds were related to
the number of events in the elastic e-p region, see
Fig.~\ref{fig:DiffContribs_B_mmX}.  These fractions were evaluated for
a number of different collimator settings, which are illustrated in
Fig.~\ref{fig:Illustration}. In addition to the collimator positions,
the distance between the catcher and the nozzle was changed once, and
the beam position was adjusted accordingly. As it was expected, the
background fraction increased when the collimator jaws came too close
to the electron beam. For a given distance between the catcher and the
nozzle, an optimum position of the collimator could be found.
%
%------------------------------------------------------------------------------------------------------------
\subsection{Handling of residual background}
\label{sec:resBG}
\begin{figure}[htbp]
  \centering
  \includegraphics[width=0.9\columnwidth]{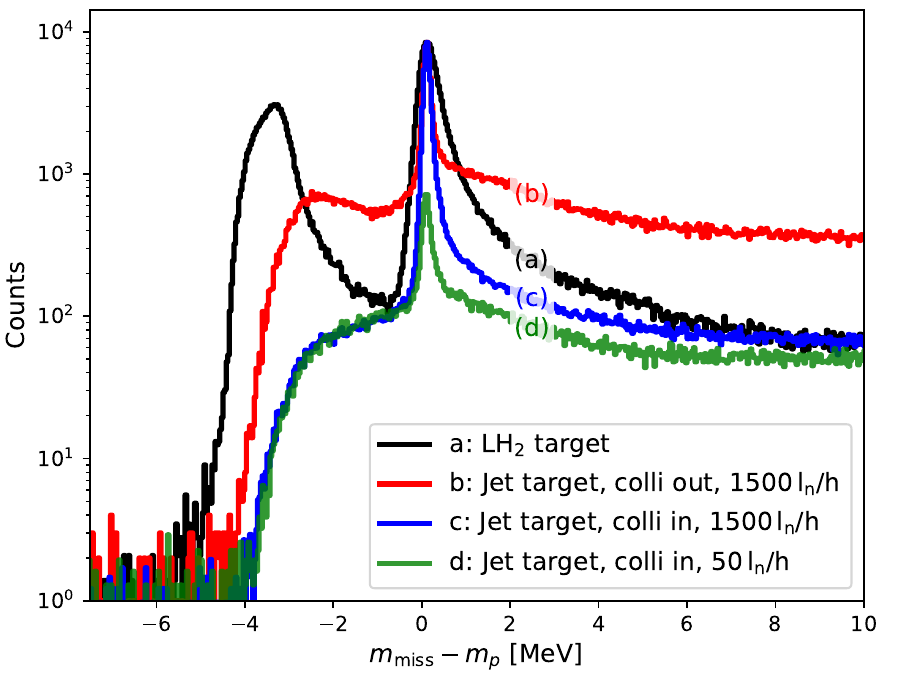}\\
  %generated: cola@a1analyse ~/jet2019/ana $ python 2020-02-06_AllIn.py
  \caption{Missing mass spectra for \unit[315]{MeV} beam energy
    measured with spectrometer~B.  The three measurements with the jet
    target were performed at $\theta_{\text B} = 15^\circ$ without the
    collimators of the beam halo blocker (colli out), with the
    collimators inserted corresponding to setting 13 (colli in), and
    with the collimators inserted and at a reduced gas flow rate. A
    soft cut on the threshold Cherenkov particle identification
    detector signal is performed for electron selection. For
    comparison, a spectrum measured with a liquid hydrogen target cell
    (black, from \cite{Bernauer:2013}) at the same beam energy and
    similar scattering angles is shown.  The gas jet target leads to a
    narrower peak and has no background from elastic scattering off
    any cell walls.}
  \label{fig:2020-02-06_AllIn_mm}
\end{figure}

%------------------------------------------------------------------------------------------------------------
By inserting the collimators, the background was suppressed by one
order of magnitude and a relatively clean e-p scattering spectrum was
achieved, see Fig.~\ref{fig:2020-02-06_AllIn_mm}.  For a
high-precision cross section measurement, i.e., with systematic
uncertainties on the sub-percent level, the residual background
contribution needs to be determined by additional measurements.

A standard background reduction technique for a liquid hydrogen target
is the measurement with an empty cell. However, a comparison of the
spectra taken with liquid hydrogen and those without provides serious
challenges. The electron beam experiences a much smaller energy loss
in the empty cell than in the filled one. For example, the mean energy
of the electrons, arriving at the backward cell wall, can differ by up
to several $\unit[]{MeV}$ with a significantly different energy
spread, compare Table~\ref{table:TargetComparison}. Since cross
sections depend strongly on the electron energy, measured spectra are
shifted and distorted. This is especially problematic as beam-cell
interactions produce a large background and its substantial tail
contributes to the e-p scattering peak region. Further, the energy
spread widens this peak, and a cut on the signal must accept a larger
fraction of the background.  One solution is to resort to
simulations~\cite{Bernauer:2013}.  One advantage of the hydrogen jet
target compared to liquid hydrogen targets is that effects from energy
loss and multiple scattering do not change significantly when the gas
flow rate is changed.  When using the gas jet target, a measurement
without any gas flow is obviously ideal to provide a pure background
measurement. In the presented study, a warm target system was avoided
because of the thermal movements that result in a different background
situation, and measurements were taken at cryogenic temperatures with
a finite gas flow rate of $q_V = \unit[50]{l_n/h}$ to minimize the
risk of a freezing nozzle in the case of any undetected vacuum
leaks. At MAGIX, a precise adjustment of the target position will be
possible during operation to accomodate any movements and measurements
will be performed without any gas flow.  The resulting spectrum in
Fig.~\ref{fig:2020-02-06_AllIn_mm} demonstrates that the rate of
elastic e-p events is significantly reduced, while the rate from
interactions of beam halo electrons with the target structure, which
dominates the spectrum outside the elastic peak region, remains
unchanged.  Although the gas flow rate is reduced by a factor of 30,
the number of counts in the elastic peak is reduced by a factor of
approximately 12 only.  The reason why this factor is lower can be
explained by the fact that no cut on the vertex position has been
applied, avoiding possible distortions of the spectra, and the event
yield associated with a scattering on the residual gas does not
decrease linearly with the gas flow, compare
Fig.~\ref{fig:PressureVsFlowRate}. The result for this factor agrees
well with the expectation based on an evaluation of data dedicated to
the density profile measurements, $12.2 \pm 1.5$.  This spectrum can
be used to model the background or to subtract it.
%, apractical example is given in \ref{sec:BGS}.
For comparison, the missing mass peak from a measurement with a liquid
hydrogen target is shown in Fig.~\ref{fig:2020-02-06_AllIn_mm} as
well.
%
%############################################################################################################
\section{Stability of the luminosity}
\label{sec:Stability}
%############################################################################################################
A sub-percent measurement of the luminosity $\mathcal L$, proportional
to beam current and areal thickness of the gas jet, is crucial for
high-precision experiments. While different techniques can be used for
the determination of the absolute value of the beam current, the areal
thickness of the gas jet can not be inferred from the measured target
parameters with high accuracy.  In addition, a small change in the
horizontal beam position of $\Delta x \approx$ \unit[$\pm$ 0.2]{mm}
can result in a significant change of the luminosity by $\Delta
{\mathcal L} \approx$ \unit[$\pm$ 2]{\%} due to the varying overlap of
the beam with the jet.
\begin{figure}[htbp]
  \centering
  \includegraphics[width=\columnwidth]{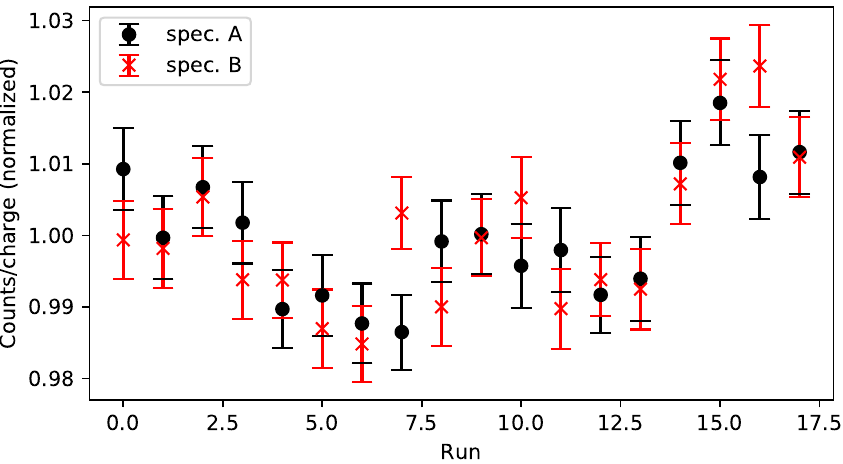}
  \includegraphics[width=\columnwidth]{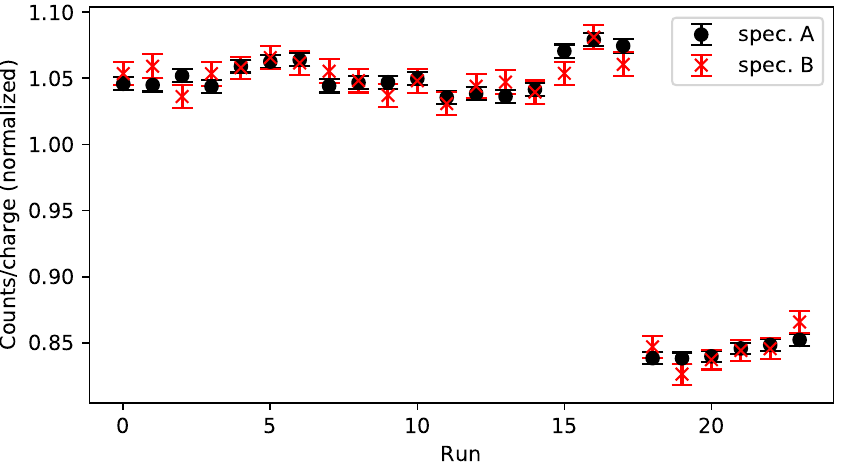}  
  \caption{ Top: Normalized event rates per run measured with
    spectrometer~A and B for the setting with B at a scattering angle
    of $\theta_{\text B} = 20^\circ$. A correlated drift of the rates
    is observable. Bottom: Normalized event rates per run measured
    with spectrometer~A and B for the setting in which the gas flow
    rate of the jet target was reduced by approximately \unit[20]{\%}
    after run 17. Both event rates follow the change in luminosity.}
  \label{fig:LVar}
\end{figure}
A simultaneous measurement of the M{\o}ller scattering cross section
could provide a relative luminosity determination.  Alternatively, if
more than one high-resolution magnetic spectrometer is available in
the experimental setup, one of them could be used at a fixed setting
for elastic e-p scattering. The achievable accuracy is limited by the
knowledge of the reference cross section. Moreover, for experiments
such as elastic form factor measurements, the data analysis can be
performed using a floating normalization, where physical constraints
are used to adjust common normalization parameters, see
Refs.~\cite{Bernauer:2010, Bernauer:2013}.

\begin{figure}[htbp]
  \centering
  \includegraphics[width=\columnwidth]{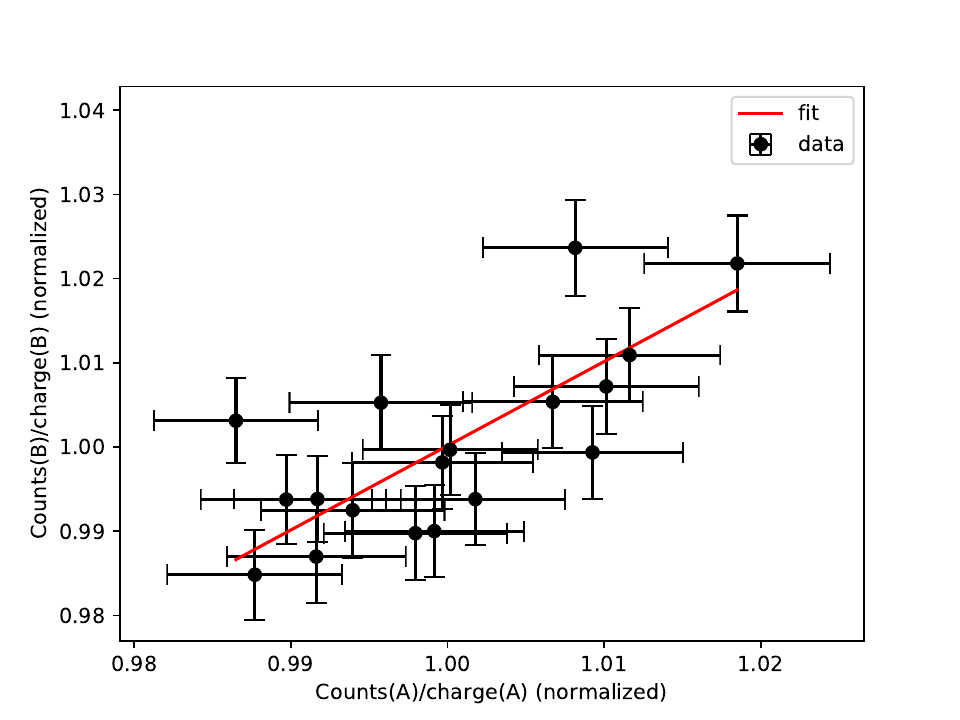}
  \caption{Correlation of the normalized event rates per run for the
    setting with spectrometer B at a scattering angle of
    $\theta_{\text B} = 20^\circ$. A linear fit through the origin was
    applied.}
  \label{fig:elastic20Mar_AvB}
\end{figure}

For different angular settings of the spectrometers, multiple runs
with a length of \unit[30]{min} each were taken over a period of
several days to study the stability of the luminosity and to test the
reliability of the relative luminosity monitoring. The number of
events in the elastic e-p scattering peak for each run was extracted
for both spectrometers by subtracting the background, correcting for
the measured data acquisition dead time, and applying a missing mass
cut at $\unit[-0.5]{MeV} < m_{\text{miss}} - m_p <
\unit[2]{MeV}$. This number was normalized by the integrated charge
from the beam current measurement using a F\"orster
probe. Figure~\ref{fig:LVar} gives an example for a setting with
spectrometer B at a scattering angle of $\theta_{\text B} = 20^\circ$.
A drift of the event rates per run beyond statistical fluctuations was
observed in both spectrometers.  Figure~\ref{fig:elastic20Mar_AvB}
demonstrates a linear correlation caused by luminosity drifts.  The
root mean square value $\text{RMS}_{\text{A/B}} \approx 0.0087$ of the
distribution of the count rate ratio from spectrometers~A and B agrees
well with the typical statistical uncertainty
$\sigma_{\text{A/B}}^{\text{typ}} \approx 0.0090$ of a single run.
In one setting the gas flow rate of the jet target was intentionally
reduced by about \unit[20]{\%}, see Fig.~\ref{fig:LVar}.  The mean
value of the ratio for the runs with a higher gas flow, \unit[0.3312
  $\pm$ 0.0030], does not deviate from the mean value for the runs
with the lower gas flow, \unit[0.3314 $\pm$ 0.0036]. The values for
all settings in this study are tabulated in
Table~\ref{tab:lumistability}.  A very important result is that no
sign of a significant systematic error contribution exceeding the
statistical uncertainty was found.  One can conclude that drifts of
the luminosity that could be on a scale of \unit[10]{\%} over several
days can be measured precisely by using one spectrometer as a
luminosity monitor.
\begin{table}[htbp]
  \caption{Luminosity study with different settings. Spectrometer A
    was fixed at a scattering angle of $\theta_{\text A} = 30^\circ$,
    while the scattering angle $\theta_{\text B}$ of spectrometer B
    was varied. For each setting $\#$ denotes the number of runs per
    setting, $D_A = (\text{max} - \text{min})/\text{mean}$ is the
    range of deviations of the event rate measured with spectrometer
    A, $\sigma_{\text{A/B}}^{\text{typ}}$ is the typical statistical
    uncertainty of the ratio of the event rates measured by the two
    spectrometers in a single run, and $\text{RMS}_{\text{A/B}}$ is
    the root mean square of its distribution.}
    \label{tab:lumistability}
  \begin{center}
  \begin{tabular}{rcccc}
    \toprule
    $\theta_{\text B}$ & $\#$ & $D_A$ & $\sigma_{\text{A/B}}^{\text{typ}}$ & $\text{RMS}_{\text{A/B}}$ \\
    \midrule
    20$^\circ$ & 18 & 0.032 & 0.00903 & 0.00868 \\
    25$^\circ$ & 47 & 0.059 & 0.00205 & 0.00217 \\
    30$^\circ$ & 73 & 0.101 & 0.00134 & 0.00128 \\
    35$^\circ$ & 82 & 0.060 & 0.00113 & 0.00143 \\
    40$^\circ$ & 82 & 0.172 & 0.00161 & 0.00134 \\
    \bottomrule
  \end{tabular}
  \end{center}
\end{table}
%
%############################################################################################################
\section{Comparison to other hydrogen targets}
\label{sec:OtherTargets}
%############################################################################################################
\begin{sidewaystable}
  \caption{Typical parameters of the hydrogen jet target for MAGIX
    compared to other hydrogen targets in current nuclear physics
    experiments at electron accelerators.  State denotes the state of
    hydrogen inside the target, setup refers to the experiment or
    setup, $T_0$ is the operating temperature, $l$ the length in beam
    direction, $\rho_{\text{areal}}$ the areal thickness,
    $I_{\text{beam}}$ the maximum beam current achievable for the
    setup, $\mathcal L$ the maximum achievable luminosity,
    $E_{\text{beam}}$ the typical beam energy, and $\Delta
    E_{\text{beam}}$ the most probable energy loss for electrons at
    the typical beam energy.  The last two columns address the
    background in terms of cell walls and hydrogen purity.  The beam
    energy loss is calculated employing the energy-dependent total
    stopping power from Ref.~\cite{ESTAR} for a full passage of the
    target including target cell walls at the front and back where in
    use. The impinging electron beam is assumed to be centered and
    point-like. The FWHM spread of the energy loss is not separately
    listed: According to Landau theory, it is of the order of
    \unit[11]{\%}. Note that the OLYMPUS and the DarkLight targets
    have side walls, and the PRad target a Kapton foil which scattered
    electrons have to pass in addition.}
  \begin {center}
    \begin{tabular}{llccccccccc}
      \toprule
      \multicolumn{1}{c}{State} &
      \multicolumn{1}{c}{Setup} &
      \multicolumn{1}{c}{$T_0$} &
      \multicolumn{1}{c}{$l$}   &
      \multicolumn{1}{c}{$\rho_{\text{areal}}$} &
      \multicolumn{1}{c}{$I_{\text{beam}}$} &
      \multicolumn{1}{c}{$\mathcal L$} &
      \multicolumn{1}{c}{$E_{\text{beam}}$} &
      \multicolumn{1}{c}{$\Delta E_{\text{beam}}$} &
      Windowless & Purity \\
      & &
      \multicolumn{1}{c}{[K]} &
      \multicolumn{1}{c}{[mm]} &
      \multicolumn{1}{c}{[$\unit[10^{18}]{\frac{atoms}{cm^{2}}}$]} &
      \multicolumn{1}{c}{[$\mu$A]} &
      \multicolumn{1}{c}{[$\unit[10^{35}]{cm^{-2}s^{-1}}$]} &
      \multicolumn{1}{c}{[MeV]} &
      \multicolumn{1}{c}{[keV]} &
      & \\
      \midrule
      gas    & gas jet, A1      & $40$ & $\sigma\approx 1$& $1$       & $20$          & $0.001$   & $195$ & $0.01$       & \checkmark & \checkmark \\
      gas    & gas jet, MAGIX   & $40$ & $\sigma\approx 1$& $2$       & $1000 - 10000$& $0.1 - 1$ & $105$ & $0.02$       & \checkmark & \checkmark \\
      gas    & PRad             & $25$ & $40$             & $2$       & $0.01$        & $0.00001$ & $2200$& $0.13$       &(\checkmark)& \checkmark \\
      gas    & DarkLight        & $293$& $600$            & $1.2 - 10$& $10000$       & $0.7 - 6$ & $100$ & $0.01 - 0.11$&(\checkmark)& \checkmark \\
      gas    & OLYMPUS          & $75$ & $600$            & $0.003$   & $65000$       & $0.01$    & $2000$& $0.0002$     &(\checkmark)& \checkmark \\\midrule
      liquid & round LH$_2$, A1 & $23$ & $20$             & $84000$   & $10$          & $52$      & $195$ & $1245$       & $\times$   & \checkmark \\
      liquid & cigar LH$_2$, A1 & $23$ & $50$             & $210000$  & $10$          & $131$     & $195$ & $2944$       & $\times$   & \checkmark \\
      liquid & LH$_2$, MUSE     & $21$ & $60$             & $251000$  & $0.0000005$   & $0.000008$& $160$ & $3345$       & $\times$   & \checkmark \\ 
      liquid & waterfall, A1    & $293$& $0.5$            & $3600$    & $20$          & $4$       & $195$ & $388$        & \checkmark & $\times$   \\\midrule
      solid  & plastic, A1      & $293$& $2$              & $16000$   & $1$           & $1$       & $195$ & $1200$       & \checkmark & $\times$   \\
      \bottomrule
    \end {tabular}
  \end {center}
  \label{table:TargetComparison}
\end{sidewaystable}
%
% DENSITIES:
% MX: measured.
% DarkLight: given in Lee et al., Table 1: (1.2-10)*10**18 at/cm**2
%OLYMPUS: no specific value in Bernauer et al.; DarkLight paper says: 3.1*10**15 cm**-2 ; Olympus PRL: approx 3*10**15cm**-2 referencing the Bernauer paper...
% PRAD: Xiong et al. says 2*10**18 atoms per cm**2; (DarkLight paper previously said: 9.9*10**17 cm**-2)
% LH(5cm): 0.0699 g/cm**3 -> 0.0699[g/cm**3] * 5 [cm] / 2.01588 [g/mol] * 6.0221409e23 [particles/mol] * 2 [H/molecule]:  0.0699 * 5 / 2.01588     * 6.0221409e23 * 2 = 208815.8e18
% LH(2cm): 0.0699 g/cm**3 -> 0.0699[g/cm**3] * 2 [cm] / 2.01588 [g/mol] * 6.0221409e23 [particles/mol] * 2 [H/molecule]:  0.0699 * 2 / 2.01588     * 6.0221409e23 * 2 = 83526.3e18
% LH(6cm,MUSE): 0.070 g/cm**3 -> %0.070 [g/cm**3] * 6 [cm] / 2.01588 [g/mol] * 6.0221409e23 [particles/mol] * 2 [H/molecule]:  0.070 * 6 / 2.01588     * 6.0221409e23 * 2 = 250937.5e18
% waterfall: Kahrau PhD p 56: 54.2mg/cm**2: 0.0542 [g/cm**2] / 18.01528 [g/mol] * 6.0221409e23 [particles/mol] * 2 [H/molecule]: 0.0542 / 18.01528 * 6.0221409e23 * 2 = 3623.6e18 
% Lupolen: 0.94g/cm**3, 2mm used in ElasticWithPlastic: 0.94 [g/cm**3] * 0.2 [cm] / 28.05 [g/mol C2H4] * 6.0221409e23 [particles/mol] * 4 [H/pattern]: 0.94 * 0.2/ 28.05 * 6.0221409e23 * 4 = 16144.9e18
%
% E-loss MUSE:
% Ignored 2cm FWHM beam spot size, ignored superinsulation foil, assumed pure Kapton windows (ignored glue)

A comparison of the main parameters of hydrogen targets used in
current nuclear physics experiments at electron accelerators such as
Darklight~\cite{Lee:2019},
OLYMPUS~\cite{Bernauer:2014, Henderson:2016},
PRad~\cite{Xiong:2019, Gasparian:2014},
MUSE~\cite{Roy:2019, Gilman:2017},
and A1 at MAMI~\cite{Bernauer:2013, Weber:2017, Ewald:2000, VoeglerFriedrich:1982, Kahrau:1999}
is given in Table~\ref{table:TargetComparison}, an overview of previous
targets can be found in \cite{Ekstrom:1995}.

Liquid hydrogen targets are common. Inside a containing target cell of
several \unit{cm} length, a sizable amount of hydrogen is provided at
cryogenic temperatures. On the one hand, the large areal thickness
allows to reach high luminosities, on the other hand, incoming
electrons as well as outgoing particles can be significantly affected
by the large thickness. An absolute areal density determination on the
percent-level is feasible for such targets by knowledge of the cell
geometry and monitoring the target parameters while taking care that
the beam load does not substantially change the density of the
hydrogen by local heating above the boiling point. However, the target
cell material itself, e.g., a metal or metal alloy, and additional
layers of superinsulation foil can be a significant source of
irreducible background.  Empty cell measurements need to be acquired
for background subtraction.

Liquid or solid state targets consisting of a compound material
including hydrogen, such as waterfalls (H$_2$O) or plastics (CH$_2$),
can be constructed without any walls. The impurities in the form of
nuclei other than hydrogen lead to irreducible background
contributions. In turn, these can be used for a relative cross section
measurement, for instance, normalizing the e-p cross section to the
e-$^{12}$C cross sections, and vice versa.  The absolute areal density
can in principle be determined on a percent-level as well, with
limitations arising from the determination of the target length in
beam direction or the knowledge of the target volume density.  In case
of solid state targets such as plastics, a change of chemical or
physical properties during the experiment could be an issue. For
instance, a physical deformation of the target shape or a variation of
the ratio of hydrogen to other atoms is possible when the target is
heated by the electron beam.

Basic features of gas targets are a small areal thickness, a high
purity, and the possibility for a windowless design. As a result, very
clean data can be obtained, with negligible effects from energy
straggling and multiple scattering, and a minimum of irreducible
background related to target contamination and cell walls.  Because of
the small areal thickness, high beam currents must be provided to
achieve competitive luminosities.  In this aspect, a windowless design
avoids the destructive heating of cell walls and allows for the
recirculation of the beam inside a storage ring or an energy recovery
linac. None of the gas targets listed in
Table~\ref{table:TargetComparison} have a window along the beam
line. The OLYMPUS and the DarkLight targets have side walls, which the
scattered electrons have to pass before detection.  In the PRad setup,
holes inside Kapton foils allow the beam to enter and exit the
hydrogen gas volume without producing background.  Based upon our
understanding, all scattered electrons except those at the smallest
angles have to pass this foil.

For high-precision experiments, it is important to correct for the
most probable energy loss of the incoming electron as well as of the
outgoing electron and other particles to improve the momentum
determination.  A reconstruction of the vertex point increases the
accuracy, since the energy loss depends on the path length in the
target material.  For a precise comparison of measurements to
simulations, the distributions of energy loss, external bremsstrahlung
and multiple scattering need to be modeled. It is evident that
windowless gas targets provide the highest precision.

%############################################################################################################
\section{Summary and outlook}
\label{sec:Summary}
%############################################################################################################
In this paper, the performance of a new cryogenic gas jet target in an
electron accelerator experiment was presented.  The target was
designed to be the centerpiece of the upcoming MAGIX electron
scattering setup at the MESA energy recovery linac. Its windowless
design facilitates a new level of precision. Previous setups have been
limited by systematic effects associated with energy straggling,
multiple scattering, and irreducible backgrounds.

The target was successfully commissioned at the spectrometer setup of
the A1 Collaboration at MAMI, and its performance was studied.  The
size of the gas jet is small, $\sigma \approx \unit[1]{mm}$, which is
ideal for use at high-resolution spectrometer setups, and its areal
thickness, $\rho_{\text{areal}} \geq \unit[10^{18}]{atoms/cm^2}$, is
small enough for beam recirculation, but large enough to provide
competitive luminosities of up to $\mathcal L \approx
\unit[10^{35}]{cm^{-2}s^{-1}}$ with the high beam currents anticipated
at MESA.

A severe issue was found related to the electron beam halo, which can
scatter off parts of the target assembly and produce significant
background contributions. As a solution, a beam halo blocker was
installed, which successfully reduced this background by one order of
magnitude. From measurements with different gas flow rates, the
residual background contribution was determined and a precise
background subtraction was performed.

\begin{figure}[htbp]
  \centering
  \includegraphics[width=0.8\columnwidth]{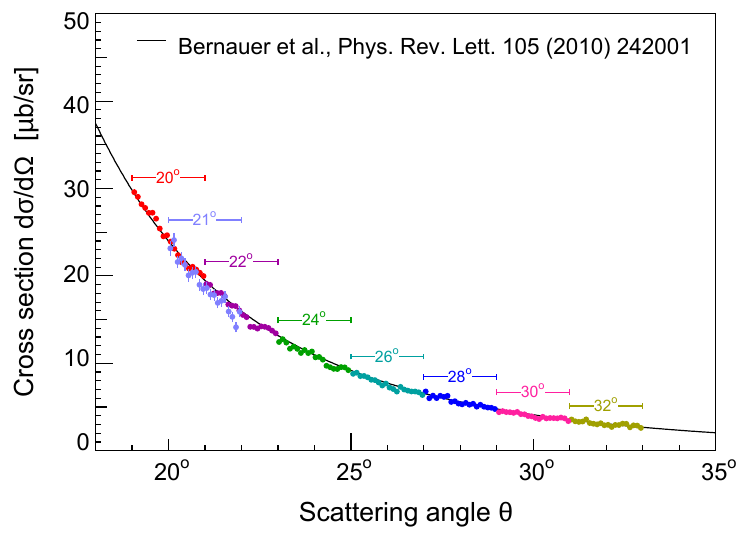}
  %generated: cola@a1analyse ~/jet2018 $ hm -o CS2018.eps CrossSection2018_bss.hm #->inksc@a1cool/work
  \caption{Cross section measurement of elastic e-p scattering during
    commissioning of the gas jet target compared to the data from
    Ref.~\cite{Bernauer:2010}.  The data were collected using eight
    different angular and magnetic field settings of spectrometer~B
    with one common normalization parameter. The central spectrometer
    angle for each of the settings is depicted in the figure.
    Spectrometer~A provided a monitoring of the luminosity.}
  \label{fig:CS2018X}
\end{figure}

Although the main motivation of the target installation at MAMI was a
test of the target system under experimental conditions, high-quality
e-p scattering data have been collected. Figure~\ref{fig:CS2018X} shows
the online analysis of the e-p cross section from the first
commissioning beam time.
The setup and data quality have improved significantly since then, and
settings dedicated to a precise proton form factor measurement have
been determined and the collected data are currently under analysis.

In the future, the consistent windowless design of the MAGIX target
will enable the detection of low-energy recoil particles such as
protons, deuterons, or $\alpha$-particles in detectors inside the
scattering chamber. Measurements with argon as target gas are planned
in the near future. The finalization of a gas recirculation will
further enable the use of expensive gases such as deuterium and
helium, while the use of oxygen is challenging but highly desirable to
study the $^{16}$O$(e,e'\alpha)^{12}$C reaction.  The possibility to
use many different gases makes the jet target ideal for the versatile
physics program with the MAGIX setup at MESA, which will come into
operation in a few years.
%
%############################################################################################################
%bss\input{CRediT}
%############################################################################################################
%
%############################################################################################################
\section*{Acknowledgment}
\addcontentsline{toc}{section}{Acknowledgment}
%############################################################################################################
We gratefully acknowledge the support from the technical staff at the
Mainz Microtron and thank the accelerator group for the excellent beam
quality.

This work was supported in part by 
the PRISMA$^+$ (Precision Physics, Fundamental Interactions and
Structure of Matter) Cluster of Excellence,
the Deutsche Forschungsgemeinschaft (DFG, German Research Foundation)
through the Collaborative Research Center 1044 and the Research
Training Group GRK 2128 AccelencE (Accelerator Science and Technology
for Energy-Recovery Linacs),
the Federal State of Rhineland-Palatinate,
the US National Science Foundation (NSF) grant 2012114,
and the European Union's Horizon 2020 research and innovation
programme, project STRONG2020, under grant agreement No 824093.
%

%############################################################################################################
%------------------------------------------------------------------------------------------------------------
% - - - - - - - - - - - - - - - - - - - - - - - - - - - - - - - - - - - - - - - - - - - - - - - - - - - - - -
%\section*{References} 
\nocite{Edwards:1993} %appears in appendix following the references
\bibliographystyle{elsarticle-num} 
\bibliography{References}

\begin{thebibliography}{10}
\expandafter\ifx\csname url\endcsname\relax
  \def\url#1{\texttt{#1}}\fi
\expandafter\ifx\csname urlprefix\endcsname\relax\def\urlprefix{URL }\fi
\expandafter\ifx\csname href\endcsname\relax
  \def\href#1#2{#2} \def\path#1{#1}\fi

\bibitem{Walecka:2001}
J.~D. Walecka, {Electron scattering for nuclear and nucleon structure}, Vol.~16
  of Camb. Monogr. Part. Phys., Nucl. Phys. Cosmol., Cambridge University
  Press, 2005.
\newblock \href {http://dx.doi.org/10.1017/CBO9780511535017}
  {\path{doi:10.1017/CBO9780511535017}}.

\bibitem{Voitsekhovsky:1979}
B.~B. Voitsekhovsky, D.~M. Nikolenko, S.~G. Popov, D.~K. Toporkov, {Measurement
  of 'radiation tail' - electron spectrum in the reaction $ep \rightarrow
  e'p\gamma$ (in Russian)}, Pisma Zh. Eksp. Teor. Fiz. 29 (1979) 105--109,
  {English version available at
  \url{http://jetpletters.ru/ps/1447/article_22033.shtml}}.

\bibitem{Hug:2016}
F.~Hug, K.~Aulenbacher, R.~G. Heine, B.~Ledroit, D.~Simon, {MESA} --- an {ERL}
  project for particle physics experiments, in: Proc. Linear Accelerator
  Conference (LINAC2016), East Lansing, MI, USA, 25 -- 30 September 2016, 2017,
  pp. 313--315.
\newblock \href {http://dx.doi.org/10.18429/JACoW-LINAC2016-MOP106012}
  {\path{doi:10.18429/JACoW-LINAC2016-MOP106012}}.

\bibitem{Hug:2020}
F.~Hug, K.~Aulenbacher, S.~Friederich, P.~Heil, R.~Heine, R.~Kempf,
  C.~Matejcek, D.~Simon, {Status of the MESA ERL Project}, in: {63rd ICFA
  Advanced Beam Dynamics Workshop on Energy Recovery Linacs}, 2020.
\newblock \href {http://dx.doi.org/10.18429/JACoW-ERL2019-MOCOXBS05}
  {\path{doi:10.18429/JACoW-ERL2019-MOCOXBS05}}.

\bibitem{Caiazza:2020}
S.~S. Caiazza, et~al., {The MAGIX focal plane time projection chamber}, J.
  Phys. Conf. Ser. 1498 (2020) 012022, {Proc. 6th International Conference on
  Micro Pattern Gaseous Detectors (MPGD2019), La Rochelle, France, 5 -- 10 May
  2019}.
\newblock \href {http://dx.doi.org/10.1088/1742-6596/1498/1/012022}
  {\path{doi:10.1088/1742-6596/1498/1/012022}}.

\bibitem{Grieser:2018}
S.~Grieser, D.~Bonaventura, P.~Brand, C.~Hargens, B.~Hetz, L.~Leßmann,
  C.~Westphälinger, A.~Khoukaz, {A cryogenic supersonic jet target for
  electron scattering experiments at MAGIX@MESA and MAMI}, Nucl. Instrum.
  Methods Phys. Res. A 906 (2018) 120--126.
\newblock \href {http://arxiv.org/abs/1806.05409} {\path{arXiv:1806.05409}},
  \href {http://dx.doi.org/10.1016/j.nima.2018.07.076}
  {\path{doi:10.1016/j.nima.2018.07.076}}.

\bibitem{Bernauer:2010}
J.~C. Bernauer, et~al., {High-precision determination of the electric and
  magnetic form factors of the proton}, Phys. Rev. Lett. 105 (2010) 242001.
\newblock \href {http://arxiv.org/abs/1007.5076} {\path{arXiv:1007.5076}},
  \href {http://dx.doi.org/10.1103/PhysRevLett.105.242001}
  {\path{doi:10.1103/PhysRevLett.105.242001}}.

\bibitem{Mihovilovic:2016}
M.~Mihovilovi\v{c}, et~al., {First measurement of proton's charge form factor
  at very low $Q^2$ with initial state radiation}, Phys. Lett. B 771 (2017)
  194--198.
\newblock \href {http://arxiv.org/abs/1612.06707} {\path{arXiv:1612.06707}},
  \href {http://dx.doi.org/10.1016/j.physletb.2017.05.031}
  {\path{doi:10.1016/j.physletb.2017.05.031}}.

\bibitem{Mihovilovic:2021}
M.~Mihovilovi\v{c}, et~al., {The proton charge radius extracted from the
  initial-state radiation experiment at MAMI}, Eur. Phys. J. A 57 (2021) 107.
\newblock \href {http://arxiv.org/abs/1905.11182} {\path{arXiv:1905.11182}},
  \href {http://dx.doi.org/10.1140/epja/s10050-021-00414-x}
  {\path{doi:10.1140/epja/s10050-021-00414-x}}.

\bibitem{Merkel:2011}
H.~Merkel, et~al., {Search for Light Gauge Bosons of the Dark Sector at the
  Mainz Microtron}, Phys. Rev. Lett. 106 (2011) 251802.
\newblock \href {http://arxiv.org/abs/1101.4091} {\path{arXiv:1101.4091}},
  \href {http://dx.doi.org/10.1103/PhysRevLett.106.251802}
  {\path{doi:10.1103/PhysRevLett.106.251802}}.

\bibitem{Merkel:2014}
H.~Merkel, et~al., {Search at the Mainz Microtron for light massive gauge
  bosons relevant for the muon g-2 anomaly}, Phys. Rev. Lett. 112 (2014)
  221802.
\newblock \href {http://arxiv.org/abs/1404.5502} {\path{arXiv:1404.5502}},
  \href {http://dx.doi.org/10.1103/PhysRevLett.112.221802}
  {\path{doi:10.1103/PhysRevLett.112.221802}}.

\bibitem{Doria:2018}
L.~Doria, P.~Achenbach, M.~Christmann, A.~Denig, P.~G{\"u}lker, H.~Merkel,
  Search for light dark matter with the {MESA} accelerator, in: {Proc. 13th
  International Conference on the Intersection of Particle and Nuclear Physics
  (CIPANP18), Palm Springs, CA, USA, 29 May -- 3 June 2018}, to appear in
  eConf, 2018.
\newblock \href {http://arxiv.org/abs/1809.07168} {\path{arXiv:1809.07168}}.

\bibitem{Doria:2019}
L.~Doria, P.~Achenbach, M.~Christmann, A.~Denig, H.~Merkel, {Dark matter at the
  intensity frontier: The new MESA electron accelerator facility}, in: {Proc.
  An Alpine LHC Physics Summit 2019 (ALPS 2019), Obergurgl, Austria, 22 -- 27
  April 2019}, Vol. 360 of PoS Proc. Sci. (ALPS2019), 2021, p.~22.
\newblock \href {http://dx.doi.org/10.22323/1.360.0022}
  {\path{doi:10.22323/1.360.0022}}.

\bibitem{Friscic:2019}
I.~Fri{\v s}{\v c}i{\'c}, T.~W. Donnelly, R.~G. Milner, {New approach to
  determining radiative capture reaction rates at astrophysical energies},
  Phys. Rev. C 100 (2019) 025804.
\newblock \href {http://arxiv.org/abs/1904.05819} {\path{arXiv:1904.05819}},
  \href {http://dx.doi.org/10.1103/PhysRevC.100.025804}
  {\path{doi:10.1103/PhysRevC.100.025804}}.

\bibitem{Holt:2019}
R.~J. Holt, B.~W. Filippone, {Impact of $^{16}$O($e,e'\alpha$)$^{12}$C
  measurements on the $^{12}$C($\alpha,\gamma$)$^{16}$O astrophysical reaction
  rate}, Phys. Rev. C 100 (2019) 065802.
\newblock \href {http://arxiv.org/abs/1908.11407} {\path{arXiv:1908.11407}},
  \href {http://dx.doi.org/10.1103/PhysRevC.100.065802}
  {\path{doi:10.1103/PhysRevC.100.065802}}.

\bibitem{Herminghaus:1976}
H.~Herminghaus, A.~Feder, K.-H. Kaiser, W.~Manz, H.~von~der Schmitt, {The
  design of a cascaded 800 MeV normal conducting CW racetrack microtron}, Nucl.
  Instrum. Methods 138 (1976) 1--12.
\newblock \href {http://dx.doi.org/10.1016/0029-554X(76)90145-2}
  {\path{doi:10.1016/0029-554X(76)90145-2}}.

\bibitem{Kaiser:2008}
K.-H. Kaiser, et~al., {The 1.5 GeV harmonic double-sided microtron at Mainz
  University}, Nucl. Instrum. Methods Phys. Res. A 593 (2008) 159--170.
\newblock \href {http://dx.doi.org/10.1016/j.nima.2008.05.018}
  {\path{doi:10.1016/j.nima.2008.05.018}}.

\bibitem{Dehn:2011}
M.~Dehn, K.~Aulenbacher, R.~Heine, H.-J. Kreidel, U.~Ludwig-Mertin,
  A.~Jankowiak, {The MAMI C accelerator: The beauty of normal conducting
  multi-turn recirculators}, Eur. Phys. J. ST 198 (2011) 19--47.
\newblock \href {http://dx.doi.org/10.1140/epjst/e2011-01481-4}
  {\path{doi:10.1140/epjst/e2011-01481-4}}.

\bibitem{Dehn:2016}
M.~Dehn, K.~Aulenbacher, H.-J. Kreidel, F.~Nillius, B.~S. Schlimme,
  V.~Tioukine, {Recent challenges for the 1.5 GeV MAMI-C accelerator at JGU
  Mainz}, in: {Proc. 7th International Particle Accelerator Conference (IPAC
  2016), Busan, Korea, 8 -- 13 May 2016}, 2016, pp. 4149--4151, {THPOY026}.
\newblock \href {http://dx.doi.org/10.18429/JACoW-IPAC2016-THPOY026}
  {\path{doi:10.18429/JACoW-IPAC2016-THPOY026}}.

\bibitem{Blomqvist:1998}
K.~I. Blomqvist, et~al., {The three-spectrometer facility at the Mainz
  microtron MAMI}, Nucl. Instrum. Meth. Phys. Res. A 403 (1998) 263--301.
\newblock \href {http://dx.doi.org/10.1016/S0168-9002(97)01133-9}
  {\path{doi:10.1016/S0168-9002(97)01133-9}}.

\bibitem{Wilhelm:1993}
W.~Wilhelm, {Entwicklung eines schnellen Elektronenstrahlwedelsystems mit
  Positionsr\"uckmeldung zur Verringerung der lokalen Aufheizung von
  Tieftemperaturtargets}, Diploma thesis, Mainz U., Inst. Kernphys. (1993).

\bibitem{Taeschner:2013}
A.~Täschner, E.~Köhler, H.-W. Ortjohann, A.~Khoukaz, {Determination of
  hydrogen cluster velocities and comparison with numerical calculations}, J.
  Chem. Phys. 139 (2013) 234312.
\newblock \href {http://dx.doi.org/10.1063/1.4848720}
  {\path{doi:10.1063/1.4848720}}.

\bibitem{Pauly:2000}
H.~Pauly, Atom, Molecule, and Cluster Beams I, Springer-Verlag Berlin
  Heidelberg, 2000.
\newblock \href {http://dx.doi.org/10.1007/978-3-662-04213-7}
  {\path{doi:10.1007/978-3-662-04213-7}}.

\bibitem{Brauksiepe:1996}
S.~Brauksiepe, et~al., {COSY-11, an internal experimental facility for
  threshold measurements}, {Nucl. Instrum. Methods Phys. Res. A} 376 (1996)
  397--410.
\newblock \href {http://dx.doi.org/10.1016/0168-9002(96)00080-0}
  {\path{doi:10.1016/0168-9002(96)00080-0}}.

\bibitem{Dombrowski:1997}
H.~Dombrowski, D.~Grzonka, W.~Hamsink, A.~Khoukaz, T.~Lister, R.~Santo, {The
  Münster cluster target for the COSY-11 experiment}, {Nucl. Phys. A} 626
  (1997) 427--433, {Proc. Third International Conference on Nuclear Physics at
  Storage Rings (STORI96), Bernkastel-Kues, Germany, 30 September -- 4 October
  1996}.
\newblock \href {http://dx.doi.org/10.1016/S0375-9474(97)00565-4}
  {\path{doi:10.1016/S0375-9474(97)00565-4}}.

\bibitem{Taschner:2011}
A.~Täschner, E.~Köhler, H.-W. Ortjohann, A.~Khoukaz, {High density cluster
  jet target for storage ring experiments}, {Nucl. Instrum. Methods Phys. Res.
  A} 660 (2011) 22--30.
\newblock \href {http://dx.doi.org/10.1016/j.nima.2011.09.024}
  {\path{doi:10.1016/j.nima.2011.09.024}}.

\bibitem{PANDA-Target-TDR:2012}
A.~Khoukaz, et~al., {Technical Design Report for the PANDA internal targets},
  {available at
  \url{https://fair-center.eu/fileadmin/fair/publications_exp/PANDA_Targets_TDR.pdf}}
  (2012).

\bibitem{Grieser:2019}
S.~Grieser, et~al., {Nm-sized cryogenic hydrogen clusters for a laser-driven
  proton source}, {Rev. Sci. Instrum.} 90 (2019) 043301.
\newblock \href {http://dx.doi.org/10.1063/1.5080011}
  {\path{doi:10.1063/1.5080011}}.

\bibitem{EPICS}
{EPICS: Experimental Physics and Industrial Control System}, available at
  \url{https://epics-controls.org/} (accessed 2020).

\bibitem{CSS}
{CSS: Control System Studio}, available at
  \url{http://controlsystemstudio.org/} (accessed 2020).

\bibitem{StreamDevice}
{StreamDevice}, available at
  \url{http://epics.web.psi.ch/software/streamdevice/} (accessed 2020).

\bibitem{ASYN}
asyndriver: Asynchronous driver support, available at
  \url{https://epics-modules.github.io/master/asyn/} (accessed 2020).

\bibitem{TI-ADS1248}
{Texas Instruments Inc.}, {ADS124x 24-Bit, 2-kSPS, Analog-To-Digital Converters
  With Programmable Gain Amplifier (PGA) For Sensor Measurement}, {available at
  \url{https://www.ti.com/lit/ds/symlink/ads1248.pdf}} (2016).

\bibitem{Brand:2019}
P.~Brand, {Investigation on the elastic ep scattering at MAMI and simulation of
  gas flows through different jet nozzle geometries}, Master's thesis,
  {M\"unster U., Inst. Kernphys.} (2019).

\bibitem{Greenshields:2009}
C.~J. Greenshields, H.~G. Weller, L.~Gasparini, J.~Reese, {Implementation of
  semi-discrete, non-staggered central schemes in a colocated, polyhedral,
  finite volume framework, for high-speed viscous flows}, {Int. J. Numer. Meth.
  Fluids} 63 (2009) 1--21.
\newblock \href {http://dx.doi.org/10.1002/fld.2069}
  {\path{doi:10.1002/fld.2069}}.

\bibitem{Peng:1976}
D.-Y. Peng, D.~B. Robinson, {A New Two-Constant Equation of State}, {Ind. Eng.
  Chem. Fundam.} 15 (1976) 59--64.
\newblock \href {http://dx.doi.org/10.1021/i160057a011}
  {\path{doi:10.1021/i160057a011}}.

\bibitem{Bernauer:2013}
J.~C. Bernauer, et~al., {Electric and magnetic form factors of the proton},
  Phys. Rev. C 90 (2014) 015206.
\newblock \href {http://arxiv.org/abs/1307.6227} {\path{arXiv:1307.6227}},
  \href {http://dx.doi.org/10.1103/PhysRevC.90.015206}
  {\path{doi:10.1103/PhysRevC.90.015206}}.

\bibitem{ESTAR}
M.~J. Berger, J.~S. Coursey, M.~A. Zucker, J.~Chang, ESTAR, PSTAR, and ASTAR:
  Computer Programs for Calculating Stopping-Power and Range Tables for
  Electrons, Protons, and Helium Ions, {available at
  \url{http://physics.nist.gov/Star}} (2020).
\newblock \href {http://dx.doi.org/10.18434/T4NC7P}
  {\path{doi:10.18434/T4NC7P}}.

\bibitem{Lee:2019}
S.~Lee, et~al., {Design and operation of a windowless gas target internal to a
  solenoidal magnet for use with a megawatt electron beam}, Nucl. Instrum.
  Methods Phys. Rev. A 939 (2019) 46--54.
\newblock \href {http://arxiv.org/abs/1903.02648} {\path{arXiv:1903.02648}},
  \href {http://dx.doi.org/10.1016/j.nima.2019.05.071}
  {\path{doi:10.1016/j.nima.2019.05.071}}.

\bibitem{Bernauer:2014}
J.~C. Bernauer, V.~Carassiti, G.~Ciullo, B.~S. Henderson, E.~Ihloff, J.~Kelsey,
  P.~Lenisa, R.~Milner, A.~Schmidt, M.~Statera, {The OLYMPUS internal hydrogen
  target}, Nucl. Instrum. Methods Phys. Res. A 755 (2014) 20--27.
\newblock \href {http://arxiv.org/abs/1404.0579} {\path{arXiv:1404.0579}},
  \href {http://dx.doi.org/10.1016/j.nima.2014.04.029}
  {\path{doi:10.1016/j.nima.2014.04.029}}.

\bibitem{Henderson:2016}
B.~S. Henderson, et~al., {Hard two-photon contribution to elastic lepton-proton
  scattering: determined by the OLYMPUS Experiment}, Phys. Rev. Lett. 118
  (2017) 092501.
\newblock \href {http://arxiv.org/abs/1611.04685} {\path{arXiv:1611.04685}},
  \href {http://dx.doi.org/10.1103/PhysRevLett.118.092501}
  {\path{doi:10.1103/PhysRevLett.118.092501}}.

\bibitem{Xiong:2019}
W.~Xiong, et~al., {A small proton charge radius from an electron--proton
  scattering experiment}, Nature 575 (2019) 147--150.
\newblock \href {http://dx.doi.org/10.1038/s41586-019-1721-2}
  {\path{doi:10.1038/s41586-019-1721-2}}.

\bibitem{Gasparian:2014}
A.~Gasparian, {The PRad experiment and the proton radius puzzle}, EPJ Web Conf.
  73 (2014) 07006.
\newblock \href {http://dx.doi.org/10.1051/epjconf/20147307006}
  {\path{doi:10.1051/epjconf/20147307006}}.

\bibitem{Roy:2019}
P.~Roy, et~al., {A Liquid Hydrogen Target for the MUSE Experiment at PSI},
  Nucl. Instrum. Meth. A 949 (2020) 162874.
\newblock \href {http://arxiv.org/abs/1907.03022} {\path{arXiv:1907.03022}},
  \href {http://dx.doi.org/10.1016/j.nima.2019.162874}
  {\path{doi:10.1016/j.nima.2019.162874}}.

\bibitem{Gilman:2017}
R.~Gilman, et~al., {Technical Design Report for the Paul Scherrer Institute
  Experiment R-12-01.1: Studying the Proton ''Radius'' Puzzle with $\mu p$
  Elastic Scattering}\href {http://arxiv.org/abs/1709.09753}
  {\path{arXiv:1709.09753}}.

\bibitem{Weber:2017}
A.~B. Weber, {First determination of the proton electric form factor at very
  small momentum transfer using initial state radiation}, Ph.D. thesis, Mainz
  U., Inst. Kernphys. (2017).

\bibitem{Ewald:2000}
I.~Ewald, {Koh\"arente Elektroproduktion von neutralen Pionen am Deuteron nahe
  der Schwelle}, Ph.D. thesis, Mainz U., Inst. Kernphys. (2000).

\bibitem{VoeglerFriedrich:1982}
N.~Voegler, J.~Friedrich, {A background-free oxygen target for electron
  scattering measurements with high beam currents}, Nucl. Instrum. Methods
  Phys. Res. 198 (1982) 293--297.
\newblock \href {http://dx.doi.org/10.1016/0167-5087(82)90266-6}
  {\path{doi:10.1016/0167-5087(82)90266-6}}.

\bibitem{Kahrau:1999}
M.~Kahrau, {Untersuchung von Nukleon-Nukleon Korrelationen mit Hilfe der
  Reaktion $^{16}$O$(e,e'pp)^{14}$C in super-paralleler Kinematik}, Ph.D.
  thesis, Mainz U., Inst. Kernphys. (1999).

\bibitem{Ekstrom:1995}
C.~Ekstrom, {Internal targets: A Review}, Nucl. Instrum. Meth. A 362 (1995)
  1--15.
\newblock \href {http://dx.doi.org/10.1016/0168-9002(95)00240-5}
  {\path{doi:10.1016/0168-9002(95)00240-5}}.

\bibitem{Edwards:1993}
L.~G. Edwards, M.~Haberbusch, {Temperature and pressure effects on capacitance
  probe cryogenic liquid level measurement accuracy}, {Contractor Report}
  190763, {NASA}, available at
  \url{https://ntrs.nasa.gov/citations/19940012048} (1993).

\end{thebibliography}
\addcontentsline{toc}{section}{References}
%############################################################################################################
~\\[0.1cm]
\appendix
\addcontentsline{toc}{section}{Appendix}
%\newpage
\noindent\textcolor{red}{These appendices are not included in the submitted version.}
%############################################################################################################
%
%############################################################################################################
\section{Background subtraction}
\label{sec:BGS}
%############################################################################################################
%------------------------------------------------------------------------------------------------------------
\subsection{Example spectra}
\begin{figure}[t]
  \centering
  \includegraphics[width=\columnwidth]{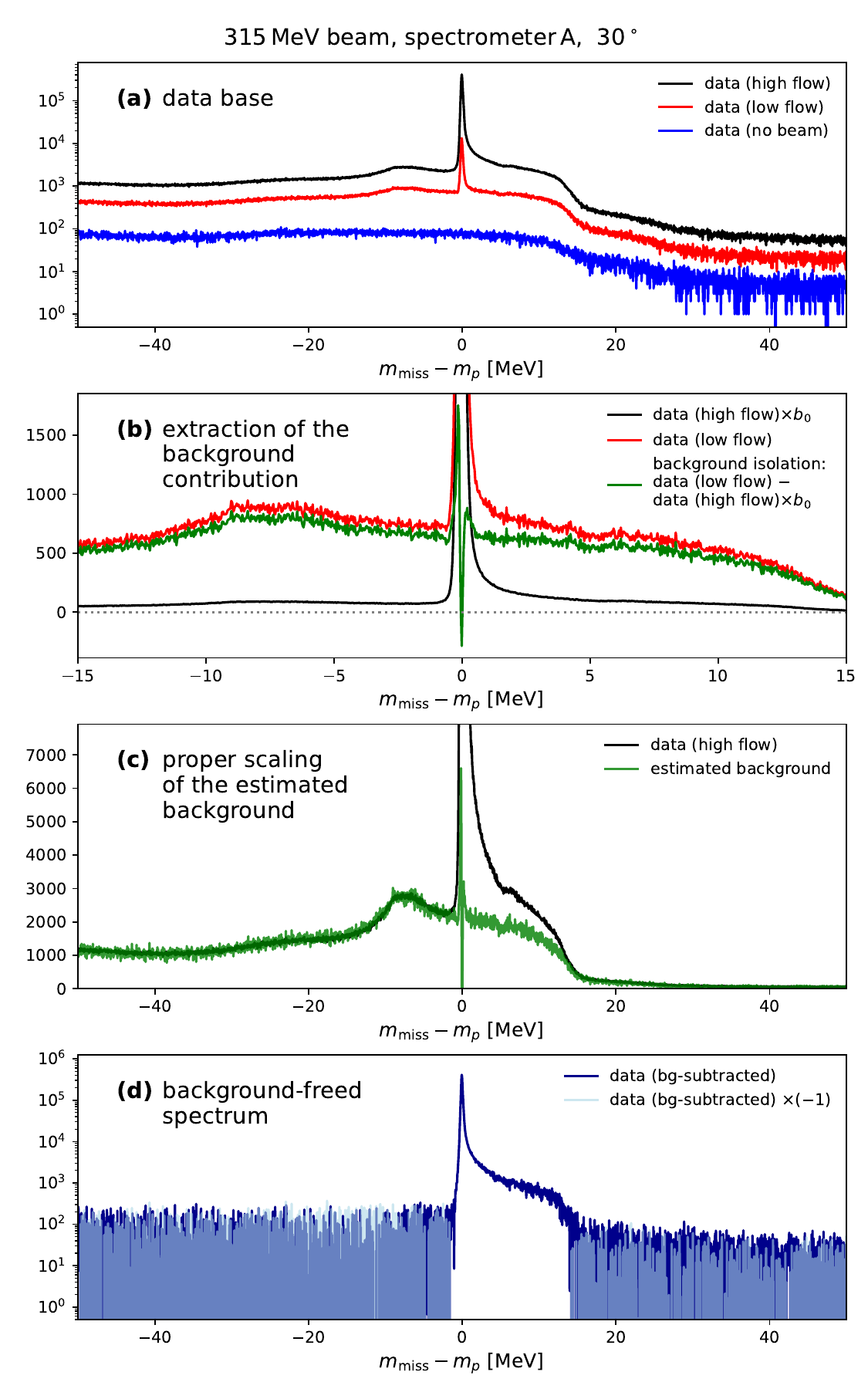}
  %generated: cola@a1analyse ~/jet2019/ana $ python 2020_BGS.py
  %no analysis cuts are applied. 
  \caption{Missing mass spectra taken for elastic e-p form factor
    measurements.  (a) Measured counts in the individual data
    sets. (b) Extraction of the beam related background spectrum. (c)
    Full background spectrum including beam related and cosmogenic
    background. (d) Background subtracted elastic e-p scattering
    spectrum. The inverted spectrum is shown in addition to visualize
    the scattering around zero aside the elastic peak in the
    logarithmic scale.  }
  \label{fig:2020_BGS_B40_A_NoCuts_Overview}
\end{figure}

%------------------------------------------------------------------------------------------------------------
Data for the elastic e-p cross section determination were taken with a
high gas flow rate of $q_V = \unit[1200]{l_n/h}$. Additional data were
taken with a low gas flow rate of $q_V= \unit[50]{l_n/h}$ and without
beam for background subtraction, see
Fig.~\ref{fig:2020_BGS_B40_A_NoCuts_Overview}\,(a).

The approximate shape of the background was deduced from the low flow
spectrum which includes a small elastic e-p peak on top of the
background. From this data the high flow spectrum was subtracted after
appropriate scaling, see
Fig.~\ref{fig:2020_BGS_B40_A_NoCuts_Overview}\,(b).  In addition to
beam related background, also cosmogenic background contributes to the
spectrum, its fraction varies for instance with the beam current and
with the data acquisition dead time.  To accommodate, we include data
measured without beam to the background model.  In
Fig.~\ref{fig:2020_BGS_B40_A_NoCuts_Overview}\,(c), the beam related
background spectrum and the cosmogenic background spectrum were scaled
and summed to match the high flow data in the background region of the
spectrum.  The background free spectrum was obtained from the high
flow data by subtraction of the full background contribution. A sharp
elastic e-p scattering peak with radiative tail remains, see
Fig.~\ref{fig:2020_BGS_B40_A_NoCuts_Overview}\,(d).
%
%------------------------------------------------------------------------------------------------------------
\subsection{Error contribution related to background subtraction}
%------------------------------------------------------------------------------------------------------------
To control target-related uncertainties at a sub-percent level,
measurements at a high gas flow rate must be accompanied by an
appropriate amount of data taken at a low gas flow rate. To estimate
the best fraction of high to low flow settings for a minimal
uncertainty, one assumes a constant beam current, constant detector
and data acquisition dead times, a pure background spectrum in the low
flow spectrum, and no cosmogenic background. Furthermore, all event
counts are assumed to be large so that the statistical uncertainty can
be estimated by the square root of the event count.

With $\dot S_{\text{high}}$ as the expected rate of signal events
measured in a high flow setting, the number of expected signal events
within a measurement time of $t_{\text{high}}$ is
$S_{\text{high}}=S_{\text{high}}(t_{\text{high}})=\dot
S_{\text{high}}\cdot t_{\text{high}}$ with a statistical uncertainty
of
\begin{equation}
    \Delta S_{\text{high}}=\sqrt{S_{\text{high}}}.
\end{equation}
If the fraction of background to signal events is $b_{\text{high}}$ in
the high flow setting, a number of $B_{\text{high}}=\dot
B_{\text{high}}\cdot t_{\text{high}}=b_{\text{high}}\cdot
S_{\text{high}}$ background events is expected in the very same
measurement.  The low flow setting provides the estimate
$B_{\text{high}}^{\text{est.}}$. This requires either a precisely
monitored beam current, so that the scale between the low flow and the
high flow setting is well known, or a precise scaling by comparison of
the background dominated part of the spectra. The background
subtracted number of signal events $S_{\text{high}}^{\text{bgs}}$ is:
\begin{equation}
    S_{\text{high}}^{\text{bgs}} \approx \underbrace{S_{\text{high}} + B_{\text{high}}}_{\text{high\ flow\ settings}} - \underbrace{B_{\text{high}}^{\text{est.}}}_{\text{low\ flow\ settings}}.
\end{equation}
Accordingly, the uncertainty of the estimate
\begin{equation}
    B_{\text{high}}^{\text{est.}} = B_{\text{low}}(t_{\text{low}}) \cdot \frac{t_{\text{high}}}{t_{\text{low}}}
\end{equation}
contributes to the total error. It is related to the statistical
precision of $B_{\text{low}}$. In addition, $B_{\text{high}}$
contributes the statistical uncertainty
\begin{equation}
    \Delta B_{\text{high}} = \sqrt{b_{\text{high}}\cdot S_{\text{high}}},
\end{equation}
which can not be reduced by supplemental background measurements.

\begin{figure}[htbp]
  \centering
  \includegraphics[width=\columnwidth]{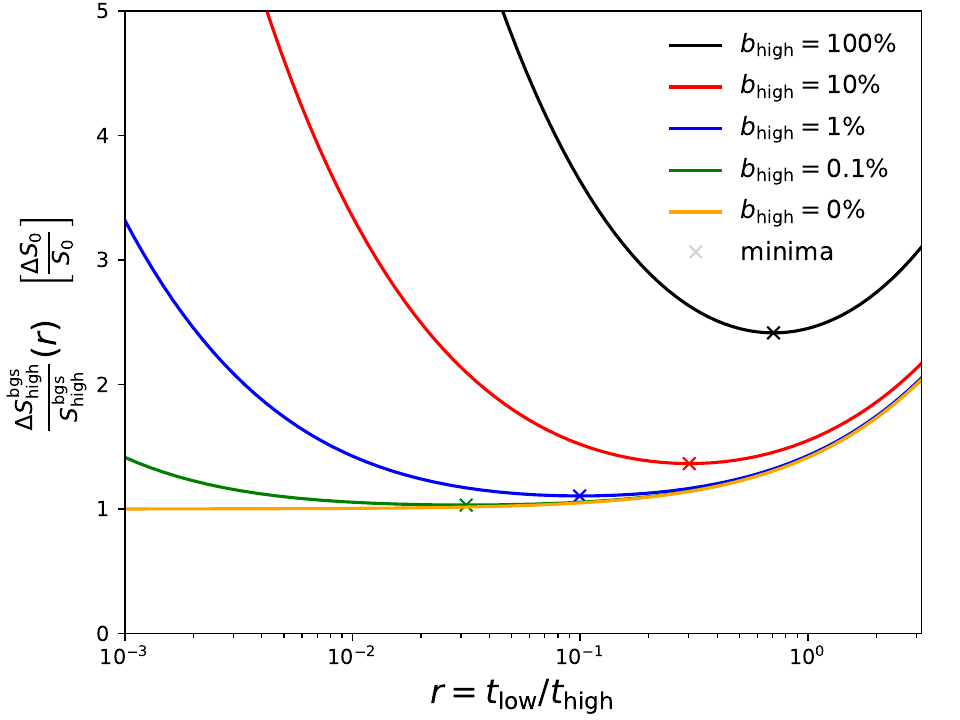}
  %generated: python work/bg_error.py
  \caption{The relative uncertainty of the signal as a function of the
    fraction of time spent for measuring at high and low flow settings
    shown for different background to signal fractions
    $b_{\text{high}}=B_{\text{high}}/S_{\text{high}}$ in the high flow
    data.}
  \label{fig:bgs_rel_error_of_r}
\end{figure}

If for an available measurement time
$t_{\text{total}}=t_{\text{high}}+t_{\text{low}}$ the same amount of
time is used for the different gas flow settings, the two
contributions of the background error are equal,
\begin{equation}
    \Delta B_{\text{high}}^{\text{est.}} \approx \Delta B_{\text{high}}, \quad \textnormal{for } t_{\text{high}} = t_{\text{low}}.
\end{equation}
The minimal error for $S_{\text{high}}^{\text{bgs}}$, however, can be
achieved for a value of
\begin{equation}
    r = \frac{t_{\text{low}}}{t_{\text{high}}},
\end{equation}
which depends on the background to signal ratio $b_{\text{high}}$.
If, for instance, the background contribution is negligible, a smaller
$r$ will provide a smaller total error.  In the hypothetical case of a
background-free measurement, one could use all available time for the
high flow setting.  With this special case as a reference,
\begin{equation}
    S_0 = S_{\text{high}}(t=t_{\text{total}}), \quad\Delta S_0 = \sqrt{S_0},
\end{equation}
and using
\begin{equation}
    t_{\text{low}}  = t_{\text{total}}\cdot \frac{1}{1+1/r}, \quad t_{\text{high}} = t_{\text{total}} \cdot \frac{1}{1+r},
\end{equation}
it follows:
\begin{flalign}
&\frac{\Delta S_{\text{high}}^{\text{bgs}}}{S_{\text{high}}^{\text{bgs}}}(r)\approx \frac{\Delta S_{\text{high}}^{\text{bgs}}(r)}{S_{\text{high}}(t_{\text{high}})}\nonumber\\
&\quad \approx \frac{\sqrt{
    (\Delta S_{\text{high}}(t_{\text{high}}))^2 +
    (\Delta B_{\text{high}}(t_{\text{high}}))^2 +
    (\Delta B_{\text{low}}(t_{\text{low}})/r)^2}
}{S_0\cdot \frac{t_{\text{high}}}{t_{\text{total}}}}\nonumber\\
&\quad  = \frac{\sqrt{ 
     \Delta S_0^2 \cdot \frac{1}{1+r}
  +  \Delta S_0^2 \cdot \frac{1}{1+r} \cdot b_{\text{high}}
  +  \Delta S_0^2 \cdot \frac{1}{1+1/r}\cdot \frac{1}{r^2} \cdot b_{\text{high}}}
}{S_0\cdot \frac{1}{1+r}}\nonumber\\
%%%&= \frac{\Delta S_0}{S_0} \cdot \sqrt{ 
%%%     (1+r)
%%%  +  (1+r) \cdot b_{\text{high}}
%%%  +  (1+r)\cdot \nicefrac{1}{r} \cdot b_{\text{high}}}\nonumber\\
&\quad = \frac{\Delta S_0}{S_0} \cdot \sqrt{1+2b_{\text{high}}+r(1+b_{\text{high}})+b_{\text{high}}/r}
\end{flalign}
The relative error with respect to the relative error for a
background-free measurement is shown in
Fig.~\ref{fig:bgs_rel_error_of_r} as a function of $r$ for different
background fractions $b_{\text{high}}$. It is minimized at an optimal
data taking ratio of
\begin{equation}
  r_{\text{opt}} = \sqrt{\frac{b_{\text{high}}}{1+b_{\text{high}}}}.
\end{equation}

\begin{figure}[htbp]
  \centering
  \includegraphics[width=\columnwidth]{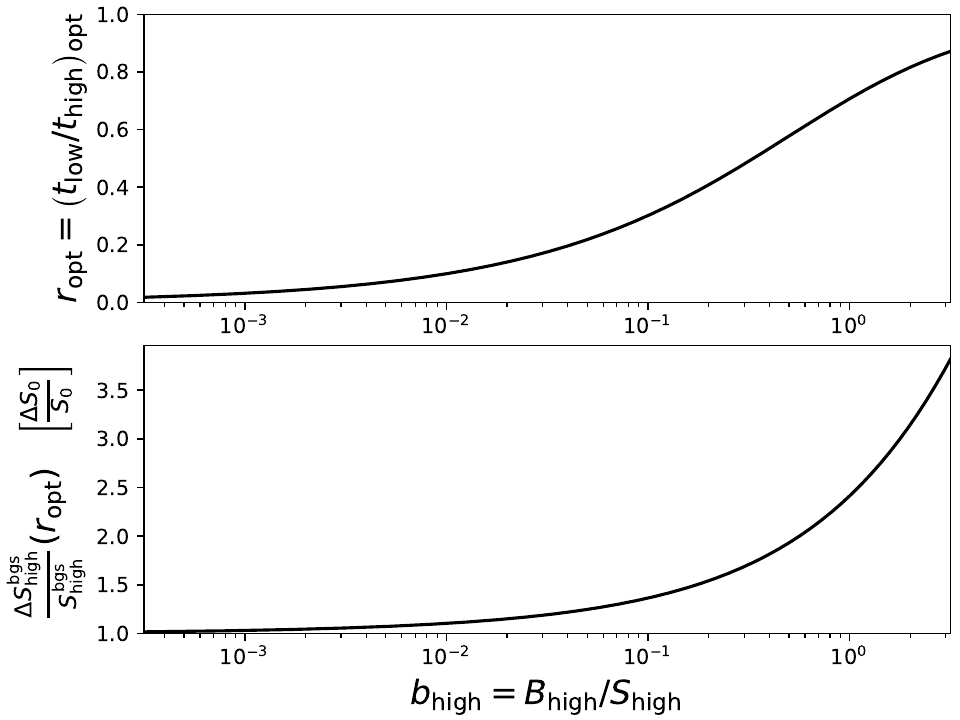}
  %generated: python work/bg_error.py
  \caption{Top: Optimal ratio of measurement times for the low and high flow settings. Bottom: Relative error with respect to the hypothetical error for a background free measurement during the available time $t_{\text{total}}$.}
  \label{fig:bgs_rel_error_of_b_and_r_opt.pdf}
\end{figure}

Fig.~\ref{fig:bgs_rel_error_of_b_and_r_opt.pdf} shows $r_{\text{opt}}$
as a function of $b_{\text{high}}$ as well as the relative error of
the measurement signal after background subtraction for the optimal
$r_{\text{opt}}$.
%
% Example: cola@a1analyse ~/jet2019/ana $ python extract_2020_BGS.py
% Cuts: CC & -2MeV < MM < 15MeV
%  S+B= 3366522
%    B= 210897.1
%=>  S= 3155624.9
%=>B/S= 0.0668

For the data presented in
Fig.~\ref{fig:2020_BGS_B40_A_NoCuts_Overview} with a cut on the
Cherenkov detector signals to remove the cosmogenic background and a
cut on the missing mass region from $\unit[-2]{MeV}$ to
$\unit[15]{MeV}$ to select the elastic e-p scattering peak, the
fraction of background events below the elastic peak is
\unit[6.7]{\%}.  From the previous considerations follows
$r_{\text{opt}} = \unit[25]{\%}$, i.e., the time spent for the low
flow setting with respect to the high flow setting should be $1:4$ in
order to achieve a minimal error.
%
%############################################################################################################
\section{Nitrogen level meter of the booster stage}
\label{sec:levelmeter}
%############################################################################################################
\begin{figure}[b!]
  \centering
  \includegraphics[width=\columnwidth]{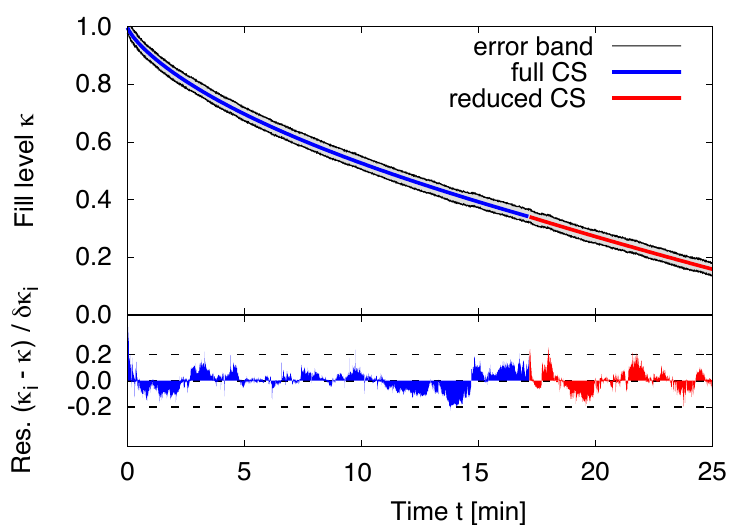}
  \caption{Typical time evolution of the liquid nitrogen (LN) fill
    level after automated refill. Upper panel: The evaporation of LN
    for precooling of the target gas evolves linearly in time. The
    cooling of the initially warmed dewar causes an additional LN
    consumption following an exponential law. Below a fill level of
    $\kappa = \unit[34]{\%}$, the heat exchanger is mounted which
    reduces the effective cross section (CS) of the LN volume. This
    results in a faster decrease of the level when assuming a constant
    evaporation per unit volume. Lower panel: The residual $\kappa_i -
    \kappa$ of a single measurement $i$ is below \unit[20]{\%} of the
    Poisson distributed counting error $\delta \kappa_i$.}
  \label{fig:level-meter}
\end{figure}
The nitrogen level meter monitors the fill level of the booster stage
and enables an automated refill during operation of the gas jet
target. It consists of two concentric stainless steel tubes forming a
cylindrical capacitor by enclosing a sensitive volume in the
interspace. Liquid nitrogen in the sensitive volume modifies the
dielectric, since its relative permittivity $\epsilon =
1.425$~\cite{Edwards:1993} is larger than that of gaseous nitrogen
$\epsilon\approx 1$. The sensitive volume can be reached by the liquid
only through small holes at the bottom. These holes are acting as a
flow impedance that limits the hydrodynamic coupling to fluctuations
during the refill process. The pressure inside the sensitive volume is
equalized through an opening at the top, that is shielded by amorphous
SiO$_2$ against moisture.

The meter capacitance is converted to an equivalent frequency by an
oscillator circuit, that cyclically charges and discharges the
cylindrical capacitor through a resistor. A symmetric oscillator with
a precision reference capacitor is thermally coupled to the main
oscillator. Both frequencies are measured by fast synchronized 32
bit-counters with an adjustable acquisition time. The fill level is
determined from the ratio of both counters, in which the dependence on
the acquisition time cancels and the dependence on oscillator specific
parameters is weak.

The statistical accuracy is optimized by oscillator frequencies in the
MHz range, low-capacitance connection cables, and signal filtering
with an adaptive-sized moving average.  In combination with the flow
impedance as an hydrodynamic low pass filter, this results in an
excellent fill level resolution, see Fig.~\ref{fig:level-meter}. The
hardware drivers are running on a Raspberry Pi and are interlinked
with the EPICS database, c.f.\ Subsect.~\ref{subsec:SlowControl}.

%############################################################################################################
\end{document}